\documentstyle[11pt,aaspp4]{article}
\def\etal{{et al.}}
\def\asca{{\it ASCA}}

\def\ex{{\it EXOSAT}}
\def\gi{{\it Ginga}}

\def\he{{\it HEAO-1}}

\def\pcmsq{{$\rm cm^{-2}$}}
\def\chisq{{$\chi^{2}$}}

\def\delchi{{$\Delta \chi^{2}$}}

\def\nh{{$N_{\rm H}$}}

\def\mdot{{$\dot{M}$}}
\def\deg{\hbox{$^{\circ}$}}
\def\Msun{\hbox{$\rm\thinspace M_{\odot}$}}
\begin{document}

\title{ASCA observations of Seyfert 1 galaxies: \\ II. Relativistic 
Iron K$\alpha$ emission}

\author {K. Nandra\altaffilmark{1,2}, 
I.M. George \altaffilmark{1, 3}, R.F. Mushotzky\altaffilmark{1}, 
T.J. Turner \altaffilmark{1, 3}, T. Yaqoob\altaffilmark{1, 3}}

\altaffiltext{1}{Laboratory for High Energy Astrophysics, Code 660,
	NASA/Goddard Space Flight Center,
  	Greenbelt, MD 20771}
\altaffiltext{2}{NAS/NRC Research Associate}
\altaffiltext{3}{Universities Space Research Association}

\slugcomment{Submitted to {\em The Astrophysical Journal}}

\begin{abstract}

We present evidence for widespread relativistic effects in the central
regions of active galactic nuclei. In a sample of 18 Seyfert 1
galaxies observed by \asca, 14 show an iron K$\alpha$ line which is is
resolved, with mean width $\sigma_{\rm K\alpha}=0.43\pm 0.12$~keV for
a gaussian profile (Full Width at Half Maximum, FWHM$\sim 50,000$~km
s$^{-1}$).  However, many of the line profiles are asymmetric.  A
strong red wing is indicative of gravitational redshifts close to a
central black hole and accretion disk models provide an excellent
description of the data.

The peak energy of the line is 6.4~keV, indicating that it arises by
fluorescence in near-neutral material.  Our fits imply a low
inclination for the disk in these Seyfert 1 galaxies, with a mean of
30\deg, consistent with orientation-dependent unification schemes.
Differences in the line profiles from source-to-source imply slight
variations in geometry, which cannot be accounted for solely by
inclination.  In most cases, we require that the line emission arises
from a range of radii.  Although a small contribution to the emission
from a region other than the disk is not ruled out, it is not
generally required and has little effect on our conclusions regarding
the disk line.  Our data are fit equally well with rotating (Kerr) and
non-rotating (Schwarzschild) black hole models.  We find a mean
spectral index in the 3-10 keV range of $<\Gamma_{3-10}>=1.91\pm 0.07$
after accounting for the effects of reflection.

Such observations probe the innermost regions of AGN,
and arguably provide the best evidence yet obtained for the existence
of super-massive black holes in the centers of active galaxies.

\end{abstract}


\section{Introduction}

\label{sec:intro}

The X-ray spectra of Seyfert 1 galaxies can contain numerous components.
The continuum has an apparent power law form, which covers several
decades in energy. This is attenuated at soft X-ray energies by
absorption both in the interstellar medium in our Galaxy
and by material local to the active galaxy. 
An emission line from iron K$\alpha$ has 
been observed at 6-7 keV in the vast majority of sources.
A hard component affects the spectrum
above $\sim 10$~keV which is probably produced by Compton
scattering (``reflection'') of the continuum in the material 
which produces the line. In the soft X-ray regime, many sources
have ``soft excess'' emission and can show emission features
from oxygen and/or iron-L.

At least three nuclear emission/absorption regions are envisaged to
account for these components. The region which emits the primary 
X-ray continuum may consist
of hot and/or relativistic electrons and/or electron-positron pairs which
Compton up-scatter softer ``seed'' photons.
Compton-thick material in a low state of ionization, which
may be identified with an accretion disk, gives rise to the iron
K$\alpha$ line, the reflection continuum and possibly the ``soft excess''.
In many cases, the local absorption is due
to a Compton-thin ``warm absorber'': highly-ionized gas 
identified by features due to oxygen and iron,
which may also produce the soft X-ray lines.

Detailed measurement of the various emission and absorption features,
and continua, can therefore yield information regarding the physical
conditions of the regions, such their dynamics, geometry and
ionization state.  Of particular interest in this regard is the iron
K$\alpha$ emission line.  These lines were first discovered with
HEAO-1 in heavily obscured Seyfert galaxies, such as NGC 4151 (Warwick
\etal\ 1989 and references therein), and a number of Narrow Emission
Line Galaxies (Mushotzky 1982). Not until the launch of \gi\ was iron
K$\alpha$ emission recognized as an important, and possibly universal,
property of Seyfert galaxies (Nandra \& Pounds 1994, hereafter NP94,
and references therein). From the time of their discovery in
relatively unabsorbed AGN, the emission lines have been assigned an
origin in material intimately associated with the accretion process,
the most promising candidates being an accretion disk, or ``blobs'' of
material surrounding the central source (Guilbert \& Rees 1988; Fabian
\etal\ 1989; Nandra \etal\ 1989; Pounds \etal\ 1990).  However, for
individual sources, the properties of the line were not generally well
determined, given the limited spectral resolution of \gi.

NP94 have discussed the properties of the 
iron K$\alpha$ line in some detail, using a sample of Seyfert 
galaxies. The mean energy was found to be $\sim 6.4$~keV, which indicates that
the most likely origin for the line
is fluorescence in near-neutral material. Given this
information, it can be deduced that the line cannot arise in
material uniformly covering the source.  The mean equivalent width of
the line of 140~eV requires a high optical depth for the material,
which would produce considerably more soft X-ray absorption in these
objects than is observed. Indeed, the equivalent width of the line is
sufficiently high to indicate that the material may be optically
thick to Compton scattering. Further evidence for a Compton-thick
region exists in the form of the ``hard tail'',
which can be produced by ``reflection''
from the line--producing region. Given this evidence, an
accretion disk is an obvious source for these features,
especially when considering that the strengths of both the
line and ``reflection'' component imply a covering 
fraction of $\sim 50$~per cent for
the material. Nonetheless, the ``blob'' geometry was not ruled out by
the \gi\ data (Bond \& Matsuoka 1993; Nandra \& George 1994).

Further impetus for the study of iron K$\alpha$ lines in Seyfert
galaxies has come from more recent \asca\ data, which have indicated
that this feature may be a key element in our understanding of
AGN. The earliest data showed good evidence that the emission lines
were resolved (Fabian \etal\ 1994; Mushotzky \etal\ 1995),
confirming tentative suggestions from the \gi\ data (NP94).
Most dramatically, high signal--to--noise data for two sources
have shown characteristic line profiles (MCG-6-30-15 Tanaka \etal\
1995; NGC 4151 Yaqoob \etal\ 1995).  The lines in these sources are extremely
broad, with FWHM implying relativistic velocities
of order 0.2c. Furthermore, there is a strong asymmetry to the
red, which is indicative of the gravitational redshifts associated
with the inner regions of an accretion disk surrounding the black
hole. Fabian \etal\ (1995) concluded that this was the most plausible
explanation for such profiles, rejecting other interpretations such as
Comptonization. Whilst it is tempting to extrapolate such a model
and apply it to all AGN, we must bear in mind that these extraordinary
profiles are relatively rare in the literature.  Further reinforcement
of the black hole/accretion disk paradigm requires confirmation of
that model for further sources. It is our intention to explore such
avenues in the present paper.

In a previous paper (Nandra \etal\ 1996b; hereafter paper I), we have
presented imaging and timing data for a sample of Seyfert 1 galaxies
observed by \asca. Here we present corresponding spectral data in the
3--10 keV regime in order to define the properties of the iron line.
In Section \ref{sec:anal} we describe our analysis techniques.
We will present the spectra in the full band in a subsequent paper
(hereafter paper III).
Following that, Section~\ref{sec:gauss} present simple
parameterizations for the iron K$\alpha$ line. In Section~\ref{sec:dl}
we test specific disk--line models.  Our aim thereby is to determine
if such models, which have successfully described the two sources
mentioned above, are generally applicable.  In Section~\ref{sec:comp},
we compare our results to previous X-ray observations.  We summarize
our findings in Section~\ref{sec:summ} and discuss them in
Section~\ref{sec:discuss}.

\section{Data analysis}

\label{sec:anal}

In the present paper, we examine the spectra of a
sample of Seyfert 1 galaxies observed by
\asca\ in the 3-10 keV band. We chose this energy range as it is
relatively free from the effects of absorption, either by cold gas or
the ionized ``warm absorber'', allowing us to attempt a definition of
the continuum shape and, particularly, the iron K$\alpha$ emission
line. Furthermore, we consider only the time-averaged spectrum
for each observation, to maximize the signal-to-noise ratio.
Our sample is listed in Table~\ref{tab:sample}. We refer
to each observation using the name of the source and, in
cases where there are multiple observations, the observation number
in brackets (e.g. NGC 4151(2) is the second observation of NGC 4151;
note that rejected observations are still assigned a number --
see paper 1 for details). The selection criteria, observation log and 
data analysis methods are presented in Paper 1. Briefly, we have 
employed data from the
US public archive at NASA/GSFC. We applied standard criteria for
rejecting poor quality data and then extracted the source spectra
using the FTOOLS/XSELECT package from both the Solid-state Imaging 
Spectrometers (SIS) and Gas Imaging Spectrometers (GIS). Background
spectra were extracted from source-free regions.

For both instruments we considered only the gain-corrected
(Pulse Invariant or PI) channels. The raw spectra were rebinned 
such that each resultant channel had at least 20 counts per bin, 
permitting us to use \chisq\ minimization for spectral fitting. 
All such fits were undertaken using the XSPEC
package (v9.0). Calibration data were taken from the HEASARC
calibration database. The SIS response matrices were released
on 1994 November 9 and the GIS matrices on 1995 March 6. Effective
areas were calculated using the standard software, which does
not include a parameterization of the azimuthal dependence of the the point
spread function. This can cause inaccuracies in the normalization,
particularly in the case where the source region is not circular.
Uncertainties in determining the source centroid can also result
in normalization errors, and the cross-calibration of the four
instruments is not known to arbitrarily high accuracy. These factors
lead to cross-calibration uncertainties between the four instruments,
but as the response of the XRT is not strongly dependent on energy, 
we do not expect them to introduce distortions into
the individual spectra. We therefore chose to fit all
four \asca\ detectors simultaneously with the same model, but allowed
the normalization of that model to be free for each detector.

Due to radiation damage, the response of the CCD detectors
is changing slowly with time (Dotani \etal\ 1995). The main
effects of this damage relevant to the analysis presented here are small
changes in the energy scale and a degradation of the energy resolution.
We have not accounted for such effects, which are most
pronounced in 4-CCD mode data. However, even in this most extreme case,
we expect the effects to be small for our data, which were taken early
in the mission. The estimated offset and resolution changes are
$<50$~eV for 4-CCD mode and are negligible for 1 and 2-CCD modes
for data taken $<1$~yr after launch (Dotani \etal\ 1995). 
Normalizations and fluxes quoted for the sources will be for the SIS0
detector. Typically, the two SIS detectors agree in flux to within
$\sim 1-3$~per cent, but are higher than the GIS by $\sim 10-20$~per cent.

\subsection{Statistical considerations}

As mentioned above, we have employed \chisq-minimization for spectral
fitting (Lampton, Margon \& Bowyer 1976). 
Although this forces us to rebin the data arbitrarily,
sometimes beyond the spectral resolution of the instrument, we
prefer this over maximum likelihood methods such as the C-statistic
(Cash 1979) because the data can be background-subtracted. This is
highly desirable in cases, such as these, where reliable background
models have not yet been developed. Although the background represents
typically only a few per cent of the total count rate in the
3-10 keV band, it can have a significant effect at high energies. 
We therefore note here the fact that we have not accounted for vignetting of
the diffuse X-ray background when calculating our background rates,
nor any other variations in the background rates over the detectors. Such
effects are not currently well-understood for \asca\ data. 
Our chosen bin-size is not
optimal for determining the goodness-of-fit; at low energies the data
are under binned compared to the resolution of the detector.
We prefer this to the alternative of over binning, which would
sacrifice spectral information essential to our determination
of the line parameters. Although detailed justification for our choice
of 20 counts per bin requires simulations beyond the scope of this
paper, a comparison of our spectra with the simulations
of Nousek (1990), suggests that using a gaussian approximation
and \chisq-minimization will introduce uncertainties of the order of
$1$~per cent to our derived parameters, smaller than the typical
statistical error. 

Once the minimum \chisq\ value is found in each fit, confidence regions
for the parameters have been determined using the prescription of
Lampton \etal\ (1976). Unless otherwise stated we have
calculated the confidence region by projection of the joint confidence
region defined by the number of ``interesting''  parameters, 
as recommended by Lampton \etal  We have 
considered the model normalizations as uninteresting. This 
is because we do not expect the origin of the
difference in the normalizations -- uncertainties in the effective
areas -- to affect the spectral form. Thus the number of
interesting parameters which we have used to calculate the \chisq\
deviates is given by $N_{\rm fr}-4$, where $N_{\rm fr}$ is the number
of free parameters in the model. Specific \delchi\ values are shown
as footnotes to the Tables. For completeness we have quoted the
best-fit normalization for one of the models in Table~\ref{tab:nl},
although we have not attempted to determine the confidence interval
for that parameter, as we consider it uninteresting. 

In most cases we quote the 68 per cent confidence limits for
the parameters, which is the same confidence level as the 1$\sigma$ 
distribution of a gaussian. Furthermore, in some
specific circumstances, we have gone on to interpret the 68 per
cent confidence limit as if it were the 1$\sigma$ error bar
of a gaussian distribution. Examples are the calculation of
the weighted mean, and the determination of the mean and dispersion
using the maximum-likelihood method. This assumes that the confidence
surface is itself gaussian, which is often
a poor approximation. However, this is unimportant, as long
as the gaussian approximation does not underestimate the true
confidence region. We believe this to be true in most
circumstances. However, in most cases the confidence region is
asymmetric, resulting in two-sided ``error bars''. In all such cases 
we use the larger uncertainty
as our 1$\sigma$ error bar. Furthermore, where the 68 per cent
confidence region overlaps the limits of the allowed parameter
values (e.g., where the inclination reaches 0\deg or 90\deg),
we assume that the 1$\sigma$ error bar is essentially infinite,
thus placing no weight on the derived value. This is a highly
conservative approach which should avoid any spurious results
arising from our use of confidence regions as gaussian error-bars.

In comparing various models for the X-ray spectra, and particularly
the iron lines, we have employed various statistical tests, including:
the acceptability of \chisq, although as noted above, our data are not
optimally binned to maximize \chisq; the F-test (e.g., Bevington 1969) 
which has been employed to determine whether additional components, or 
parameters, provide a significant improvement; the likelihood ratio 
(Edwards 1972; Mushotzky 1982) which has been used to determine whether 
models with the same number of free parameters are more or less likely
to produce the observed data.  We have adopted the 95 per cent
confidence level in assessing whether or not various features are
real, but we bear in mind that for a sample of 23 (or more often 22)
observations, this will typically result in one spurious detection
each time the sample is considered as a whole. 

\section{Iron K$\alpha$ emission: simple parameterizations}

\label{sec:gauss}

We first attempted some simple spectral fits to the \asca\ data,
without reference to a specific, physical model.

\subsection{Narrow lines}
\label{sec:nl}

\begin{figure}
\plotone{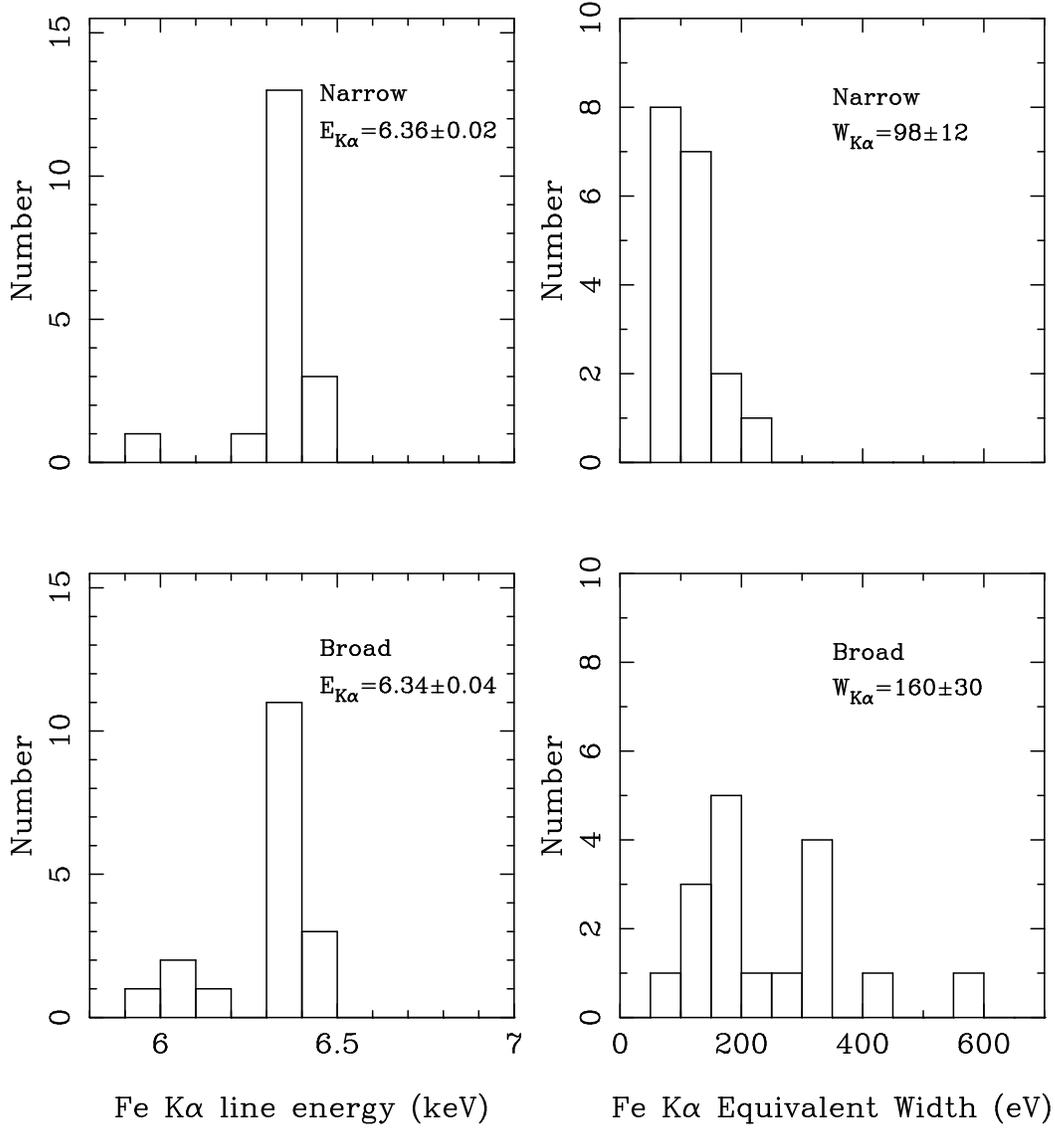}
\caption{Histograms of the line energy (left panels) and
the equivalent width (right panels) 
for the sources in our sample. Where multiple observations exist the
weighted mean value has been adopted. The upper panels show the values
obtained when the line is assumed to be narrow relative to the instrumental
response (Table~\ref{tab:nl}), whereas the lower plots show the values 
when the line width is left as a free parameter (Table~\ref{tab:bl}). 
In both cases, the mean line energy is found to be very close
to the 6.4 keV expected of neutral iron, even being consistent with
a mild redshift. The equivalent widths in the broad-line fits
are typically $\sim 50$~per cent larger than for a narrow line.
\label{fig:lparms}}
\end{figure}

Our initial fit was of a power-law with a 
gaussian component with width, $\sigma=10$~eV, 
to represent the iron line. This width is smaller than the
instrumental response and therefore is effectively ``narrow'' for our
purposes. In practice, this fit will model any line emission up to the
width of the spectral response of $\sim 120$~eV (FWHM) at 6 keV.  The
gaussian provides a significant improvement in the fit in all 18
sources, confirming that iron K$\alpha$ emission is extremely common
in Seyfert 1 galaxies.  The details of these narrow--line fits are
shown in Table~\ref{tab:nl}. The energies quoted are in the rest-frame
throughout. Three of the fits are unacceptable 
at $>95$~per cent confidence: NGC 4151(2), NGC 4151(5), MCG-6-30-15(1). 
The mean values of the parameters derived from these fits are shown 
in Table~\ref{tab:means}. That table
shows the unweighted mean and standard deviation (column 1), the
weighted mean and uncertainty (column 2), and in columns 3 and 4
estimates of the mean and dispersion of the parent population,
calculated using the maximum likelihood method of Maccacaro \etal\
(1988).  We consider the latter two values to be most appropriate
for determining the average properties of the sample, and quote
these in the text. A histogram of the line energies and equivalent widths is
shown in the upper panels of Fig.~\ref{fig:lparms}.  
Confirming and extending the \gi\
results, our \asca\ observations show a strong preference for iron in
a relatively low state of ionization, with the mean energy of 
$<E_{\rm K\alpha}>=6.36\pm 0.02$~keV, which is even indicative 
of a small net redshift (but only at $\sim 90$~per cent confidence). 
There is no measured dispersion in these values, with an upper
limit of $60$~eV; the large deviation in the case of NGC 6814 is due
its large measurement error, as it is a factor $>10$ weaker than all other
sources. The lack of dispersion is remarkable; any systematic deviations
in the energy scale due to residual calibration errors 
(see Section~\ref{sec:anal}) must be smaller
than this. Also note that the differences in the {\it observed-frame} 
energy between the sources is $\sim 300$~eV.
We find a mean ionization state $<$Fe {\sc xvi} 
($E_{\rm K\alpha}<6.41$) at the 99 per cent confidence
level.  The mean equivalent width of $98\pm 12$~eV for these narrow
lines is only $\sim 70$~per cent of the corresponding value
derived by \gi\ (140 eV; NP94).  An examination of the residuals makes clear
why this is the case.  In a number of sources the line flux clearly
extends beyond the bounds of the spectral response function,
indicating the line has a significant width. The wider response of
\gi\ would result in more of this flux being modeled in a fit what
what is effectively a delta-function profile. Such a profile is
clearly not appropriate for our higher-resolution \asca\ data, and we
now go on to describe this more quantitatively.

\subsection{Broad lines}
\label{sec:bl}

\begin{figure}
\plotone{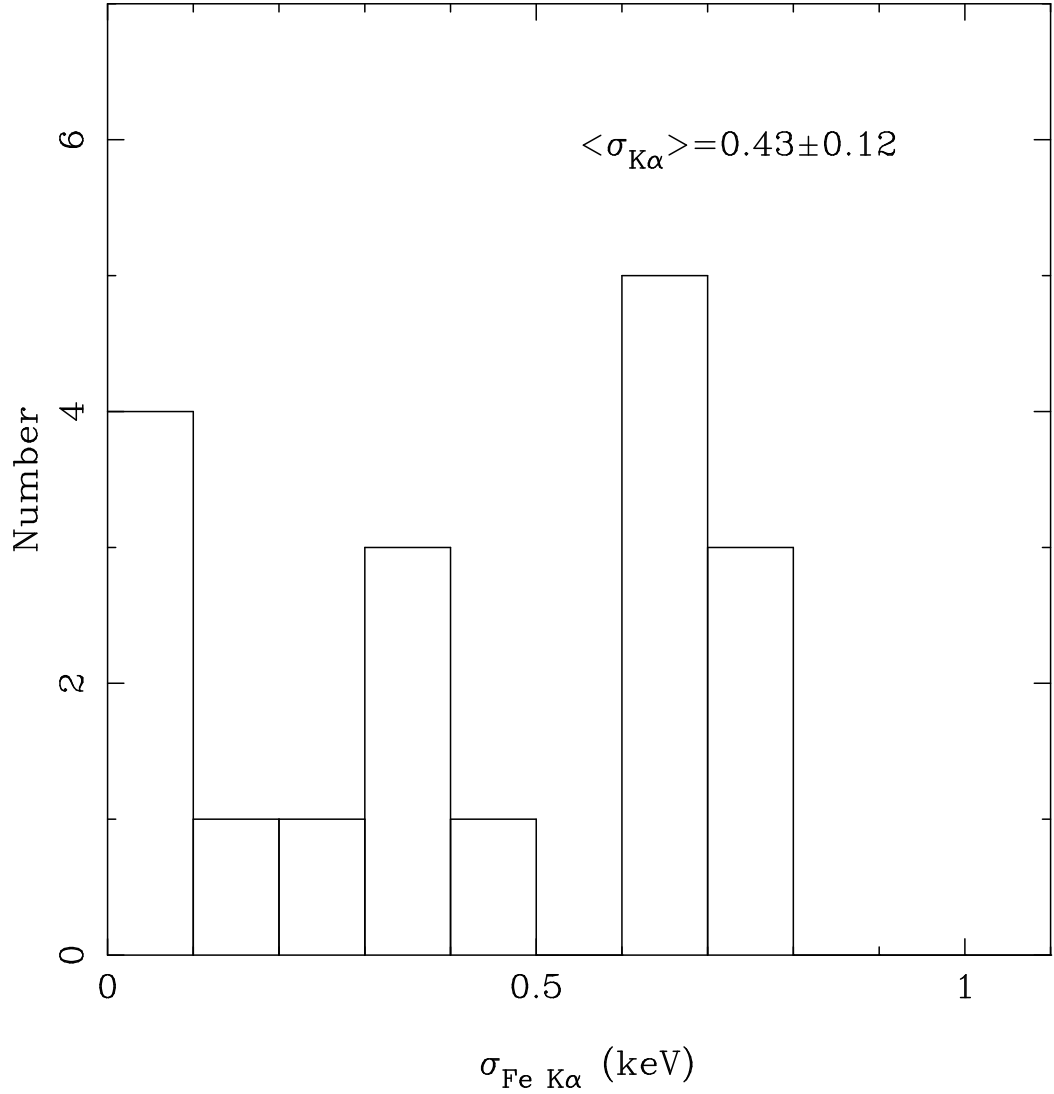}
\caption{Histogram of the line widths obtained from the broad-gaussian fit.
In the cases where no significant improvement is obtained 
when $\sigma_{\rm K\alpha}$ is a free parameter,
the point has been added to the
lowest bin. In 14 of the 18 sources the line is resolved, with
$<\sigma_{\rm K\alpha}>=0.43\pm 0.12$, which corresponds to
a FWHM$\sim 50000$~km s$^{-1}$. This implies highly relativistic motions
if the width is due to velocity broadening.
\label{fig:sigma}}
\end{figure}

Some \gi\ spectra of Seyfert galaxies showed a significant improvement
in the fit when the line width was allowed to be a free parameter
(NP94). Due to the nature of the \gi\ spectral
response this implied huge line widths of order $1$~keV in the
cases where significant broadening was indicated. This led
NP94 to speculate that the apparent broadening might
be spurious, instead reflecting unmodelled continuum components.
Improved spectral resolution is the key to removing this ambiguity.
Indeed, early \asca\ observations showed that, at least in some cases,
the iron line was significantly broadened (e.g. Mushotzky \etal\ 1995).
We tested this for our sources by allowing the width of the
gaussian $\sigma$ to be a free parameter. The
results of these fits are shown in Table~\ref{tab:bl}.
18/23 observations and 14/18 sources showed a significant 
improvement, with only two fits
just unacceptable at the 95 per cent confidence
level: NGC 4151(2) and MCG-6-30-15(1). In cases where the line width is
not detected, we have shown only the 90~per cent
upper limit to the line width. 
These are all consistent
with the mean value of $<\sigma_{\rm K\alpha}>=0.43 \pm 0.12$~keV
(Table~\ref{tab:means}). This corresponds to a FWHM of 
approximately $47,000$~km s$^{-1}$. A histogram of the line widths
is shown in Fig.~\ref{fig:sigma}.

Once again, the energy of these broad lines strongly indicates
material in a low ionization state is the source of the line, with a
mean energy $<E_{\rm K\alpha}>=6.34 \pm 0.04$~keV.  The 99 per cent
confidence upper limit to the energy of 6.41 keV again corresponds to
an ionization state for iron $<$Fe {\sc xvi} --  above this
ionization state the line rises sharply in energy (Makishima 1985).  
If the material were in collisional ionization equilibrium, this corresponds
to a temperature of $\lesssim 3 \times 10^{6}$~K (Arnaud \& Raymond
1992). Of course, any
material close to the central source is more likely to be photoionized
than collisionally ionized, in which case the temperature is probably
much lower. 

The mean equivalent width of these broad lines is 
$<W_{\rm K\alpha}>=160\pm 30$~eV.
This value is consistent with the mean determined by \gi\ for a 
line of $\sigma=0.1$~keV, of $140\pm 20$~eV. As mentioned above, a fit with
a line which is narrow compared to the instrumental response of a given
instrument will tend to model any line flux which has a width up to
the spectral resolution. For the case of \gi\ this is
equivalent to $\sigma \sim 0.5$~keV (Turner \etal\ 1989). This 
corresponds roughly to the mean width of the lines determined 
in our broad-gaussian fits. In this respect, then, the equivalent 
widths determined by \asca\ and \gi\ show a remarkable consistency.

\subsection{Line Profiles}

\label{sec:prof}

At least in some sources, the iron K$\alpha$ line is not well modeled by
a symmetrical profile such as a gaussian. We have
attempted to determine the line profiles
in a relatively model-independent way, with the results
show in Fig~\ref{fig:profiles}. These were determined
with a fit to the 3-10 keV SIS data only with a power law, but excluding the
iron band between 5-7 keV. We then multiplied the resultant data/model
ratio by the best-fit continuum model. This results in an estimate
of the line profile as seen in the observed frame, deconvolved
from the instrumental response.
Almost all of the sources show a clear excess,
with broadening far greater than the instrumental response. It is also
evident that a number of sources show skewed, asymmetric profiles. This could
represent multiple line components (for instance, there may be both broad
and narrow components). However, many sources show the characteristic
black hole/accretion disk profile (Fabian \etal\ 1989;
Laor 1991; Matt \etal\ 1992), in which case 
they may be predominantly one component. 

To demonstrate this further
we have created a composite line profile for all 18 sources. This was
achieved from by transforming the above-mentioned data/model ratios 
into the rest frame of each source and taking the mean ratio
in bins of width 40~eV. 
To estimate the mean line profile from these plots,
we have multiplied by a continuum model defined by the mean values
for the sample. 
This is shown in Fig~\ref{fig:mother}a.
In Fig~\ref{fig:mother}b we show the same plot excluding MCG-6-30-15
and NGC 4151, to investigate whether these sources dominate the
mean profile.
Several things are apparent from Fig.~4:

\begin{itemize}
\item
The line profiles are clearly consistent, showing
that our results are not biased by the inclusion of 
MCG-6-30-15 and NGC 4151.
\item
The profiles are extremely broad $\sim 2$~keV Full-width 
at Zero Intensity (FWZI),
which corresponds to velocities of order 0.3c.
\item
As in many of the individual sources the profile consists of a 
relatively narrow core, with an underlying broad component.
\item
The core peaks at an energy remarkably close to 6.4~keV.
\item
The broadening is primarily to the red, with relatively little
flux blue-ward of 6.4 keV and specifically no strong component
due to helium-like or hydrogen-like iron (6.7-6.9 keV).
\end{itemize}

To illustrate these more quantitatively, we have modeled the profiles
with two gaussians, following Tanaka \etal\ (1995). The resulting fit
is shown in Fig.~\ref{fig:mother}a. The centroid energies of the core
are 6.38~keV in both cases, with a width of $\sigma=0.1$~keV.  The
broad component is strongly redshifted with centroid energy $\sim
6$~keV and width of $\sigma=0.7$~keV and carries $\sim 75$~per cent of
the total flux.  The dotted line in Fig.~\ref{fig:mother}b shows the
instrumental response. Whilst some artificial broadening could result
from radiation damage to the CCDs, particularly in 4-CCD mode, these
residual calibration uncertainties of $\sim$few per cent in this
regime are much smaller than the observed deviations ($\sim 10-20$~per
cent in the red wing).  To demonstrate this, we show in
Fig.~\ref{fig:cal}, the residuals for our Seyfert sample compared to
those for the Coma Cluster and also the Dark Earth (essentially the
particle background of the SIS; Gendreau 1995). All three plots show
an emission line, but only the Seyferts show the red wing, indicating
that this feature is not due to a persistent calibration error.


It is also highly unlikely that emission lines from any any other element
contribute significantly to the emission in the 5-7 keV band. The
other elements with K-shell energies (V, Cr, Mn, Co) in this range
have abundances and fluorescence yields which would lead to line
intensities of a factor $>100$ less than iron, to which we attribute
the entire line flux. Iron line blends alone cannot account for the observed
widths in most cases, as the line flux extends up to $\sim 1.5$~keV
below the energy expected from neutral iron. Compton down-scattering
can, in principle, redshift the line sufficiently,
but many workers have argued against this mechanism
(Mushotzky \etal\ 1995; Tanaka \etal\ 1995; Fabian \etal\ 1995).
The dramatic redshifts and broadening expected due to special
and general relativistic effects close to a black hole
provide a plausible explanation for the observed line
profiles. One class of models which has been successfully applied
to \asca\ observation of MCG-6-30-15 and NGC 4151 is
that of an accretion disk (Tanaka \etal\ 1995;
Yaqoob \etal\ 1995). The other sources apparently
exhibit rather similar profiles and we now consider such models.

\section{Accretion Disk modeling}

\label{sec:dl}

We will refer to a number of parameters
in the following sections which characterize the disk line models:
the inner radius of the disk, $R_{\rm i}$, the outer radius, $R_{\rm o}$,
the inclination of the axis of rotation with respect to the 
line-of-sight, $i$, the rest energy $E_{\rm K \alpha}$ and 
the line normalization $I_{\rm K \alpha}$.
We excluded NGC 6814 from the following analysis as the 
signal-to-noise ratio is so much lower than that of the other sources,
leaving 22 observations of 17 sources.

\subsection{Schwarzschild model}

\label{sec:dl-sch}

Our first test of the black hole/accretion disk hypothesis was using
the model of Fabian \etal\ (1989). Their disk--line
model assumes a Schwarzschild geometry and computes the line profile
only and not the strength. 
No specific geometry for the X-ray source is assumed, and therefore
the line emissivity is parameterized by a power law as a
function of radius, $R^{-q}$. The value of $q$ can be a free parameter.
For a point-like X-ray source located above the center of the disk, we expect
$q\sim 0$ in the inner regions, steepening to $q\sim 3$ at large radii.
However, different emissivity profiles may be relevant, for example if
there are multiple X-ray sources, or if the source is not point-like
compared to the line-emitting region. 

The line profile has a strong
dependence on $q$, as for values $>2$, the emission is concentrated in the
inner disk, where extreme gravitational and Doppler effects operate. In these
circumstances, $R_{\rm o}$ can rarely be constrained. Conversely, if the outer
disk is the dominant source of line emission, poor or no constraints
can be placed on $R_{\rm i}$. We have therefore chosen to fix the
inner and outer radii of the disk at 6 $R_{\rm g}$ and 1000 $R_{\rm g}$
respectively (where $R_{\rm g}$ is the gravitational radius of the black hole)
and have used $q$ as a parameterization not only 
of the geometry of the X-ray source, but of the accretion disk itself.
We allow the inclination of the disk to be a free parameter and also
the normalization of the line. We will compare
the strength of the lines to the disk model later. 
Finally, we assume $E_{\rm K \alpha}=6.4$~keV in the
rest frame. As we showed in Tables
\ref{tab:nl} and \ref{tab:bl}, and Figs~\ref{fig:profiles}
and \ref{fig:mother} both the peak and 
mean line energies lie close to this value in 
all cases, with no evidence for a helium-like or hydrogen-like line.

The results of these disk--line fits are shown in Table~\ref{tab:dl}.
There is a substantial improvement in the fit compared to the broad
gaussian model in a large number of cases, due to the
asymmetry of the profiles. All of the disk line
fits are acceptable at $<95$~per cent confidence. The total \chisq\ for all
22 observations was 16762.4 for the broad line fit, and 16627.3 for
the disk lines (\delchi=135.1) which is highly
significant as measured by the likelihood ratio 
(see Section~\ref{sec:anal}), 
which is shown in Table~\ref{tab:chisq}. The mean
power-law index is similar to the broad-line fits $<\Gamma_{3-10}>=1.78 \pm
0.07$. However, the line equivalent width is rather larger, at 
$<W_{\rm K\alpha}>=290\pm
50$~eV.  Typically, a low inclination ($\sim$face-on) disk is
preferred in our fits. The mean inclination angle is $<i>=
29\pm 3$~degrees. If the accretion disk were oriented randomly, we might
expect a mean inclination of $\sim 60$\deg, which is not preferred in
our fits. We comment on this later. The mean value of $q$, which
parameterizes the geometry of the system is $<q>=2.5 \pm 0.4$. This is in
good agreement with the value expected for a point-like X--ray source
above a flat disk, as discussed by Matt, Perola \& Piro (1991), and is
therefore supportive of the disk model. Histograms
of $i$ and $q$ are shown in Fig.~\ref{fig:dl}. 

Likelihood analysis reveals
no evidence for a significant dispersion in the values of $q$ obtained
in our fits, which suggests a common geometry for the
sources. However, our conservative error prescription, and the
assumption that the 68~per cent confidence limit equates to a
1$\sigma$ error bar makes it difficult to rule out such an hypothesis
based on this analysis. A more sensitive method of determining whether
or not there are differences in geometry from source-to-source (in
addition to differences in inclination) is to fit the data explicitly
with $q$ fixed at the mean value, and compare the \chisq\ value so
obtained with that when $q$ was left free. We have performed such
fits, and these are tabulated in Table~\ref{tab:dl-25}. It is clear
from column 5 of Table~\ref{tab:dl-25}, which shows the F-statistic
for the addition of $q$ as a free parameter, that many sources prefer
a value of $q$ different to the mean. Indeed 11 of 22 observations
and 8/17 sources show an improvement at $>95$~per cent confidence. 
As an ensemble, the reduction in \chisq\ is 96.1 for the addition 
of 22 additional free parameters,
significant at $>95$~per cent confidence.  This
implies that for this model,
the differences in line profiles cannot be accounted for
solely by difference in inclination;
assuming that the whole of the emission line arises in an
accretion disk, we infer that the X-ray source/accretion disk geometry
is different from source to source.

The above model assumes that the line profile consists of
emission integrated over a range of radii, each of which
gives rise to a characteristic line shape. Variations
in $q$ correspond to different weightings over the disk. 
An alternative explanation for the range of observed profiles
is that the emission is dominated by a single radius
in each source, but that this characteristic radius
is different in each case. In order to investigate this
possibility we have repeated the disk line model, but
setting $R_{\rm i}=R_{\rm o}$ and allowing that radius to be free.
$q$ is clearly irrelevant in this case. The \chisq\ values
for these fits are shown in Table~\ref{tab:chisq}. For the sample
as a whole they are considerably worse than those with a range
of radii with an increase in \chisq of 106.5 for the same
number of degrees of freedom. In only two observations
is there an appreciable improvement (i.e., a significant {\it reduction}
in \chisq), NGC 4051 and MCG-6-30-15(1). 

\subsubsection{Rest energy of the line}

In the presence of substantial gravitational and Doppler effects
inferred above, the energy derived from a gaussian fit may not reflect
the true rest energy of the line. For example, it is conceivable that
gravitational redshift might cause a line which was originally
helium--like, with an energy of 6.7~keV, to appear at a lower energy.
Given the remarkable consistency of our gaussian fits with an energy
of 6.4~keV and the lack of any strong emission above that energy
(Fig.~\ref{fig:mother}) this seems rather contrived. As shown by Matt
\etal\ (1992), for an ensemble of disk lines, the mean energy always
lies close to the rest energy, for a variety of emissivity laws. This
is due to the fact that the peak energy is dominated by the
low-inclination profiles, which are of highest intensity (George \&
Fabian 1991).  We have investigated this further by constructing summed
model profiles for an ensemble of accretion disks with inclinations
distributed uniformly in $\cos i$. For $q<3$, the peak energy for
the ensemble is redshifted by $<1$ per cent of the rest energy.
Therefore, unless the emission originates entirely within the central
regions, the peak energy in Fig~\ref{fig:mother} strongly
indicates a near-neutral origin for the emission line.
This is further illustrated in Fig.~\ref{fig:compmodels} which shows the
summed model profile for the disk line fits shown in Table~\ref{tab:dl}.

Nonetheless, we have gone on
to explore whether our observed profiles are consistent with a
helium--like line distorted by the gravitational
and Doppler effects in the inner disk. Our first test was with a model
identical to that used in the fits of Table~\ref{tab:dl}, i.e.
$R_{\rm i}=6$, $R_{\rm o}=1000$, $q$ and $i$ free, but with the energy
of the disk line fixed at 6.7 keV in the rest frame. These fits were
considerably worse than with a 6.4~keV line. The total 
\chisq\ was 16737.2 for 16973 d.o.f., compared to
16627.3 in the near-neutral case. We infer the latter to be $>1000$ times
more likely (Table~\ref{tab:chisq}). We have investigated
whether this might be due to our assumption of $R_{\rm o}=1000$.
At large radii, the lack of any significant gravitational
or Doppler shifts results in emission close to the rest energy 
of the line, yet we observe little emission around
6.7~keV (Fig.~\ref{fig:profiles}; Fig.~\ref{fig:mother}). 
Relaxing
the constraint on $R_{\rm o}$ with the helium-like disk line
does improve the fits.
However, these fits were still slightly worse (total \chisq=16633.2), than
those with the 6.4 keV line and a fixed outer radius 
(total \chisq=16628.2), despite the additional 22 free parameters. 
However, examination on a case-by-case basis suggests that
3C 120 shows the most marked difference in \chisq. Removal of
that outlier would leave the 6.7 keV/free $R_{\rm o}$ fits showing
a slight improvement over those at 6.4 keV/fixed $R_{\rm o}$, 
but not a significant one. 

As a further test, we have 
allowed $R_{\rm o}$ to be free in the 6.4 keV fits, for
comparison with the helium--like model. 
Our suspicion that $R_{\rm o}$ could not be constrained by
these data was largely confirmed by this analysis, with only 
four observations, NGC 4051, NGC 4151(2), MCG-6-30-15(1) and
Mrk 841(1), showing a significant improvement
at $>95$~per cent confidence. The lack of other detections is
likely to be due primarily to signal-to-noise ratio, the degeneracies
inherent in the disk line models and the fact that the bulk of
the line emission is produced in the central regions.
The total \chisq\ for those fits was 16597.0, indicating that
a 6.4 keV line is $>1000$ times more likely (Table~\ref{tab:chisq})
than a 6.7~keV line, even if the outer radius is left free. 
That conclusion holds even if 3C 120 is removed from 
consideration.

Relaxing the constraint on the inner radius
is unlikely to improve the fits which assume helium-like iron,
as we have fixed the inner radius at the last stable orbit
in the Schwarzschild metric. This maximizes the gravitational
effects, which would tend to bring any highly ionized emission
closer to the observed values.

For the four line models tested thus far (narrow gaussian, broad
gaussian, disk line with fixed $R_{\rm o}$ and disk line with free
$R_{\rm o}$), we have shown that a 6.7~keV line is strongly
disfavored.  However, this analysis does not preclude the possibility
that more complex models can be constructed which conspire to produce
an apparent rest energy of 6.4~keV. Particularly, we have assumed
that the line can be described by a single component.  We now
explore the possibility that an additional, narrow line is
present.

\subsubsection{Narrow line contribution} 

It is possible that there is a contribution
to the emission line from another region.  Differing relative
contributions from that region and the disk could then confuse our
results from the disk line fits. A promising candidate is the
obscuring torus hypothesized in Seyfert 1/2 unification schemes, which
could produce a significant contribution in some circumstances (Krolik
\& Kallman 1987; Ghisselini, Haardt \& Matt 1993; Krolik, Madau \& 
Zycki 1993). A relatively narrow line might also be produced from the optical
``broad line'' region or even the warm absorber. 

We have tested for the presence of such a feature by adding a narrow
gaussian, at a rest energy of $6.4$~keV to the disk-line models
described in Table~\ref{tab:dl}.  This provided a significant
improvement in the fit for 6 observations: NGC 3516, NGC 3783(1), NGC
4151(2,4,5) and NGC 5548.  The improvement for the ensemble is
\delchi=68.9 for the 22 degrees of freedom, which is formally highly
significant. The improvement is primarily due to NGC 4151 - without
that source the improvement for the sample would not be significant
(\delchi=32.3 for 19 degrees of freedom).  This is noteworthy, as NGC
4151 is the only heavily--absorbed source in our sample; conceivably
the line-of--sight column could be contributing to the emission line
in this case (see Weaver \etal\ 1994 and Yaqoob \etal\ 1995 for
detailed discussion of the emission line in NGC 4151).  In the other
cases where the narrow line provides a significant improvement, this
may simply reflect the fact that we have over-restricted the parameter
space for the disk-line model.  Apparently there is no requirement for
a narrow line contribution in the majority of our sources, although
the upper limits are not restrictive (typically $\sim 100$~eV,
consistent with predictions for the torus; Ghisselini \etal\ 1993).
The mean equivalent width of the narrow component for the sample
is only $\sim 30$~eV. 

Even if there is such a narrow component, it has little effect on our
conclusions regarding the remainder of the line. This is illustrated
in Fig.~\ref{fig:compmodels} which shows the summed model profiles with
and without the narrow line. To quantify the changes we show in
Fig.~\ref{fig:dl-nl} the change in the disk line
parameters ($W_{\rm K\alpha}$, $i$, $q$) when the narrow line is added
to the model. Obviously, the equivalent width of the disk line
decreases when the narrow line is added, but typically only by a few
tens of eV.  There effect on the inclinations is negligible; our fits
are driven to low inclinations by the lack of significant blue flux,
which is not changed by the narrow component.  There is a small effect
on the emissivity index, with the the mean value of $q$ from the
disk-line-plus-narrow-line fits was $q=2.8$, slightly steeper than
that without the narrow line, but not significantly so.  We have also
explored whether the presence of a narrow line can explain the {\it
apparent} difference in geometries implied by the range of $q$ values
in Table~\ref{tab:dl}.

Fixing $q$ at the mean value of 2.8 resulted a significantly worse fit
in 9/22 of the observations.  Therefore, despite the possible
contribution from a molecular torus, or other near-neutral plasma with
low velocity, we still conclude that either the geometry of the X-ray
source and/or the material which produces the broad, skewed line,
differs from source to source. 

The one conclusion which does depend critically on the presence or
absence of a narrow line is that regarding the rest energy of the
disk line. We showed above that a near-neutral origin is considerably
more likely than a helium-like line if the entire emission line arises
from the disk. In the presence of a narrow line, we can no longer
constrain the energy. We find the fits with a 6.4~keV disk line 
are not significantly better than with a helium-like line, if we allow
a narrow line of typical equivalent width $\sim 40$~eV at 6.4~keV.
This is unsurprising, as the main constraint on the rest energy of
the disk line comes from the energy of the peak, which in this case
is modeled with the narrow line. However, given the lack of conclusive
evidence for a narrow line component in addition to the disk line, a
rest energy of 6.4~keV must still be considered more likely and we
have assumed such an energy in the fits described below.

\subsubsection{Effects of the reflection continuum}

We next explored the effects of Compton reflection on the spectral
parameters. As mentioned in the introduction, the reflection
continuum is commonly observed in the \gi\ spectra (NP94) and is expected
if the material which produces the emission line is optically thick
(Lightman \& White 1988; George \& Fabian 1991; Matt \etal\ 1991).
This is is almost certainly the case, given the high equivalent widths.

Although the reflection component is expected to be 
relatively weak in the \asca\ band, it may have some influence
on the derived line and continuum parameters (e.g., Mushotzky \etal\ 1995;
Weaver \etal\ 1995; Cappi \etal\ 1996).
We therefore added a reflection component to the power-law and disk-line, 
to investigate the possible effects on these parameters. We
employed a model assuming the abundances and cross-sections
of Morrison \& McCammon (1983) and elastic scattering, following Basko (1978).
The fact that the \asca\ spectra extend only to $\sim 10$~keV makes
the latter approximation reasonable, and computationally expedient.
We assumed the reflection spectrum from a neutral slab
subtending a solid angle of 2$\pi$ steradian at an X-ray source
located above the slab. Fits allowing the normalization of the reflection
component to be free confirmed our suspicions that typically the 
reflection continuum
can be neither detected nor constrained for these low-redshift AGN, even
when all four instruments are employed. However, our hypothesis is that
the line arises by fluorescence in an optically-thick accretion disk.
Simple atomic physics then
dictates that the fluorescence must be accompanied by a Compton
scattered continuum.

We have therefore repeated our fits with the disk line, including a fixed
reflection component, assuming an inclination as derived from the
disk line model, and an isotropic X-ray source. Note that these
fits have no additional free parameters.
Including such a component can, in principle, allow
additional constraint to be placed on the disk inclination and may affect
the disk--line parameters. However, in practice the constraints were no
better than without the reflection component included. The mean values
of $q$, $i$ and $W_{\rm K\alpha}$ (Table~\ref{tab:means})
were all consistent in these fits compared to those
when a pure power-law continuum was used.
We did find that differences in the line equivalent
width tended to be systematically lower, however, 
with a mean value of $230\pm 60$~eV with reflection included
(see Table~\ref{tab:means}; c.f. $290\pm 50$~eV without reflection). 
The reflection component also affects our
determination of the continuum (see below).
Including the reflection component in the fit resulted in a reduction of the
total \chisq\ (i.e. for all sources) of 43.9, which implies that for the sample
as a whole, the
reflection model is considerably more likely than a simple power-law
(Table~\ref{tab:chisq}).

\subsection{Kerr model}

The previous section has demonstrated that relativistic
effects in an accretion disk orbiting a Schwarzschild 
black hole provide a most agreeable explanation for 
the line profiles. Those fits suggest that the emission
is concentrated close to the central black hole. In such
an extreme environment the details of the black-hole metric
may even become important.
Laor (1991) has presented line profiles for a maximally--rotating
black hole in the Kerr metric. It was indeed demonstrated that
substantial differences in the line profile result, due primarily
to the fact that the innermost stable orbit is closer to the
central point mass ($1.23\ R_{\rm g}$) and hence the gravitational
and Doppler effects more intense.

We have tested such a model on
our data including the reflection component, as described above. 
This time, we have fixed the inner radius at the last stable orbit
for a Kerr hole. We fix
the outer radius at $400\ R_{\rm g}$, the maximum value allowed in the
current implementation of this model in XSPEC. 
The parameters are tabulated in Table~\ref{tab:laor}.
The \chisq\ for these fits (Table~\ref{tab:chisq}) are 
remarkably similar to those obtained with the
Schwarzschild model given in Table~\ref{tab:dl-ref}.
The total \chisq\ values imply that the Schwarzschild model
is $\sim 3$ times more likely than the Kerr model,
but we do not consider this to be statistically significant. 
The similarity is probably a consequence of the
fact that our fits prefer a relatively low inclination angle.
The best--fitting line profiles 
assuming each metric are shown in Fig.~\ref{fig:model} for 2 sources with 
well defined profiles. They are indeed rather similar, but the Kerr
models have a more pronounced red wing, due to the increased gravitational
redshift within $6\ R_{\rm g}$.
The mean parameter values for the Kerr fits are entirely
consistent with those obtained for the Schwarzschild model
(Table~\ref{tab:means}). 

We conclude that a Kerr black hole is not required by our data.
Higher-resolution and signal-to-noise ratio are required
to determine the spin of the black hole definitively.

\subsection{The X-ray continuum}

\label{sec:cont}

Whilst our primary goal in considering the 3-10 keV spectra of these
Seyfert galaxies was to define the iron-line properties of the sample,
these data also allow us to define the continuum in this band.
For a number of the
the emission line models mentioned above, we have tabulated the
mean power-law indices, which are shown
in Table~\ref{tab:means}. As can be seen from this Table, the mean power
law slope is not strongly dependent on the {\it line} model, ranging from
$<\Gamma_{3-10}>=1.75$ for a narrow line, 
to $<\Gamma_{3-10}>=1.79$ for a disk--line 
model. However, although
the presence of a reflection continuum causes minimal change in the
derived line parameters
there is a substantial effect
on the continuum. In the fits tabulated in Table~\ref{tab:dl-ref},
the mean spectral slope when reflection is included
is $<\Gamma_{3-10}>=1.91\pm 0.10$ (Fig.~\ref{fig:ref}; 
Table~\ref{tab:means}), $\Delta\Gamma=0.12$ steeper than with
a power-law continuum.

Conceivably there might also be other spectral components
which can effect the continuum. The most obvious of these
are low energy absorption and iron K-edge absorption, both
of which have been reported for some \gi\ spectra (NP94). We have
tested for the presence of each of these in our sample.
Aside from the well-known case of NGC 4151 (for which we have already
included an absorption column), we find no evidence for
soft X-ray absorption. This justifies {\it post facto}
our choice of the 3-10 keV range for determination of
the line parameters (note that neutral columns $\lesssim 10^{22}$~\pcmsq
have little effect above 3~keV). Addition of an iron K-edge does
provide a significant improvement (at $\sim 95$~per cent
confidence) in 4 observations of 3 sources (NGC 3516 -- see 
Fig.~\ref{fig:model}; NGC 3783(1);
NGC 3783(2) and NGC 4151(2)). 
A plausible origin for the edge
is in the warm absorber mentioned in the introduction. However
even when such an edge is detected, it has minimal effect on
the line and continuum parameters and thus can be 
disregarded for the purposes of our analysis. 

\section{Comparison with previous results}

\label{sec:comp}

Although a number of the observations presented here have been
published previously, detailed discussion of the emission-line
properties has been relatively rare.  However, where comparisons can
be made we find consistency with our fits. Specifically, our results
are entirely consistent with those presented by Ptak \etal\ for NGC
3227, Mihara \etal\ for NGC 4051, Weaver \etal\ (1994) and Yaqoob
\etal\ (1995) for NGC 4151, Reynolds \etal\ (1995b) and Tanaka \etal\
(1995) for
MCG-6-30-15, Mushotzky \etal\ for IC4329A and NGC 5548, Guainazzi
\etal\ for NGC 7469 and MCG-2-58-22 (Weaver \etal\ 1995). We find some
disagreement in interpretation for NGC 3783, where George \etal\
(1995) claimed no evidence for broadening of the line. We consider
this to be due to the fact that George \etal\ fixed the photon index
in the fit. Similarly, Leighly \etal\ (1996) claim no evidence for broadening
of the line in Mrk 766, but did not allow the energy of the line to
vary from 6.4 keV. Fig~\ref{fig:profiles} indicates both narrow and
broad components to the line in Mrk 766.  We suggest that Leighly et
al. were measuring only the narrow component.

The results for the \asca\ sample are also remarkably consistent
with the lower-resolution \gi\ data. NP94 found a mean energy of
$6.37\pm 0.07$~keV and equivalent width $140\pm 20$ for a ``narrow''
gaussian. As mentioned above, the latter is most directly comparable
to that of our broad line fits, given the width of the \gi\ response,
and is entirely consistent with those fits.
The mean spectral index from \gi\ was
$1.95 \pm 0.05$ when the effects of reflection and the warm absorber
were accounted for (NP94).

\section{Summary of results}

\label{sec:summ}

In this paper, we have found:
\begin{itemize}
\item
emission at iron K$\alpha$ is detected in all 23 observations
of our 18 sources.
\item
the line to be resolved in 80 per cent of the sources, with a
mean width corresponding to a FWHM of 47,000 km s$^{-1}$ 
assuming a gaussian profile.
\item
the profiles are often asymmetric, with a peak at $\sim 6.4$~keV
in rest--frame
and a broad wing extended to the red. The FWZI of the mean line
profile corresponds to a velocity of $\sim 0.3$c. 
\item
all of the profiles are consistent with that expected from an accretion
disk orbiting a central black hole. Both Schwarzschild and Kerr metrics
model the data equally well.
\item
a mean inclination of $i\sim 30$\deg. With a random distribution
we expect a mean of $60$~\deg.
\item
an emissivity function of $R^{-q}$ between $R_{\rm i}=6\ R_{\rm g}$ and
$R_{\rm o}=1000\ R_{\rm g}$, with $q\sim 2.5$. This
demonstrates that the line emission 
is concentrated in the inner disk. However, the data are not consistent
with a universal emissivity law.
\item
assuming the whole of the line arises from the disk, fits with a 
rest-energy of 6.7 keV are strongly disfavored. It is most likely 
that the line arises by fluorescence in near-neutral material.
\item
no strong evidence for a narrow component at 6.4 keV in addition
to the disk line, but with upper limits typically $\sim 100$~eV. 
The presence or absence of the narrow component affects neither 
the values of $q$ and $i$, nor the inferred differences in $q$,
but removes our constraints on the rest energy of the disk line.
\item
a Compton--``reflection'' component, which is expected to accompany the line,
is likely to be present, but does not affect the inferred line profiles.
\item
a mean equivalent width of $230$~eV when reflection is included
\item 
the underlying continuum in our sample to be characterized by a power law 
with a photon index $\Gamma = 1.9$, with a dispersion of $0.15$.
\end{itemize}

\section{Discussion}

\label{sec:discuss}

Since the discovery of active galactic nuclei, 
most workers have assumed that they are powered by accretion onto
a black hole. Such a model apparently presents
the only reasonable explanation for the observed phenomena
(Rees 1984 and references therein). 
However, as a black hole is not directly observable, one
is only able to infer its existence through the influence of
its physical properties: mass and spin. The velocities of
gas observed at $\sim 0.1$~pc from the nucleus of
some galaxies strongly indicate
the presence of huge masses in the central regions (e.g.
Ford \etal\ 1994; Miyoshi \etal\ 1995), with the only viable 
model appearing to
be a black hole. However, detection of the characteristic
relativistic effects of the black hole requires data which
probe much smaller scales, closer to the event horizon.
The velocities here should begin to approach the speed of light
and, furthermore, the gravitational field of the black hole
will cause time-dilation effects resulting in gravitational
redshift. In order to measure such phenomena, discrete spectral features
are necessary. The characteristic broadening and distortion of
those features due to the relativistic motions represent
arguably the most compelling test of the black hole paradigm.
The iron K$\alpha$ profile observed by Tanaka \etal\ (1995) 
in MCG-6-30-15 represents strong evidence for such a
hypothesis in that object. The similarity of the profile
in NGC 4151 (Yaqoob \etal\ 1995) tempts one to extrapolate
this model to  Seyfert galaxies as a class. But is such
a generalization appropriate?  

In this paper, we have presented a uniform analysis of a sample of
Seyfert 1 galaxies including MCG-6-30-15 and NGC~4151. We found
results consistent with previous work for MCG-6-30-15 and NGC~4151,
but have further demonstrated that asymmetric line profiles are
widespread. Fits to a specific physical model for the central regions
of the AGN, that of an accretion disk orbiting in the
Schwarzschild metric,
provide significant improvements over a simple
gaussian. This supports the assertion that the line is produced very
close to a central black hole. The average emissivity function implies that
$\sim$50~per cent of the line emission originates within
$\sim$20~$R_g$, and $\sim$80~per cent within $\sim$100~$R_g$. Special
and general relativistic effects thus afford a most satisfying
explanation for the observed broadening. Furthermore, the lack of any
strong blue shifts in our sample argues for a planar geometry, such as
a disk (e.g., Fabian \etal\ 1995). However, detailed calculations of
the line profiles expected in other geometries are beyond the scope of
this paper.

Detailed interpretation of the line profiles is model dependent.
However, the distribution of parameters for the sample can be compared
to those predicted by specific disk models.  For example, if these
Seyfert 1 galaxies were viewed at all inclinations, we would expect a
mean of 60\deg\ for the sample. Therefore, our disk line fits imply
that we tend to observe the Seyfert 1s at a more face-on inclination,
on average at $30$~degrees. This may be reconciled with Seyfert 1/2
unification schemes (e.g. Antonucci \& Miller 1985). In these models,
all Seyfert galaxies are surrounded by a thick, molecular torus which
obscures the line of sight in Seyfert 2 galaxies. If the torus axis
were aligned with that of the disk, we would not then expect to see
any edge--on Seyfert 1 galaxies.  The broad lines would be obscured by
the torus for lines of sight with $i>i_{op}$, the opening angle of the
torus. Based on our mean disk inclination, we would estimate
$cos{i_{op}}\sim (1-\cos{<i>}/2$), implying an opening angle of $\sim
45$\deg. This is in excellent agreement with estimates based on the
observations of ionization cones (Pogge 1989 and references therein)
in the optical. Although our data did not show strong evidence for a
narrow emission line -- expected from the torus -- neither did they
exclude a narrow component with the predicted strength (Ghisselini
\etal\ 1993; Krolik \etal\ 1993).  Conclusive evidence requires data
with higher spectral resolution and signal-to-noise ratio.  However,
the presence of a narrow component has little effect on the derived
parameters for the disk line, particularly $q$ and $i$.  An
alternative explanation for the restricted range in $i$ is that the
disk itself -- if geometrically thick -- obscures the edge-on
lines--of--sight.

Assuming that the disk line accounts for all of the emission at
iron $K\alpha$, the requirement for a rest-energy of 6.4~keV 
implies that the line arises by
fluorescence excited by the power law continuum. This raises the question 
as to how the disk can remain relatively neutral when exposed to such intense 
illumination. The immediate conclusion is that the density of the fluorescing 
material is extremely high. Photoionized accretion disks have been
discussed by Ross \& Fabian (1993) and Matt, Fabian \& Ross (1993).
For a standard $\alpha$-disc, we expect a density, n, at 
$20 R_{\rm g}$, of (Matt \etal\ 1993):

$$
n = 1.2\times 10^{15} \frac{\eta^{2}_{0.06}}
{\dot{M}_{0.1}^{2}M_{8}\alpha_{0.01}}\   cm^{-3}
$$

where \mdot$=0.1$\mdot$_{0.1}$ is the accretion rate in units of the Eddington
rate, $M=10^{8}M_{8}$ is the black hole mass in solar masses, 
$\eta=0.06\eta_{0.06}$ is the energy conversion efficiency and
$\alpha=0.01\alpha_{0.01}$ is the dimensionless viscosity parameter. 
The ionization parameter is approximately:

$$
\xi = \frac{L_{\rm X}}{nR^{2}} \sim 
\frac{L_{44} \dot{M_{0.1}}^{2}\alpha_{0.01}}{M_{8} \eta_{0.06}^{2}}
$$

where $L_{\rm X}=10^{44}L_{44}$ is the ionizing luminosity.  Below
$\xi\sim300$ a rest energy close to 6.4 keV is predicted for the line
(Matt \etal\ 1993). Thus even in the central regions, we typically
expect the mean ionization state for the disk to be low. However,
presumably the surface layers, where the bulk of the fluorescence
occurs, could be more highly ionized than is indicated by the above
relations. Assuming fixed energy fractions between the X--ray and
intrinsic disk emission, we infer that $\xi\propto$\mdot$^{3}$ and
thus that the accretion rate is the dominant parameter in determining
the ionization state of the disk. This implies that the \mdot\ values
in these Seyfert galaxies are sub-critical. Interestingly, for some
quasars, in which the accretion rates may be higher, there is evidence
that the peak of the emission line is higher, indicative of more
ionized iron (Nandra \etal\ 1996a).

Our fits to the Schwarzschild model have also shown that the
measured line profiles are not compatible with a universal emissivity
function, even when the inclination and a possible contribution of a narrow
line from another region are allowed to vary. This might be a consequence
of slight differences in geometry from source to source. For example, if
the true inner and outer radii of the disk are different to those
assumed, this can be compensated for in the fit by changing the 
apparent emissivity function. Alternatively, as we are dealing with a 
fluorescence line, the line emissivity depends upon the X--ray 
illumination, which could change from source to source.
For example, with a point-like X--ray source located above the disk, 
the line profile depends upon the height of the source 
(Matt \etal\ 1992). Our data are only well-fit by the disk line 
model if we allow the emission to be distributed over a range of radii. 
Emission at a single radius cannot account for the observed profiles in
the majority of cases. This argues against models where the
height of the X-ray source is very small, where only the inner
disk is illuminated. Furthermore, models in which the disk presents
significant solid angle to the X-ray source only at a small number of
radii are difficult to reconcile with our data. One might envisage such a 
scenario if the X-ray source were at the inner edge of the disk itself
or if the very innermost regions of the disk were puffed up by
radiation pressure. 

A simplistic ``one zone'' model for the X-ray source may not
be appropriate. A currently-popular model is of multiple regions
or ``hot spots'' covering the surface of the disk (e.g., Stern \etal\ 1995). 
The mean emissivity profile is harder to calculate in such a case, as
it depends on the distribution of hot spots within the X-ray producing
region and the variation of their density with radius. By observation,
we conclude that in such a model, this latter quantity
should vary approximately as $R^{-2.5}$. The challenge
is now for theoretical models to reproduce such a relationship and, one
would hope, relate it to the energy--generation mechanism for the X-rays.

An area where our observational data come into mild conflict
with the conventional wisdom regarding disk reflection
is that of the line equivalent width. The detailed calculations
of George \& Fabian (1991) and Matt \etal\ (1991) predicted
a value from the disk of $W_{\rm K\alpha} = 140$~eV for a continuum 
with $\Gamma=1.9$ and an inclination of $30\deg$.
This is a factor $1.2-2$ less than our observed mean of $230\pm 60$~eV. 
An obvious way to resolve this discrepancy is to revise our assumptions about
the metal abundances. The required increase in equivalent width
can be achieved by an enhancement in the iron abundance
of a factor 1.3-3.5 over the value of Morisson \& McCammon (1983)
which are usually employed (see Fig. 16 of George \& Fabian 1991). 
Indeed, using the updated solar abundances
of Anders \& Grevesse (1989), Reynolds, Fabian \& Inoue (1995a)
find that equivalent widths of $\sim 250$~eV could be achieved
without any enhancement in iron. The possible contribution of
$\sim 30$~eV of a narrow line from another region would also ease
this discrepancy somewhat.

The mean iron abundance derived here argues that the material being
accreted is relatively metal-rich, consistent with recent broad-line
models for quasars (Hamann \& Ferland 1993). Since the central 
regions of most spiral galaxies have similar metallicities, 
but the abundances drop with increasing radius 
(McCall, Rybski \& Shields 1985; Shields, Skillman \& Kennicutt 1991). 
This tends to imply that the accretion fuel is produced by stellar mass-loss 
relatively close to the nucleus and is not brought in from 
further out in the galaxy. The required accretion rate is
\mdot$ = 3 \times 10^{-2} L_{44} \eta_{0.06}$~\Msun yr$^{-1}$ 
(here $L_{44}$ represents the accretion luminosity, rather than the 
ionizing luminosity). The mass-loss rate is expected to be
\mdot$\sim 10^{-2} M_{10}$~\Msun yr$^{-1}$, where $M_{10}$ is the mass
of stars in units of $10^{10}$~\Msun, which seems plausible.

Our fits have also allowed us to define the continuum shape in
the 3-10 keV band. We find a mean slope of $\sim 1.8$ with a
power-law plus line model. This lies between the values determined
by \he\ and \ex\ (using a single power-law; Mushotzky 1984;
Turner \& Pounds 1989) and that determined from
\gi, where a more complex model was employed, including 
a warm absorber and reflection. When we include the reflection component
in our \asca\ fits, we find $\Gamma = 1.9$, in
excellent agreement with the \gi\ data, which were
arguably more appropriate for continuum determination. These results indicate
that the reflection component has a substantial effect on the 
determination of the continuum using \asca\ data in the 3-10 keV band,
but that the warm absorber does not. We will return to this latter point
in paper III. Our data have implications for the physical models of
the X-ray continuum in AGN. For example, the disk/corona model of 
Haardt \& Maraschi (1993) has been successful at explaining the mean 
intrinsic slope for Seyfert galaxies of $\Gamma=1.9$. However, 
Stern \etal\ (1995) have pointed out that this model has some difficulties 
in explaining a {\it range} of observed slopes, and suggest a solution
in which a number of distinct active regions illuminate the disk.
Our data confirm the range of slopes observed by \gi, and thus lend
support to this hypothesis.

\acknowledgements We thank the \asca\ team for their operation of the
satellite, the \asca\ GOF and Keith Gendreau at NASA/GSFC for
discussions regarding the calibration and assistance in data analysis,
Giorgio Matt and Ari Laor for making their disk line models available
and Michael Akritas, Director of the SCCA operated at the Department
of Statistics, Penn State University for helpful discussions.  We are
also grateful to the referee, Jules Halpern, for a number of
insightful comments.  We acknowledge the financial support of the
National Research Council (KN) and Universities Space Research
Association (IMG, TJT, TY). This research has made use of the Simbad
database, operated at CDS, Strasbourg, France; of the NASA/IPAC
Extragalactic database, which is operated by the Jet Propulsion
Laboratory, Caltech, under contract with NASA; and of data obtained
through the HEASARC on-line service, provided by NASA/GSFC.

\clearpage

\begin{deluxetable}{l l c c c c c}

\tablecaption{The \asca\ Seyfert 1 sample. \label{tab:sample}}

\tablehead{
\colhead{Name} & \colhead{RA} &  \colhead{DEC} & 
\colhead{Redshift} & \colhead{\nh (Gal)$^{a}$} & 
\colhead{N$^{b}_{\rm obs}$/N$_{\rm tot}$} & $\log L_{\rm X}^{c}$ \\
& \colhead{(1950)} & \colhead{(1950)} & \colhead{(V93)} & 
\colhead{$10^{20}$~\pcmsq} & & erg s$^{-1}$ 
}

\startdata
Mrk 335     & 00 03 45.1 &  19 55 29  & 0.025 & 4.0  & 1/1 & 43.42 \nl
Fairall 9   & 01 21 51.0 & -59 03 54  & 0.046 & 3.0  & 1/1 & 44.26 \nl
3C 120      & 04 30 31.5 &  05 14 59  & 0.033 & 12.3 & 1/1 & 44.34 \nl
NGC 3227    & 10 20 46.9 &  20 07 06  & 0.003 & 2.2  & 1/1 & 42.01 \nl
NGC 3516    & 11 03 22.8 &  72 50 18  & 0.009 & 2.9  & 1/1 & 43.43 \nl
NGC 3783    & 11 36 33.0 & -37 27 42  & 0.009 & 8.9  & 2/2 & 43.25 \nl
NGC 4051    & 12 00 36.3 &  44 48 35  & 0.002 & 1.3  & 1/1 & 41.56 \nl
NGC 4151    & 12 08 00.9 &  39 40 59  & 0.003 & 2.1  & 3/6 & 42.97 \nl
Mrk 766     & 12 15 55.7 &  30 05 25  & 0.012 & 1.6  & 1/1 & 43.08 \nl
NGC 4593    & 12 37 04.6 & -05 04 11  & 0.009 & 2.0  & 1/1 & 43.06 \nl
MCG-6-30-15 & 13 33 01.4 & -34 02 27  & 0.008 & 4.1  & 2/2 & 43.07 \nl
IC 4329A    & 13 46 28.2 & -30 03 42  & 0.016 & 10.4 & 1/1 & 43.94 \nl
NGC 5548    & 14 15 43.6 &  25 21 58  & 0.017 & 1.7  & 1/1 & 43.76 \nl
Mrk 841     & 15 01 36.2 &  10 37 53  & 0.036 & 2.2  & 2/2 & 43.82 \nl
NGC 6814    & 19 39 55.7 & -10 26 35  & 0.006 & 9.8  & 1/3 & 41.28 \nl
Mrk 509     & 20 41 26.1 & -10 54 16  & 0.035 & 4.2  & 1/1 & 44.38 \nl
NGC 7469    & 23 00 44.4 &  08 36 17  & 0.017 & 4.8  & 1/3 & 43.60 \nl
MCG-2-58-22 & 23 02 07.2 & -08 57 20  & 0.047 & 3.4  & 1/1 & 44.14 \nl

\tablenotetext{a}{Galactic H{\sc i} column density from 21cm measurements}
\tablenotetext{b}{Number of observations for which
adequate data were obtained/Total number of observations}
\tablenotetext{c}{Mean 2-10 keV luminosity}

\tablerefs{V93: Veron-Cetty \& Veron (1993);
\nh\ values are from Elvis, Lockman \& Wilkes (1989);
Stark \etal\ (1992) or the HEASARC online service.}

\enddata
\end{deluxetable}

\clearpage

\begin{deluxetable}{lllllllr}
\tablecaption{Narrow line fits \label{tab:nl}}
\tablehead{
\colhead{Name} & \colhead{A$^{a}$} & \colhead{$\Gamma^{b}_{3-10}$} & 
\colhead{E$^{c}_{\rm K \alpha}$} & \colhead{$I^{d}_{\rm K\alpha}$} & 
\colhead{W$^{e}_{\rm K\alpha}$} & \colhead{\chisq$_{\rm NL}$/d.o.f.} & 
\colhead{$F^{f}_{2}$} \\ 
 & & &
\colhead{(keV)} & &
\colhead{(eV)} & & 
}

\startdata
Mrk 335        & 0.29 & $1.87\pm0.10$ & $6.37^{+0.16}_{-0.16}$ &  
  $1.2^{+0.7}_{-0.8}$ & $110^{+60}_{-70}$ & 227.9/252 & 5.0 \nl
Fairall 9      & 0.60 & $1.83\pm0.06$ & $6.33^{+0.15}_{-0.08}$ & 
  $2.9^{+0.9}_{-1.0}$ & $120^{+40}_{-40}$ & 640.6/610 & 15.0\nl
3C 120         & 1.31 & $1.79\pm0.03$ & $6.37^{+0.05}_{-0.05}$ & 
  $3.5^{+0.6}_{-1.7}$ & $70^{+10}_{-30}$ & 881.9/950 & 15.9 \nl
NGC 3227       & 0.48 & $1.52\pm 0.04$ & $6.42^{+0.04}_{-0.04}$ & 
  $3.5^{+0.9}_{-0.9}$ & $120^{+30}_{-30}$ & 885.3/920 & 30.2 \nl
NGC 3516       & 2.03 & $1.74 \pm 0.03$ & $6.37^{+0.03}_{-0.03}$ & 
  $7.9^{+2.1}_{-1.3}$ & $90^{+20}_{-10}$ & 1055.3/1004 & 42.8 \nl
NGC 3783(1)    & 1.03 & $1.57\pm0.05$ & $6.38^{+0.04}_{-0.04}$ & 
  $8.3^{+1.8}_{-1.8}$ & $150^{+30}_{-30}$ & 800.8/788 & 34.2 \nl
NGC 3783(2)    & 1.11 & $1.51\pm 0.05$ & $6.16^{+0.14}_{-0.06}$ &
  $5.7^{+2.1}_{-2.1}$ & $80^{+30}_{-30}$ & 715.1/743 & 27.4 \nl
NGC 4051       & 0.73 & $1.92\pm 0.06$ & $6.44^{+0.07}_{-0.07}$ & 
  $2.8^{+0.7}_{-1.2}$ & $140^{+35}_{-60}$ & 672.4/657 & 13.3 \nl
NGC 4151(2)    & 3.75 & $1.42\pm 0.12$ & $6.40^{+0.02}_{-0.03}$ &
  $29^{+5}_{-7}$      & $110^{+20}_{-30}$ & 760.4/693 & 63.2 \nl
NGC 4151(4)    & 6.31 & $1.59\pm0.07$ & $6.36^{+0.05}_{-0.05}$ &
  $32^{+9}_{-6}$      & $90^{+30}_{-20}$ & 1409.1/1401 & 47.2 \nl
NGC 4151(5)    & 7.73 & $1.65\pm0.05$ & $6.37^{+0.02}_{-0.02}$ &  
  $31^{+6}_{-5}$      & $90^{+20}_{-20}$ & 1348.6/1228 & 45.0 \nl
Mrk 766        & 0.75 & $2.00\pm0.05$ & $6.37^{+0.03}_{-0.03}$ & 
  $2.6^{+1.0}_{-0.7}$ & $140^{+50}_{-40}$ & 584.3/682 & 21.2\nl 
NGC 4593       & 0.92 & $1.78\pm0.05$ & $6.35^{+0.05}_{-0.05}$ & 
  $3.1^{+1.3}_{-1.1}$ & $90^{+40}_{-30}$ & 827.4/868 & 12.7 \nl
MCG-6-30-15(1) & 1.64 & $1.94\pm0.05$ & $6.26^{+0.07}_{-0.10}$ & 
  $3.2^{+1.4}_{-1.3}$ & $70^{+30}_{-30}$ & 943.4/845 & 9.0\nl
MCG-6-30-15(2) & 1.11 & $1.81\pm0.05$ & $6.38^{+0.07}_{-0.10}$ & 
  $3.8^{+0.9}_{-2.0}$ & $100^{+20}_{-50}$ & 829.7/802 & 9.7\nl
IC 4329A       & 1.95 & $1.70\pm0.03$ & $6.32^{+0.05}_{-0.05}$ & 
  $7.3^{+1.4}_{-1.6}$ & $80^{+20}_{-20}$ & 1297.7/1249 & 38.5\nl
NGC 5548       & 1.24 & $1.75\pm0.04$ & $6.39^{+0.04}_{-0.04}$ &
  $4.6^{+1.3}_{-1.3}$ & $90^{+30}_{-30}$ & 973.6/1000 & 21.6 \nl
Mrk 841(1)     & 0.38 & $1.84\pm0.09$ & $6.37^{+0.12}_{-0.02}$ & 
  $2.4^{+0.9}_{-0.8}$ & $170^{+60}_{-60}$ & 378.1/368 & 7.0\nl
Mrk 841(2)     & 0.21 & $1.57\pm0.10$ & $6.53^{+0.07}_{-0.16}$ & 
  $1.4^{+1.0}_{-0.9}$ & $120^{+90}_{-80}$ & 181.9/203 & 4.7 \nl
NGC 6814(1)    & 0.02 & $1.54\pm0.26$ & $5.9^{+0.6}_{-0.2}$ & 
  $0.4^{+0.3}_{-0.3}$  & $230^{+170}_{-170}$ & 68.1/76 & 4.1 \nl
\tablebreak
Mrk 509        & 1.21 & $1.76\pm0.05$ & $6.37^{+0.07}_{-0.07}$ & 
  $3.5^{+0.7}_{-2.4}$ & $70^{+10}_{-50}$ & 819.0/896 & 7.4 \nl
NGC 7469(2)    & 0.76 & $1.78\pm 0.07$ & $6.39^{+0.06}_{-0.05}$ &
  $3.8^{+1.6}_{-1.5}$ & $130^{+50}_{-50}$ & 501.1/451 & 19.7\nl
MCG-2-58-22    & 0.29 & $1.58 \pm 0.08$ & $6.33^{+0.18}_{-0.07}$ &
  $2.6^{+0.8}_{-1.8}$ & $150^{+100}_{-50}$ & 408.4/388 & 7.8\nl
\enddata

\tablenotetext{a}{Power law flux at 1 keV; 
  $10^{-3}$ ph cm$^{-2}$ s$^{-1}$ keV$^{-1}$}
\tablenotetext{b}{Power-law index}
\tablenotetext{c}{Energy of the emission line in the rest-frame of the source}
\tablenotetext{d}{Flux of the line; $10^{-5}$ ph cm$^{-2}$ s$^{-1}$}
\tablenotetext{e}{Line equivalent width}
\tablenotetext{f}{F-statistic for the addition of two free parameters}
\tablecomments{Fits were undertaken 
in the 3-10 keV range. Absorption is by Galactic \nh\ only, except in the case
of NGC 4151, where the column was left free. 
Error bars on the fit parameters are 68 per
cent confidence limits for 3 interesting parameters (\delchi=$3.5$).}

\end{deluxetable}

\clearpage

\begin{deluxetable}{llllrr}

\tablecaption{Broad line fits
\label{tab:bl}
}

\tablehead{
\colhead{Name} & \colhead{$\Gamma_{3-10}$} & 
\colhead{E$_{\rm K \alpha}$} & \colhead{W$_{\rm K\alpha}$} & 
\colhead{$\sigma^{a}_{\rm K\alpha}$} & \colhead{$F^{b}_{1}$}
}

\startdata
Mrk 335        & \nodata & \nodata & \nodata & $<1.1$ & 0.9 \nl
Fairall 9      & $1.91^{+0.08}_{-0.08} $ & $6.36^{+0.14}_{-0.14}$ & 
               $320^{+130}_{-80}$ & $0.35^{+0.18}_{-0.11}$ & 21.7\nl
3C 120         & $1.89^{+0.08}_{-0.06}$ & $6.39^{+0.20}_{-0.20}$ &
               $330^{+200}_{-130}$ & $0.74^{+0.34}_{-0.27}$ & 39.9 \nl
NGC 3227       & $1.52^{+0.08}_{-0.03}$ & $6.30^{+0.12}_{-0.12}$ &
               $180^{+60}_{-50}$ & $0.25^{+0.12}_{-0.10}$ & 7.6 \nl
NGC 3516       & $1.83^{+0.05}_{-0.05}$ & $6.12^{+0.13}_{-0.15}$ & 
               $340^{+130}_{-110}$ & $0.64^{+0.22}_{-0.19}$ & 40.1 \nl
NGC 3783(1)    & $1.70^{+0.11}_{-0.08}$ & $6.11^{+0.17}_{-0.18}$ & 
               $510^{+490}_{-190}$ & $0.71^{+0.35}_{-0.28}$ &19.3 \nl
NGC 3783(2)    & $1.60^{+0.04}_{-0.08}$ & $5.93^{+0.26}_{-0.39}$ &
               $360^{+300}_{-180}$ & $0.70^{+0.48}_{-0.36}$ & 24.3 \nl
NGC 4051       & \nodata & \nodata & \nodata & $<1.4$ & 1.8 \nl
NGC 4151(2)    & $1.44^{+0.14}_{-0.13}$ & $6.39^{+0.05}_{-0.05}$ &
               $130^{+60}_{-40}$ & $0.10^{+0.08}_{-0.10}$ & 4.3 \nl
NGC 4151(4)    & $1.57^{+0.10}_{-0.11}$ & $6.14^{+0.13}_{-0.18}$ & 
               $340^{+160}_{-110}$ & $0.56^{+0.25}_{-0.19}$ & 52.6 \nl
NGC 4151(5)    & $1.58^{+0.09}_{-0.11}$ & $5.92^{+0.12}_{-0.17}$ & 
               $330^{+140}_{-90}$ & $0.64^{+0.22}_{-0.15}$ & 50.1 \nl
Mrk 766        & $2.16^{+0.15}_{-0.11}$ & $6.03^{+0.24}_{-0.25}$ & 
               $550^{+390}_{-230}$ & $0.72^{+0.36}_{-0.25}$ & 16.0 \nl
NGC 4593       & \nodata & \nodata & \nodata & $<1.1$ & 3.4\nl
MCG-6-30-15(1) & $2.04^{+0.10}_{-0.07}$ & $6.06^{+0.22}_{-0.21}$ & 
               $320^{+110}_{-90}$ & $0.62^{+0.24}_{-0.18}$ & 27.3 \nl
MCG-6-30-15(2) & $1.95^{+0.12}_{-0.08}$ & $5.80^{+0.25}_{-0.27}$ & 
               $460^{+280}_{-140}$ & $0.88^{+0.43}_{-0.23}$ & 32.1 \nl
IC 4329A       & $1.71^{+0.03}_{-0.03}$ & $6.34^{+0.07}_{-0.05}$ & 
               $110^{+40}_{-30}$ & $0.14^{+0.12}_{-0.08}$ & 10.7 \nl
NGC 5548       & $1.79^{+0.05}_{-0.05}$ & $6.41^{+0.16}_{-0.14}$ & 
               $170^{+60}_{-40}$ & $0.34^{+0.19}_{-0.12}$ & 5.8 \nl
Mrk 841(1)     & $2.00^{+0.20}_{-0.15}$ & $6.15^{+0.25}_{-0.28}$ & 
               $580^{+400}_{-290}$ & $0.63^{+0.35}_{-0.32}$ & 6.9 \nl
Mrk 841(2)     & \nodata & \nodata & \nodata & $<\infty$ & 0.9 \nl
NGC 6814(1)    & $1.9^{+2.0}_{-0.5}$ & $6.3^{+0.7}_{-0.4}$ & 
               $1100^{+2100}_{-800}$ & $0.6^{+0.8}_{-0.4}$ & 8.9 \nl
Mrk 509        & $1.82^{+0.09}_{-0.07}$ & $6.35^{+0.25}_{-1.02}$ & 
               $210^{+190}_{-130}$ & $0.47^{+0.50}_{-0.30}$ & 14.2 \nl
NGC 7469(2)    & $1.84^{+0.12}_{-0.10}$ & $6.39^{+0.24}_{-0.21}$ &
               $280^{+190}_{-170}$ & $0.37^{+0.31}_{-0.37}$ & 3.7 \nl
MCG-2-58-22    & \nodata & \nodata & \nodata & $<0.85$ & 1.7 \nl
\enddata 

\tablenotetext{a}{Width of the gaussian (keV)}
\tablenotetext{b}{F statistic for the addition of $\sigma$ as a free
parameter}

\tablecomments{Fits with a power-law plus broad-line to the \asca\ data
in the 3-10 keV range. Absorption is by Galactic \nh\ only, except in the case
of NGC 4151, where the column was left free. 
Error bars on the fit parameters are 68 per
cent confidence limits for 4 interesting parameters (\delchi=$4.7$).}

\end{deluxetable}

\clearpage

\begin{deluxetable}{lllllr}

\tablecaption{Disk line fits; Schwarzschild geometry; free emissivity
\label{tab:dl}
}

\tablehead{
\colhead{Name} & \colhead{$\Gamma_{3-10}$} & 
\colhead{W$_{\rm K\alpha}$} & \colhead{$i^{a}$} &  
\colhead{$q^{b}$} & \delchi$^{c}$
}

\startdata
Mrk 335        & $1.90^{+0.17}_{-0.13}$ & $230^{+280}_{-170}$ & 
  $23^{+67}_{-23}$ & $2.2^{+\infty}_{-\infty}$ & 1.3  \nl
Fairall 9      & $1.90^{+0.10}_{-0.08}$ & $350^{+180}_{-130}$ & 
  $45^{+45}_{-21}$ & $1.9^{+0.5}_{-1.7}$ & -0.8 \nl
3C 120         & $1.87^{+0.06}_{-0.05}$ & $290^{+110}_{-70}$ & 
  $60^{+30}_{-14}$ & $2.2^{+0.3}_{-0.5}$ & -3.0  \nl
NGC 3227       & $1.55^{+0.05}_{-0.05}$ & $210^{+80}_{-80}$ & 
  $20^{+10}_{-10}$ & $2.0^{+0.4}_{-0.8}$ & 4.4 \nl
NGC 3516       & $1.80^{+0.04}_{-0.04}$ & $330^{+90}_{-80}$ & 
  $27^{+4}_{-5}$ & $2.7^{+0.7}_{-0.4}$ & 21.5 \nl
NGC 3783(1)    & $1.64^{+0.09}_{-0.06}$ & $370^{+330}_{-110}$ & 
  $21^{+10}_{-7}$ & $2.5^{+\infty}_{-0.4}$ & 10.2 \nl
NGC 3783(2)    & $1.58^{+0.07}_{-0.07}$ & $430^{+140}_{-120}$ & 
  $32^{+3}_{-13}$ & $5.2^{+\infty}_{-1.8}$ & 8.0 \nl
NGC 4051       & $2.00^{+0.08}_{-0.08}$ & $460^{+130}_{-230}$ & 
  $33^{+5}_{-13}$ & $4.4^{+\infty}_{-2.1}$ & 13.0 \nl
NGC 4151(2)    & $1.43^{+0.14}_{-0.13}$ & $190^{+90}_{-50}$ & 
  $18^{+3}_{-8}$ & $2.0^{+0.4}_{-2.0}$ & 3.9 \nl
NGC 4151(4)    & $1.36^{+0.11}_{-0.11}$ & $550^{+90}_{-120}$ & 
  $33^{+2}_{-8}$ & $5.0^{+1.9}_{-1.3}$ & 19.8 \nl
NGC 4151(5)    & $1.56^{+0.08}_{-0.09}$ & $320^{+70}_{-60}$ & 
  $17^{+6}_{-5}$ & $2.8^{+0.4}_{-0.3}$ & 28.5 \nl
Mrk 766        & $2.13^{+0.11}_{-0.08}$ & $610^{+210}_{-200}$ & 
  $35^{+2}_{-4}$ & $3.7^{+1.9}_{-1.0}$ & 7.9 \nl
NGC 4593       & $1.83^{+0.08}_{-0.07}$ & $240^{+140}_{-140}$ & 
  $45^{+45}_{-45}$ & $2.2^{+0.7}_{-\infty}$ & -0.3 \nl
MCG-6-30-15(1) & $2.04^{+0.07}_{-0.06}$ & $430^{+160}_{-160}$ & 
  $33^{+4}_{-4}$ & $3.1^{+1.1}_{-0.6}$ & 7.8 \nl
MCG-6-30-15(2) & $1.91^{+0.08}_{-0.06}$ & $450^{+140}_{-140}$ & 
  $34^{+3}_{-5}$ & $4.2^{+2.0}_{-1.1}$ & 3.0 \nl
IC 4329A       & $1.72^{+0.03}_{-0.03}$ & $150^{+60}_{-40}$ & 
  $22^{+15}_{-22}$ & $2.1^{+0.4}_{-1.0}$ & 2.6 \nl
NGC 5548       & $1.79^{+0.05}_{-0.05}$ & $200^{+70}_{-90}$ & 
  $47^{+43}_{-41}$ & $1.9^{+0.6}_{-0.7} $ & -0.6 \nl
Mrk 841(1)     & $1.95^{+0.14}_{-0.13}$ & $510^{+260}_{-220}$ & 
  $27^{+7}_{-8}$ & $2.6^{+6.0}_{-2.8}$ & 6.4 \nl
Mrk 841(2)     & $1.62^{+0.16}_{-0.13}$ & $500^{+310}_{-290}$ & 
  $38^{+2}_{-12}$ & $10^{+\infty}_{-\infty}$ & 3.6 \nl
NGC 6814(1)    & \nodata & \nodata & \nodata & \nodata & \nodata  \nl
Mrk 509        & $1.81^{+0.11}_{-0.07}$ & $170^{+160}_{-90}$ & 
  $41^{+49}_{-32}$ & $2.1^{+0.7}_{-2.0}$ & -1.4 \nl
NGC 7469(2)    & $1.80^{+0.10}_{-0.09}$ & $210^{+310}_{-130}$ & 
  $19^{+71}_{-19}$ & $1.8^{+1.0}_{-\infty}$ & -1.8 \nl
MCG-2-58-22    & $1.60^{+0.09}_{-0.11}$ & $160^{+440}_{-80}$ & 
  $71^{+19}_{-71}$ & $-10^{+15}_{-\infty}$ & 0.3 \nl

\tablenotetext{a}{Inclination of the disk to the line--of--sight}
\tablenotetext{b}{The emissivity varies as $R^{-q}$}
\tablenotetext{c}{Difference in \chisq compared to the broad-line fit
(\chisq$_{\rm BL}-$\chisq$_{\rm S}$)}

\tablecomments{Absorption is by Galactic \nh\ only, except in the case
of NGC 4151, where the column was left free. 
Error bars on the fit parameters are 68 per
cent confidence limits for 4 interesting parameters (\delchi=$4.7$).}

\enddata
\end{deluxetable}

\clearpage

\begin{deluxetable}{llllrr}

\tablecaption{Disk line fits; Schwarzschild geometry; q=2.5
\label{tab:dl-25}
}

\tablehead{
\colhead{Name} & \colhead{$\Gamma_{3-10}$} & 
\colhead{W$_{\rm K\alpha}$} & \colhead{$i$} &  
$F^{a}_{1}$ & \chisq$_{\rm S25}$/d.o.f
}

\startdata
Mrk 335        & $1.92^{+0.14}_{-0.11}$ & $280^{+160}_{-150}$ & 
  $24^{+13}_{-24}$ & 0.2 & 227.9/252 \nl
Fairall 9      & $1.91^{+0.08}_{-0.07}$ & $390^{+130}_{-120}$ & 
  $32^{+12}_{-8}$ & 3.8 & 623.3/610 \nl
3C 120         & $1.89^{+0.05}_{-0.04}$ & $350^{+90}_{-100}$ & 
  $59^{+10}_{-13}$ & 3.8 & 852.7/950 \nl
NGC 3227       & $1.56^{+0.04}_{-0.05}$ & $260^{+60}_{-60}$ & 
  $23^{+8}_{-6}$ & 6.6 & 879.9/920 \nl
NGC 3516       & $1.80^{+0.03}_{-0.04}$ & $290^{+50}_{-40}$ & 
  $26^{+4}_{-4}$ & 1.4 & 994.0/1004 \nl
NGC 3783(1)    & $1.64^{+0.05}_{-0.05}$ & $370^{+70}_{-70}$ & 
  $21^{+5}_{-5}$ & 0.0 & 771.4/788 \nl
NGC 3783(2)    & $1.55^{+0.06}_{-0.05}$ & $250^{+60}_{-70}$ & 
  $9^{+11}_{-9}$ & 10.6 & 694.3/743 \nl
NGC 4051       & $1.97^{+0.06}_{-0.06}$ & $310^{+80}_{-110}$ & 
  $27^{+7}_{-11}$ & 3.1 & 660.9/657 \nl
NGC 4151(2)    & $1.41^{+0.14}_{-0.12}$ & $260^{+60}_{-50}$ & 
  $20^{+5}_{-5}$ & 5.6 & 758.2/694 \nl
NGC 4151(4)    & $1.52^{+0.08}_{-0.07}$ & $290^{+50}_{-40}$ & 
  $21^{+5}_{-6}$ & 6.1 & 1343.8/1400 \nl
NGC 4151(5)    & $1.60^{+0.07}_{-0.07}$ & $250^{+40}_{-30}$ & 
  $15^{+4}_{-5}$ & 7.4 & 1275.1/1227 \nl
Mrk 766        & $2.09^{+0.08}_{-0.07}$ & $440^{+110}_{-120}$ & 
  $35^{+5}_{-5}$ & 9.2 & 570.7/682 \nl
NGC 4593       & $1.84^{+0.06}_{-0.06}$ & $280^{+120}_{-100}$ & 
  $45^{+12}_{-11}$ & 1.1 & 825.5/868 \nl
MCG-6-30-15(1) & $2.02^{+0.05}_{-0.05}$ & $320^{+70}_{-100}$ & 
  $34^{+5}_{-5}$ & 4.5 & 910.8/845 \nl
MCG-6-30-15(2) & $1.87^{+0.06}_{-0.05}$ & $260^{+100}_{-80}$ & 
  $33^{+8}_{-25}$ & 14.4 & 809.2/802 \nl
IC 4329A       & $1.72^{+0.03}_{-0.02}$ & $180^{+40}_{-30}$ & 
  $26^{+7}_{-8}$ & 4.1 & 1289.1/1249 \nl
NGC 5548       & $1.79^{+0.05}_{-0.04}$ & $220^{+90}_{-60}$ & 
  $39^{+10}_{-11}$ & 4.5 & 973.0/1000 \nl
Mrk 841(1)     & $1.95^{+0.10}_{-0.11}$ & $500^{+140}_{-150}$ & 
  $27^{+6}_{-6}$ & 0.1 & 364.8/368 \nl
Mrk 841(2)     & $1.61^{+0.12}_{-0.11}$ & $250^{+160}_{-160}$ & 
  $30^{+9}_{-15}$ & 4.4 & 181.4/203 \nl
NGC 6814(1)    & \nodata & \nodata & \nodata & \nodata  & \nodata \nl
Mrk 509        & $1.81^{+0.06}_{-0.05}$ & $210^{+120}_{-100}$ & 
  $40^{+48}_{-24}$ & 2.0 & 809.4/894 \nl
NGC 7469(2)    & $1.86^{+0.10}_{-0.09}$ & $370^{+180}_{-}$ & 
  $45^{+11}_{-29}$ & 1.7 &  500.7/451 \nl
MCG-2-58-22    & $1.62^{+0.11}_{-0.08}$ & $310^{+170}_{-170}$ & 
  $41^{+10}_{-15}$ & 0.5 & 406.8/388 \nl

\tablenotetext{a}{F statistic for the addition of $q$ as a free parameter}

\tablecomments{Absorption is by Galactic \nh\ only, except in the case
of NGC 4151, where the column was left free.  Error bars on the fit
parameters are 68 per cent confidence limits for 3 interesting
parameters (\delchi=$3.5$).}

\enddata
\end{deluxetable}

\clearpage

\begin{deluxetable}{lllllll}

\tablecaption{Disk line fits; Schwarzschild geometry; with reflection
\label{tab:dl-ref}
}

\tablehead{
\colhead{Name} & \colhead{$\Gamma_{3-10}$} & 
\colhead{W$_{\rm K\alpha}$} & \colhead{$i$} &  
\colhead{$q$} & \colhead{\delchi$^a$} & \colhead{Flux$^b$} \\
}

\startdata
Mrk 335        & $2.05^{+0.14}_{-0.23}$ & $230^{+310}_{-190}$ & 
  $22^{+68}_{-22}$ & $2.3^{+\infty}_{-\infty}$ & 1.1 & $0.95\pm 0.12$ \nl
Fairall 9      & $2.02^{+0.10}_{-0.18}$ & $320^{+240}_{-110}$ & 
  $46^{+44}_{-27}$ & $1.9^{+0.5}_{-2.1}$ & 0.6 & $2.05\pm 0.27$  \nl
3C 120         & $1.97^{+0.06}_{-0.14}$ & $260^{+240}_{-70}$ & 
  $60^{+30}_{-13}$ & $2.3^{+0.4}_{-0.6}$ & 0.0 & $4.69\pm 0.61$ \nl
NGC 3227       & $1.67^{+0.06}_{-0.05}$ & $190^{+80}_{-60}$ & 
  $20^{+8}_{-14}$ & $2.0^{+0.5}_{-1.0}$ & 3.9 & $2.64\pm 0.33$ \nl
NGC 3516       & $1.93^{+0.04}_{-0.04}$ & $320^{+120}_{-90}$ & 
  $27^{+6}_{-7}$ & $2.8^{+\infty}_{-0.4}$ & -2.5 & $6.82\pm 1.03$ \nl
NGC 3783(1)    & $1.77^{+0.08}_{-0.06}$ & $590^{+130}_{-140}$ & 
  $35^{+2}_{-18}$ & $8.5^{+\infty}_{-4.6}$ & 2.9 & $5.12\pm 0.67$ \nl
NGC 3783(2)    & $1.70^{+0.06}_{-0.07}$ & $400^{+120}_{-120}$ & 
  $32^{+3}_{-15}$ & $6.4^{+\infty}_{-2.6}$ & 2.0 & $6.12\pm 0.80$ \nl
NGC 4051       & $2.11^{+0.08}_{-0.08}$ & $420^{+160}_{-210}$ & 
  $34^{+3}_{-14}$ & $5.1^{+\infty}_{-2.6}$ & 0.6 & $2.13\pm 0.28$ \nl
NGC 4151(2)    & $1.60^{+0.13}_{-0.13}$ & $150^{+90}_{-70}$ & 
  $17^{+12}_{-17}$ & $1.8^{+0.6}_{-\infty}$ & 5.6 & $24.3\pm 3.1$ \nl
NGC 4151(4)    & $1.54^{+0.10}_{-0.11}$ & $490^{+100}_{-90}$ & 
  $33^{+2}_{-18}$ & $5.4^{+3.0}_{-1.5}$ & 2.3 & $27.8\pm 3.6$ \nl
NGC 4151(5)    & $1.72^{+0.09}_{-0.08}$ & $290^{+70}_{-60}$ & 
  $17^{+5}_{-5}$ & $2.9^{+0.3}_{-0.3}$ & 1.1 & $32.6\pm 4.2$ \nl
Mrk 766        & $2.25^{+0.10}_{-0.09}$ & $580^{+200}_{-200}$ & 
  $34^{+3}_{-3}$ & $3.9^{+2.9}_{-1.0}$ & 1.4 & $1.95\pm 0.25$ \nl
NGC 4593       & $1.91^{+0.09}_{-0.05}$ & $90^{+70}_{-50}$ & 
  $0^{+79}_{-0}$ & $1.5^{+1.3}_{-\infty}$ & 5.1 & $3.38\pm 0.44$ \nl
MCG-6-30-15(1) & $2.16^{+0.06}_{-0.07}$ & $400^{+150}_{-220}$ & 
  $33^{+3}_{-5}$ & $3.4^{+1.3}_{-0.8}$ & 3.2 & $4.69\pm 0.61$ \nl
MCG-6-30-15(2) & $2.03^{+0.07}_{-0.07}$ & $410^{+140}_{-140}$ & 
  $34^{+3}_{-6}$ & $4.4^{+3.0}_{-1.1}$ & -0.4 & $3.83\pm 0.50$ \nl
IC 4329A       & $1.84^{+0.03}_{-0.03}$ & $120^{+60}_{-40}$ & 
  $17^{+14}_{-17}$ & $2.1^{+2.8}_{-1.1}$ & 3.3 & $8.12\pm 1.06$ \nl
NGC 5548       & $1.89^{+0.04}_{-0.05}$ & $80^{+30}_{-30}$ & 
  $0^{+76}_{-0}$ & $-9.0^{+11.5}_{-\infty} $ & 6.2 & $4.71\pm 0.61$\nl
Mrk 841(1)     & $2.07^{+0.13}_{-0.13}$ & $470^{+430}_{-220}$ & 
  $27^{+7}_{-9}$ & $2.6^{+\infty}_{-1.3}$ & 1.0 & $1.29\pm 0.17$ \nl
Mrk 841(2)     & $1.74^{+0.16}_{-0.13}$ & $440^{+310}_{-280}$ & 
  $38^{+2}_{-16}$ & $10^{+\infty}_{-\infty}$ & -1.7 & $1.07\pm 0.14 $ \nl
NGC 6814(1)    & \nodata & \nodata & \nodata & \nodata & \nodata 
  & $0.13\pm 0.02$\nl
Mrk 509        & $1.92^{+0.06}_{-0.07}$ & $110^{+100}_{-70}$ & 
  $27^{+63}_{-27}$ & $2.2^{+1.2}_{-\infty}$ & 1.6 & $4.58\pm 0.59$ \nl
NGC 7469(2)    & $1.91^{+0.09}_{-0.08}$ & $120^{+60}_{-60}$ & 
  $0^{+89}_{-0}$ & $-9.0^{+10.7}_{-\infty}$ & 4.1 & $2.78\pm 0.36$ \nl
MCG-2-58-22    & $1.70^{+0.16}_{-0.18}$ & $140^{+160}_{-80}$ & 
  $46^{+44}_{-46}$ & $-4.8^{+\infty}_{-\infty}$ & 2.0 & $1.48\pm 0.19$ \nl

\tablenotetext{a}{Reduction in \chisq\ relative to the model with a
Schwarzschild disk line and power law, i.e. \chisq$_{\rm S}$-\chisq$_{\rm SR}$}
\tablenotetext{b}{Flux in the 2-10 keV band in units
of $10^{-11}$ erg cm$^{-2}$ s$^{-1}$, based on the model described
in this table, but corrected for absorption. The error bar corresponds
to the absolute calibration of the SIS detectors of $\sim 13$~per cent}

\tablecomments{Absorption is by Galactic \nh\ only, except in the case
of NGC 4151, where the column was left free. 
Error bars on the fit parameters are 68 per
cent confidence limits for 4 interesting parameters (\delchi=$4.7$).}

\enddata
\end{deluxetable}

\clearpage

\begin{deluxetable}{lllllr}

\tablecaption{Disk line fits; Kerr geometry; with reflection
\label{tab:laor}
}

\tablehead{
\colhead{Name} & \colhead{$\Gamma_{3-18}$} & 
\colhead{W$_{\rm K\alpha}$} & \colhead{$i$} &  
\colhead{$q$} & \colhead{\delchi$^{a}$}
}

\startdata
Mrk 335        & $2.04^{+0.15}_{-0.24}$ & $250^{+390}_{-220}$ & 
  $22^{+68}_{-22}$ & $2.2^{+1.2}_{-\infty}$ & -0.6 \nl
Fairall 9      & $1.93^{+0.14}_{-0.10}$ & $360^{+200}_{-150}$ & 
  $89^{+1}_{-49}$ & $1.3^{+0.9}_{-7.2}$ & -0.1 \nl
3C 120         & $2.01^{+0.05}_{-0.04}$ & $670^{+90}_{-120}$ & 
  $88^{+2}_{-1}$ & $2.7^{+0.4}_{-0.2}$ & 5.3 \nl
NGC 3227       & $1.67^{+0.05}_{-0.05}$ & $180^{+110}_{-70}$ & 
  $21^{+7}_{-21}$ & $1.8^{+0.6}_{-5.6}$ & -1.8 \nl
NGC 3516       & $1.92^{+0.04}_{-0.03}$ & $380^{+90}_{-90}$ & 
  $26^{+3}_{-4}$ & $2.6^{+0.3}_{-0.2}$ & 4.9 \nl
NGC 3783(1)    & $1.77^{+0.06}_{-0.06}$ & $550^{+190}_{-190}$ & 
  $26^{+5}_{-7}$ & $2.7^{+0.3}_{-0.4}$ & -1.6 \nl
NGC 3783(2)    & $1.67^{+0.08}_{-0.07}$ & $630^{+360}_{-390}$ & 
  $40^{+12}_{-40}$ & $3.9^{+1.7}_{-0.9}$ & -0.8 \nl
NGC 4051       & $2.10^{+0.08}_{-0.07}$ & $440^{+340}_{-200}$ & 
  $25^{+12}_{-4}$ & $2.8^{+0.8}_{-0.5}$ & 0.2 \nl
NGC 4151(2)    & $1.60^{+0.13}_{-0.13}$ & $120^{+40}_{-30}$ & 
  $9^{+18}_{-9}$ & $-10^{+12.1}_{-\infty}$ & -3.2 \nl
NGC 4151(4)    & $1.58^{+0.14}_{-0.14}$ & $470^{+190}_{-170}$ & 
  $24^{+5}_{-7}$ & $2.8^{+0.3}_{-0.4}$ & -0.6 \nl
NGC 4151(5)    & $1.57^{+0.17}_{-0.15}$ & $540^{+220}_{-210}$ & 
  $21^{+5}_{-11}$ & $3.0^{+0.3}_{-0.3}$ & 0.0 \nl
Mrk 766        & $2.25^{+0.10}_{-0.10}$ & $860^{+470}_{-370}$ & 
  $36^{+8}_{-7}$ & $3.0^{+0.8}_{-0.4}$ & -0.5 \nl
NGC 4593       & $1.91^{+0.16}_{-0.14}$ & $90^{+200}_{-50}$ & 
  $0^{+56}_{-0}$ & $-0.6^{+3.2}_{-\infty}$ & -0.5 \nl
MCG-6-30-15(1) & $2.15^{+0.07}_{-0.06}$ & $480^{+260}_{-200}$ & 
  $34^{+5}_{-6}$ & $2.8^{+0.5}_{-0.5}$ & -0.5 \nl
MCG-6-30-15(2) & $2.02^{+0.08}_{-0.07}$ & $730^{+540}_{-280}$ & 
  $34^{+16}_{-9}$ & $3.3^{+1.9}_{-0.4}$ & 2.9 \nl
IC 4329A       & $1.84^{+0.03}_{-0.03}$ & $110^{+80}_{-40}$ & 
  $10^{+13}_{-10}$ & $1.8^{+0.7}_{-5.8}$ & 0.5 \nl
NGC 5548       & $1.89^{+0.04}_{-0.04}$ & $90^{+120}_{-40}$ & 
  $10^{+80}_{-10}$ & $0.5^{+1.9}_{-\infty} $ & -1.7 \nl
Mrk 841(1)     & $2.07^{+0.14}_{-0.11}$ & $590^{+390}_{-270}$ & 
  $26^{+8}_{-5}$ & $2.5^{+0.5}_{-0.7}$ & -1.1 \nl
Mrk 841(2)     & $1.59^{+0.16}_{-0.13}$ & $140^{+120}_{-110}$ & 
  $90^{+0}_{-90}$ & $-10^{+16.3}_{-\infty}$ & -2.9 \nl
NGC 6814(1)    & \nodata & \nodata & \nodata & \nodata & \nodata \nl
Mrk 509        & $1.84^{+0.14}_{-0.09}$ & $230^{+340}_{-140}$ & 
  $89^{+1}_{-89}$ & $1.8^{+1.0}_{-6.0}$ & 0.8 \nl
NGC 7469(2)    & $1.93^{+0.09}_{-0.09}$ & $170^{+230}_{-100}$ & 
  $20^{+70}_{-20}$ & $1.4^{+1.0}_{-\infty}$ & -0.2 \nl
MCG-2-58-22    & $1.72^{+0.09}_{-0.10}$ & $140^{+120}_{-80}$ & 
  $26^{+64}_{-26}$ & $-10.0^{+15.3}_{-\infty}$ & 0.0 \nl

\tablenotetext{a}{Reduction in \chisq\ relative to the model with a
Schwarzschild disk line and reflection continuum, 
i.e. \chisq$_{\rm SR}$-\chisq$_{\rm KR}$}

\tablecomments{Absorption is by Galactic \nh\ only, except in the case
of NGC 4151, where the column was left free. 
Error bars on the fit parameters are 68 per
cent confidence limits for 4 interesting parameters (\delchi=$4.7$).}

\enddata
\end{deluxetable}

\clearpage

\begin{deluxetable}{lrrrrrrl}

\tablecaption{
Comparison of \chisq\ values
\label{tab:chisq}
}

\tablehead{
\colhead{Name} & \colhead{\chisq$_{\rm BL}$} & \colhead{\chisq$_{\rm S}$} &
\colhead{\chisq$_{\rm S67}$} & \colhead{\chisq$_{\rm SS}$} & 
\colhead{\chisq$_{\rm SR}$} & \colhead{\chisq$_{\rm KR}$} & d.o.f
}

\startdata
Mrk 335        & 227.0  & 225.7  & 226.3  & 225.3  & 224.6  & 225.2  & 251 \nl
Fairall 9      & 618.6  & 619.4  & 625.5  & 626.9  & 618.8  & 618.9  & 609 \nl
3C 120         & 846.3  & 849.3  & 868.9  & 862.7  & 849.3  & 844.0  & 949 \nl
NGC 3227       & 878.0  & 873.6  & 890.3  & 875.9  & 869.7  & 871.5  & 919 \nl
NGC 3516       & 1014.1 & 992.6  & 994.5  & 1011.3 & 995.1  & 990.2  & 1003\nl
NGC 3783(1)    & 781.6  & 771.4  & 775.8  & 771.4  & 768.5  & 770.1  & 787 \nl
NGC 3783(2)    & 692.4  & 684.4  & 683.1  & 693.2  & 682.4  & 683.2  & 742 \nl
NGC 4051       & 670.8  & 657.8  & 656.4  & 645.0  & 657.2  & 657.0  & 656 \nl
NGC 4151(2)    & 755.9  & 752.1  & 763.8  & 757.1  & 746.5  & 749.7  & 693 \nl
NGC 4151(4)    & 1357.7 & 1337.9 & 1331.6 & 1347.5 & 1335.6 & 1336.2 & 1399\nl
NGC 4151(5)    & 1295.9 & 1267.4 & 1280.9 & 1282.1 & 1266.3 & 1266.3 & 1226\nl
Mrk 766        & 570.9  & 563.0  & 562.2  & 577.2  & 563.0  & 563.5  & 681 \nl
NGC 4593       & 824.2  & 824.5  & 831.7  & 824.5  & 819.4  & 819.9  & 867 \nl
MCG-6-30-15(1) & 913.8  & 906.0  & 905.5  & 898.6  & 902.8  & 903.3  & 844 \nl
MCG-6-30-15(2) & 797.7  & 794.7  & 792.1  & 797.9  & 795.1  & 792.2  & 801 \nl
IC 4329A       & 1287.5 & 1284.9 & 1296.2 & 1306.9 & 1281.6 & 1281.1 & 1248\nl
NGC 5548       & 968.0  & 968.6  & 974.2  & 974.1  & 962.4  & 964.1  & 999 \nl
Mrk 841(1)     & 371.1  & 364.7  & 369.9  & 359.4  & 362.3  & 363.4  & 367 \nl
Mrk 841(2)     & 181.1  & 177.5  & 180.9  & 181.8  & 179.2  & 182.1  & 202 \nl
NGC 6814(1)    &\nodata &\nodata &\nodata &\nodata &\nodata &\nodata \nl
Mrk 509        & 806.2  & 807.6  & 816.7  & 809.7  & 806.0  & 805.2  & 893 \nl
NGC 7469(2)    & 497.0  & 498.8  & 503.3  & 498.4  & 494.7  & 494.9  & 450 \nl
MCG-2-58-22    & 406.6  & 406.3  & 407.4  & 407.8  & 404.3  & 404.3  & 387 \nl
\cutinhead{Total Values}
Value (-16000) & 762.4  & 628.2  & 737.2  & 734.7  & 584.4  & 586.3  & 973\nl
Probability    & 0.87   & 0.97   & 0.90   & 0.90   & 0.98   & 0.98 &\nodata\nl
\tablebreak
\cutinhead{Relative likelihoods$^{a}$}
\chisq$_{\rm BL}$    &\nodata & $>10^{3}$ & $>10^{3}$ & $>10^{3}$ & 
  $>10^{3}$ & $>10^{3}$ & \nodata \nl
\chisq$_{\rm S}$    & $<10^{-3}$ &\nodata & $<10^{-3}$ & $<10^{-3}$ & 
  $>10^{3}$ & $>10^{3}$ & \nodata \nl
\chisq$_{\rm S67}$    & $<10^{-3}$ & $>10^{3}$ &\nodata & 0.29 
  & $>10^{3}$ & $>10^{3}$ & \nodata \nl
\chisq$_{\rm SS}$    & $<10^{-3}$ & $>10^{3}$ & 3.5 &\nodata & 
  $>10^{3}$ & $>10^{3}$ & \nodata \nl
\chisq$_{\rm SR}$    & $<10^{-3}$ & $<10^{-3}$ & $<10^{-3}$ & $<10^{-3}$ &
  \nodata & 0.39 & \nodata \nl
\chisq$_{\rm KR}$    & $<10^{-3}$ & $<10^{-3}$ & $<10^{-3}$ & $<10^{-3}$ & 
  2.6 &\nodata & \nodata \nl
\enddata

\tablenotetext{a}{Likelihood of the model in the column heading
compared to the row, based on the total \chisq\ values.}

\tablecomments{The subscripts used are: BL - broad gaussian
(Table~\ref{tab:bl}); S - Schwarzschild model (Table~\ref{tab:dl}); 
S67 - Schwarzschild model with a rest energy of 6.7~keV; 
SS - Schwarzschild model at a single radius;
SR - Schwarzschild with reflection (Table~\ref{tab:dl-ref}); 
KR - Kerr model with reflection (Table~\ref{tab:laor})
}

\end{deluxetable}

\clearpage

\begin{deluxetable}{lccccc}

\tablecaption{Mean parameters for the Seyfert 1 galaxies
\label{tab:means}}

\tablehead{
\colhead{Parameter}  & \colhead{Unit} & \colhead{$\mu$}$^{a}$ & 
\colhead{$\overline{\mu}^{b}$} & \colhead{$<\mu>^{c}$} & 
\colhead{$\sigma_{i}^{d}$}
}

\startdata
\cutinhead{Narrow Line Fits, 3-10 keV}
$\Gamma_{3-10}$              & --  & $1.74\pm 0.14$ & $1.74\pm 0.01$ 
  & $1.75 \pm 0.05$ & $0.12\pm 0.04$ \nl
$E_{K\alpha}$      & keV & $6.34\pm 0.12$ & $6.36\pm 0.01$ 
  & $6.36\pm 0.02$ & $<0.06$ \nl
$W_{K\alpha}$     &  eV & $115\pm 39$ & $98 \pm 8$ 
  & $98 \pm 12$ & $<24$ \nl
\cutinhead{Broad Line Fits, 3-10 keV}
$\Gamma_{3-10}$              & --  & $1.80\pm 0.17$ & $1.76\pm 0.02$ 
  & $1.78\pm 0.06$ & $0.13\pm 0.05$ \nl
$E_{K\alpha}$      & keV & $6.29\pm 0.14$ & $6.34\pm 0.02$ 
  & $6.34\pm 0.04$ & $<0.09$ \nl
$W_{K\alpha}$     & eV  & $290\pm 240$ & $150\pm 20$ 
  & $160\pm 30$ & $<90$ \nl
$\sigma_{K\alpha}$ & keV & $0.40\pm 0.28$ & $0.39\pm 0.06$ 
  & $0.43\pm 0.12$ & $<0.29$ \nl
\cutinhead{Disk Line, free q}
$\Gamma_{3-10}$              & --  & $1.80\pm 0.17$ & $1.76\pm 0.02$ 
  & $1.79\pm 0.07$ & $0.15\pm 0.05$  \nl
$W_{K\alpha}$     &  eV & $310\pm 140$ & $280 \pm 30$ 
  & $290 \pm 50$ & $<140$ \nl
$i$                   & $\deg$  & $35\pm 15$ & $29\pm 2$ 
  & $29\pm 3$ & $<9$  \nl
$q$                   & --  & $1.9\pm 3.2$ & $2.5\pm 0.2$ 
  & $2.5\pm 0.4$ & $<0.8$  \nl
\cutinhead{Disk Line, free q, reflection}
$\Gamma_{3-10}$              & --  & $1.92\pm 0.17$ & $1.88\pm 0.02$ 
  & $1.91\pm 0.07$ & $0.15\pm 0.05$  \nl
$W_{K\alpha}$     &  eV & $270\pm 160$ & $180 \pm 20$ 
  & $230 \pm 60$ & $110\pm 50$ \nl
$i$                   & $\deg$  & $26\pm 17$ & $29\pm 2$ 
  & $29\pm 5$ & $<11$  \nl
$q$                   & --  & $1.1\pm 4.5$ & $2.8\pm 0.3$ 
  & $2.8\pm 0.4$ & $<0.9$  \nl
\tablebreak
\cutinhead{Kerr model, free q, reflection}
$\Gamma_{3-10}$              & --  & $1.90\pm 0.17$ & $1.88\pm 0.02$ 
  & $1.89\pm 0.07$ & $0.15\pm 0.05$  \nl
$W_{K\alpha}$     &  eV & $320\pm 230$ & $210 \pm 30$ 
  & $270 \pm 60$ & $130\pm 60$ \nl
$i$                   & $\deg$  & $34\pm 28$ & $28\pm 3$ 
  & $28\pm 4$ & $<8$  \nl
$q$                   & --  & $1.3\pm 3.1$ & $2.8\pm 0.2$ 
  & $2.8 \pm 0.2$ & $<0.4$  \nl
\enddata

\tablenotetext{a}{Unweighted mean and standard deviation}
\tablenotetext{b}{Weighted mean}
\tablenotetext{c}{Expectation allowing for measurement errors (see text)}
\tablenotetext{d}{Intrinsic dispersion of distribution (see text)}

\tablecomments{Error bars are 68 per cent confidence limits. Upper
limits are 90 per cent confidence.}

\end{deluxetable}

\clearpage

\begin{figure}
\epsscale{0.5}
\epsfysize=0.4\textwidth
\plotone{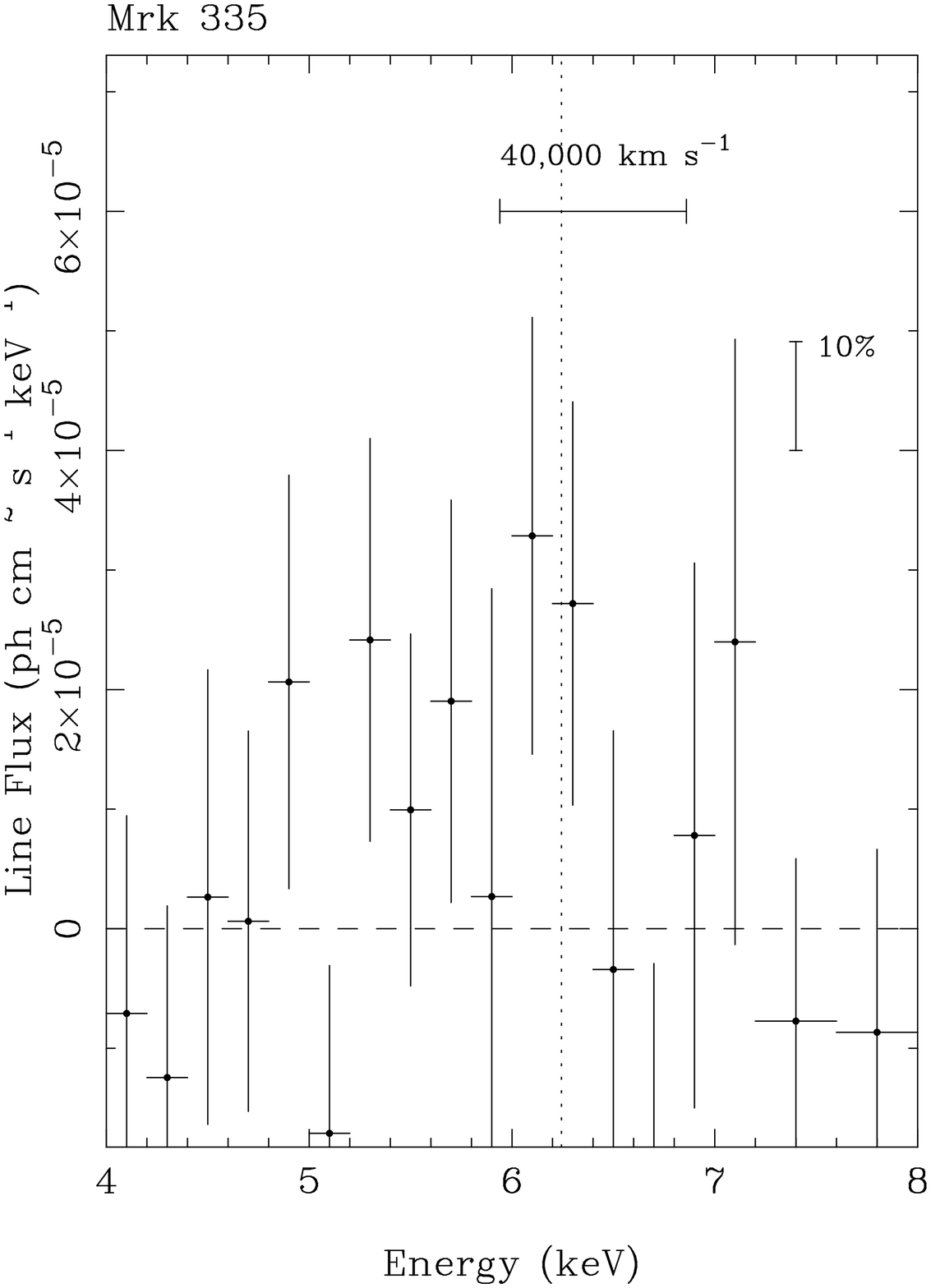}
\epsfysize=0.4\textwidth
\plotone{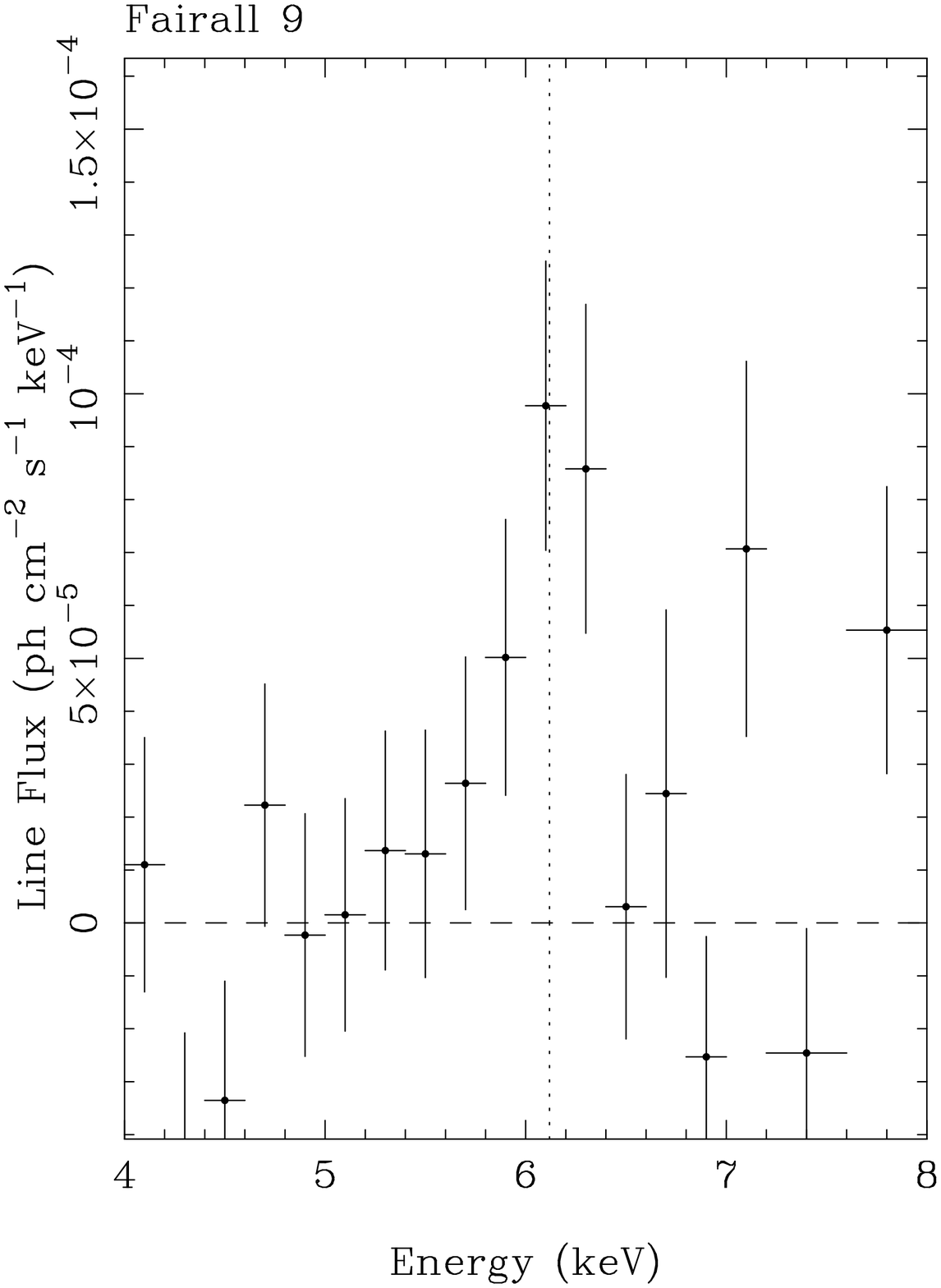}
\end{figure}

\begin{figure}
\epsscale{0.5}
\epsfysize=0.4\textwidth
\plotone{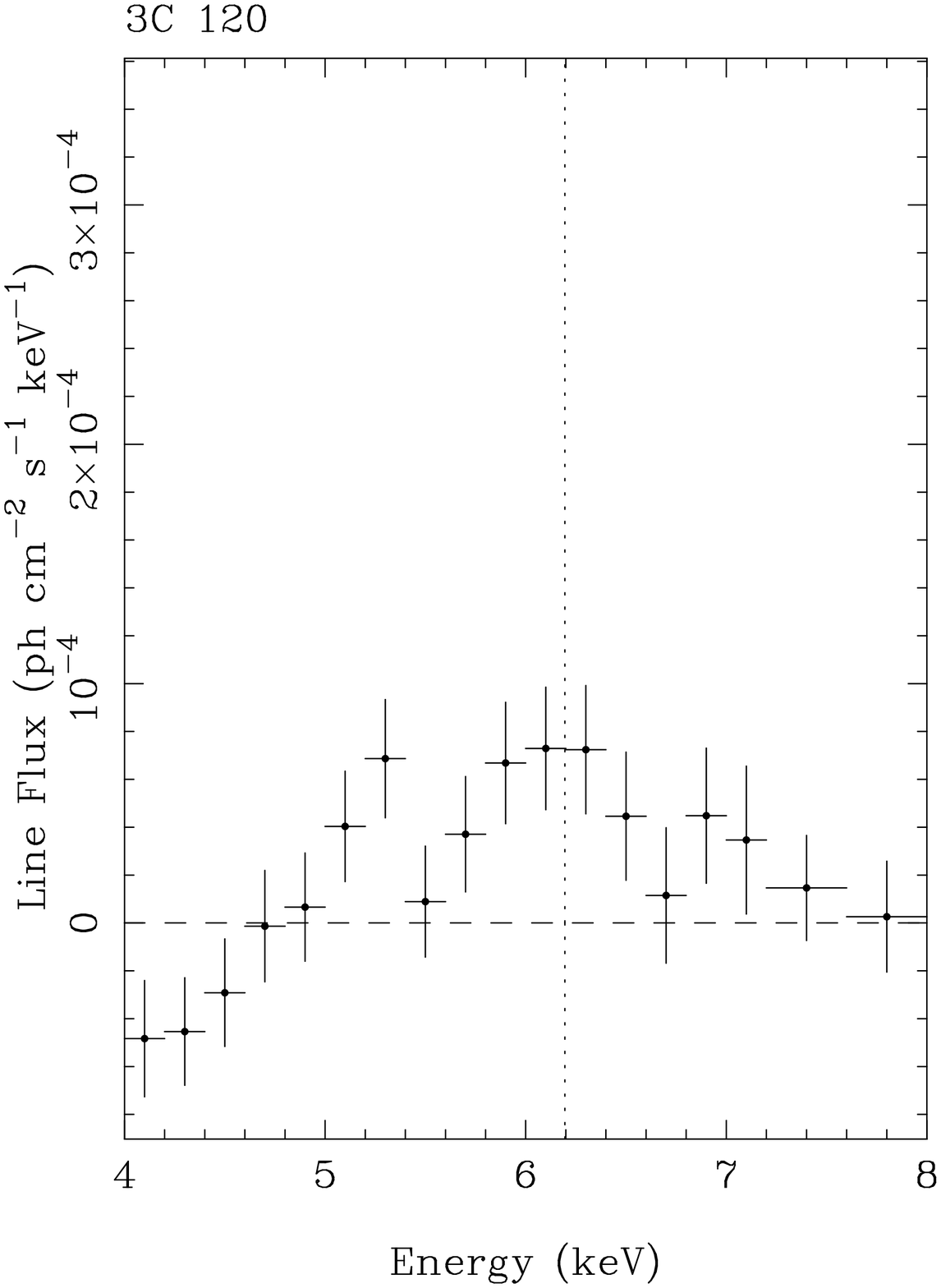}
\epsfysize=0.4\textwidth
\plotone{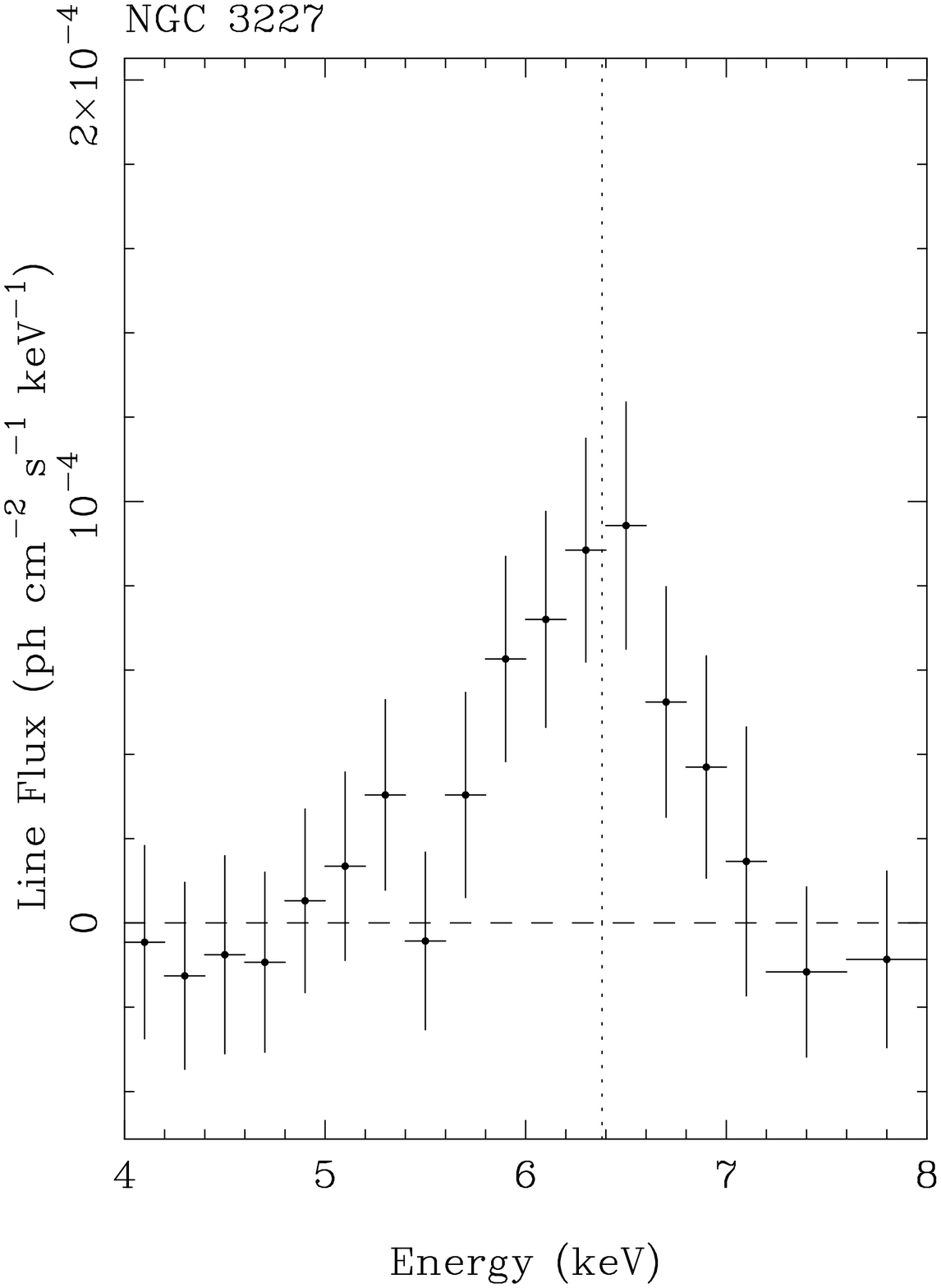}
\end{figure}

\begin{figure}
\epsscale{0.5}
\epsfysize=0.4\textwidth
\plotone{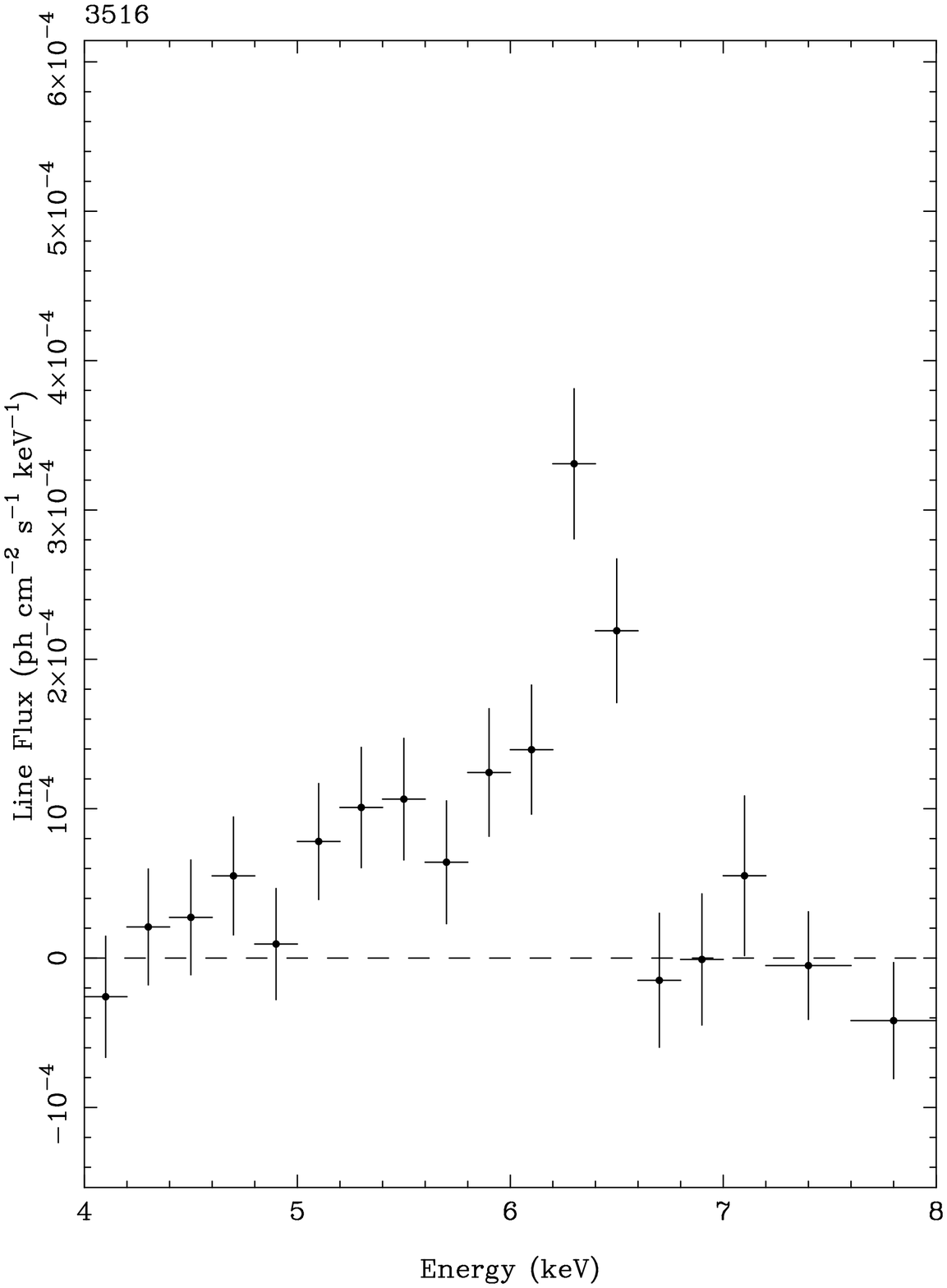}
\epsfysize=0.4\textwidth
\plotone{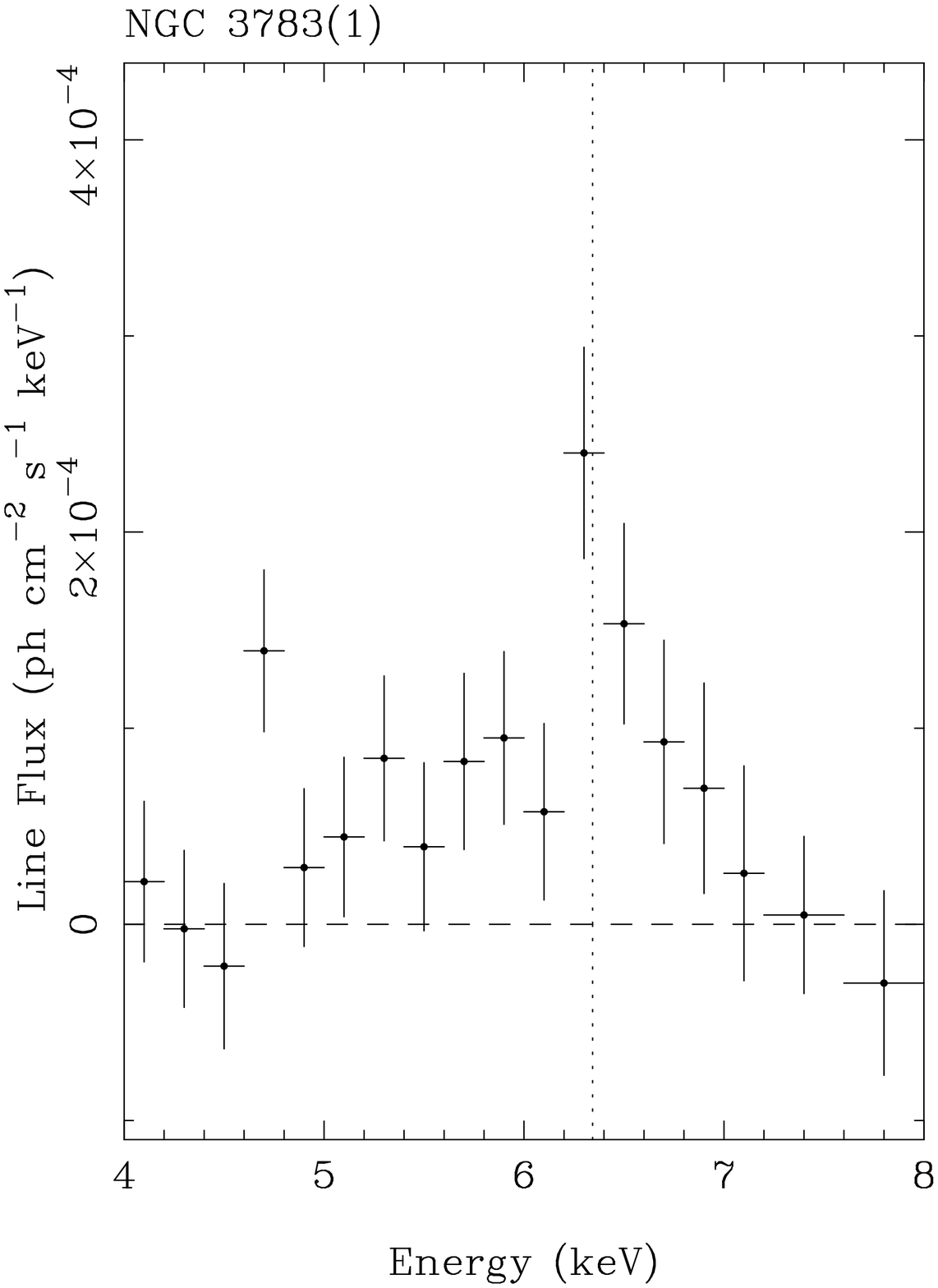}
\end{figure}

\begin{figure}
\epsscale{0.5}
\epsfysize=0.4\textwidth
\plotone{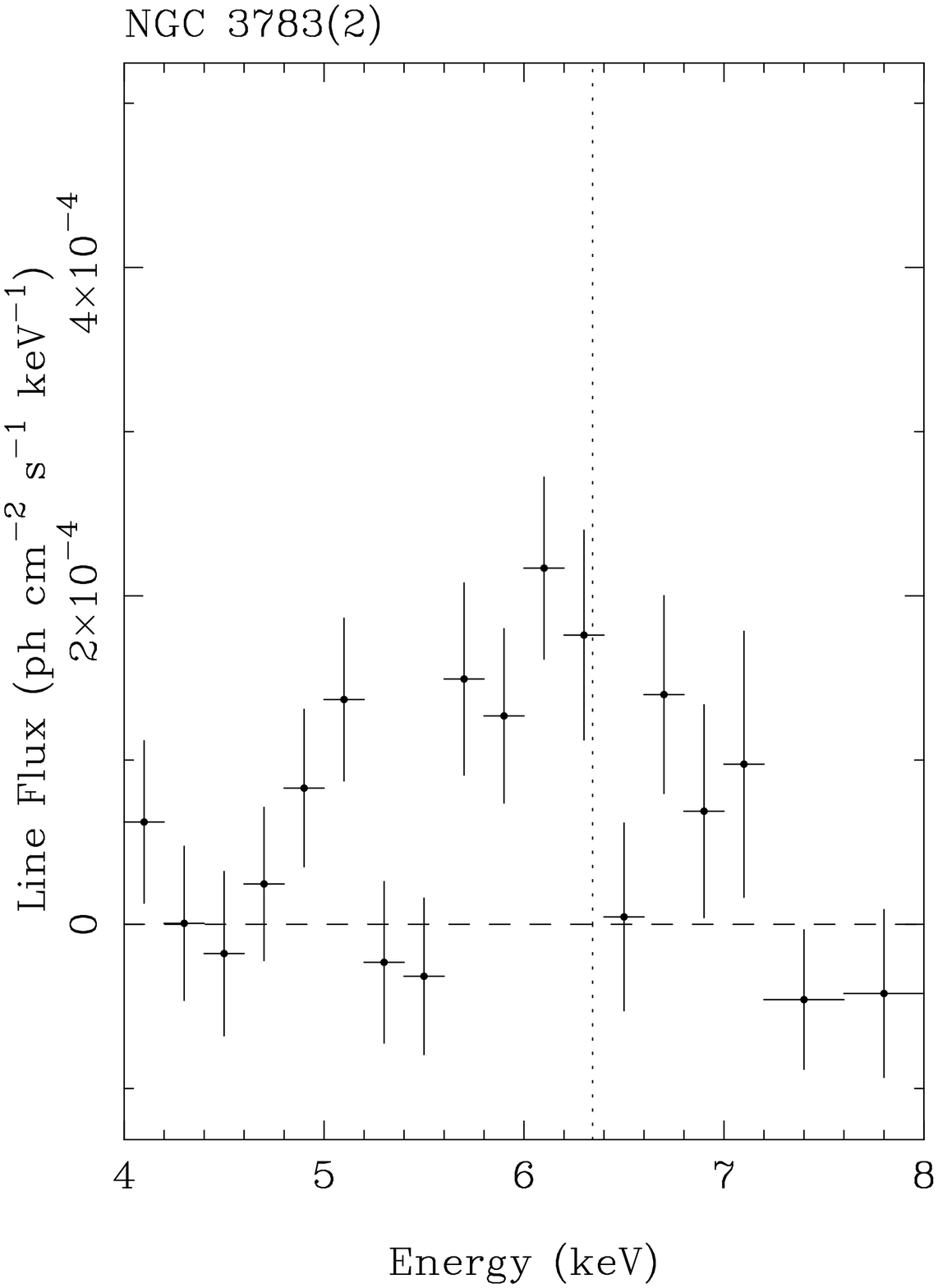}
\epsfysize=0.4\textwidth
\plotone{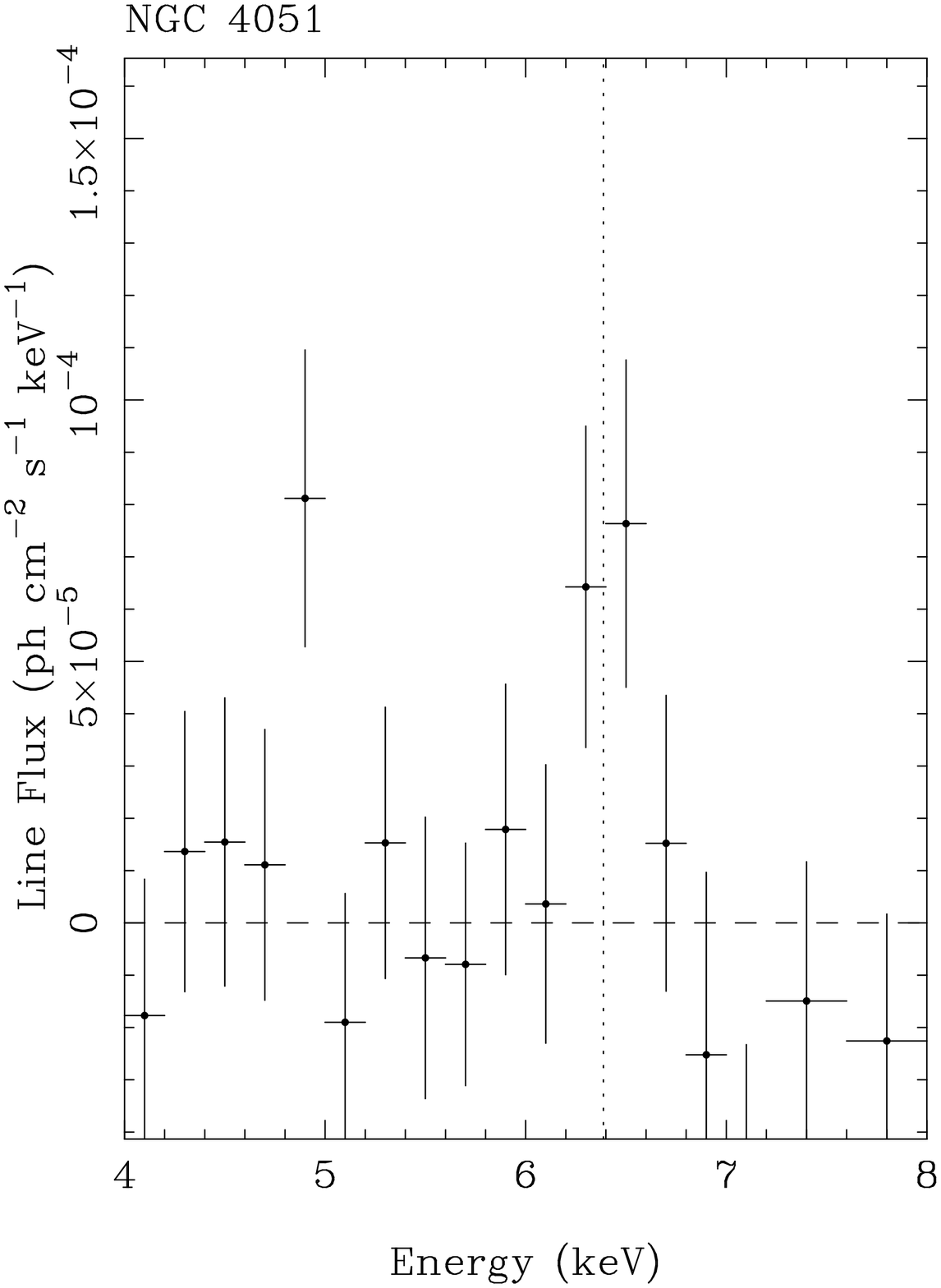}
\end{figure}

\begin{figure}
\epsscale{0.5}
\epsfysize=0.4\textwidth
\plotone{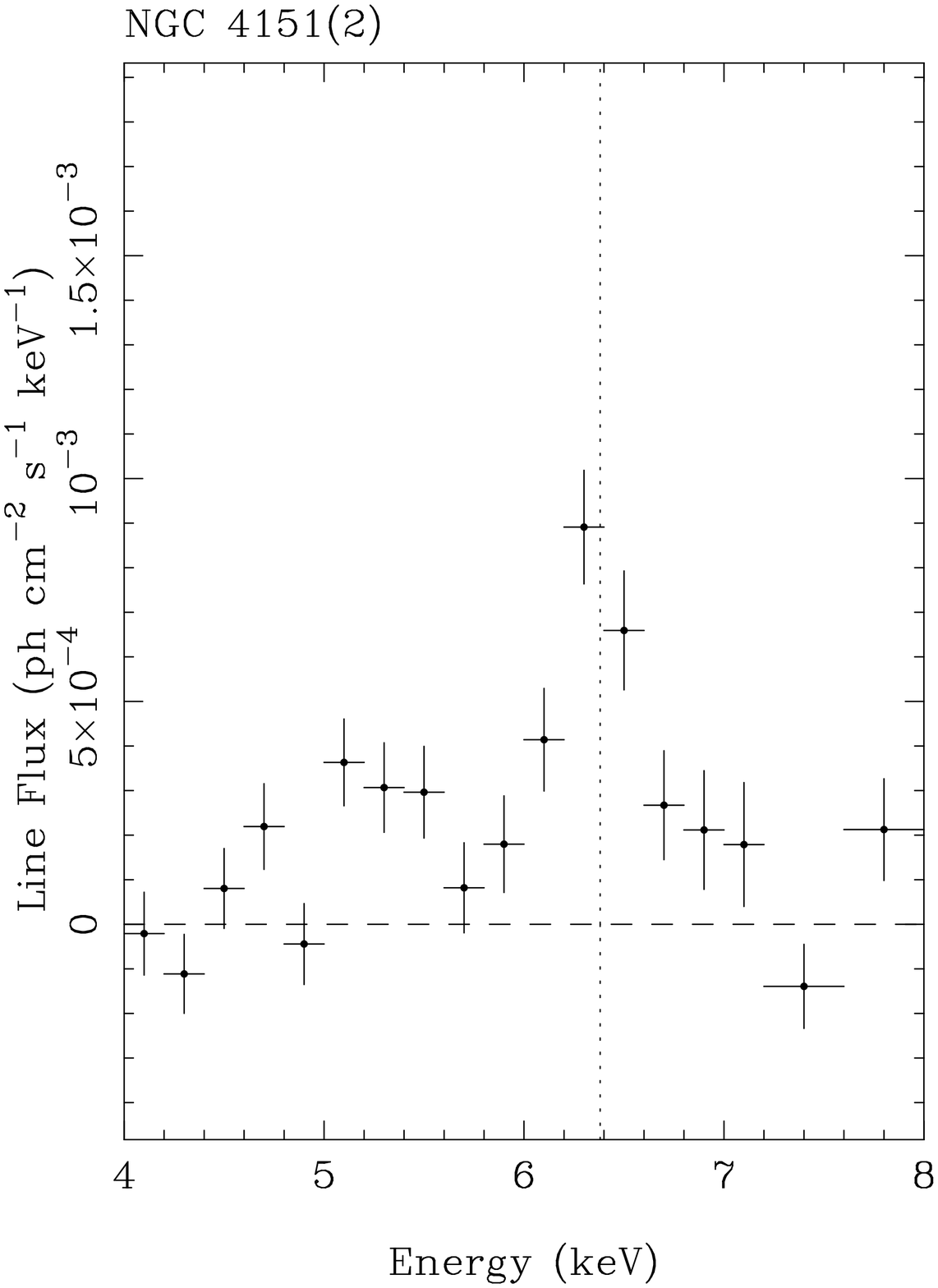}
\epsfysize=0.4\textwidth
\plotone{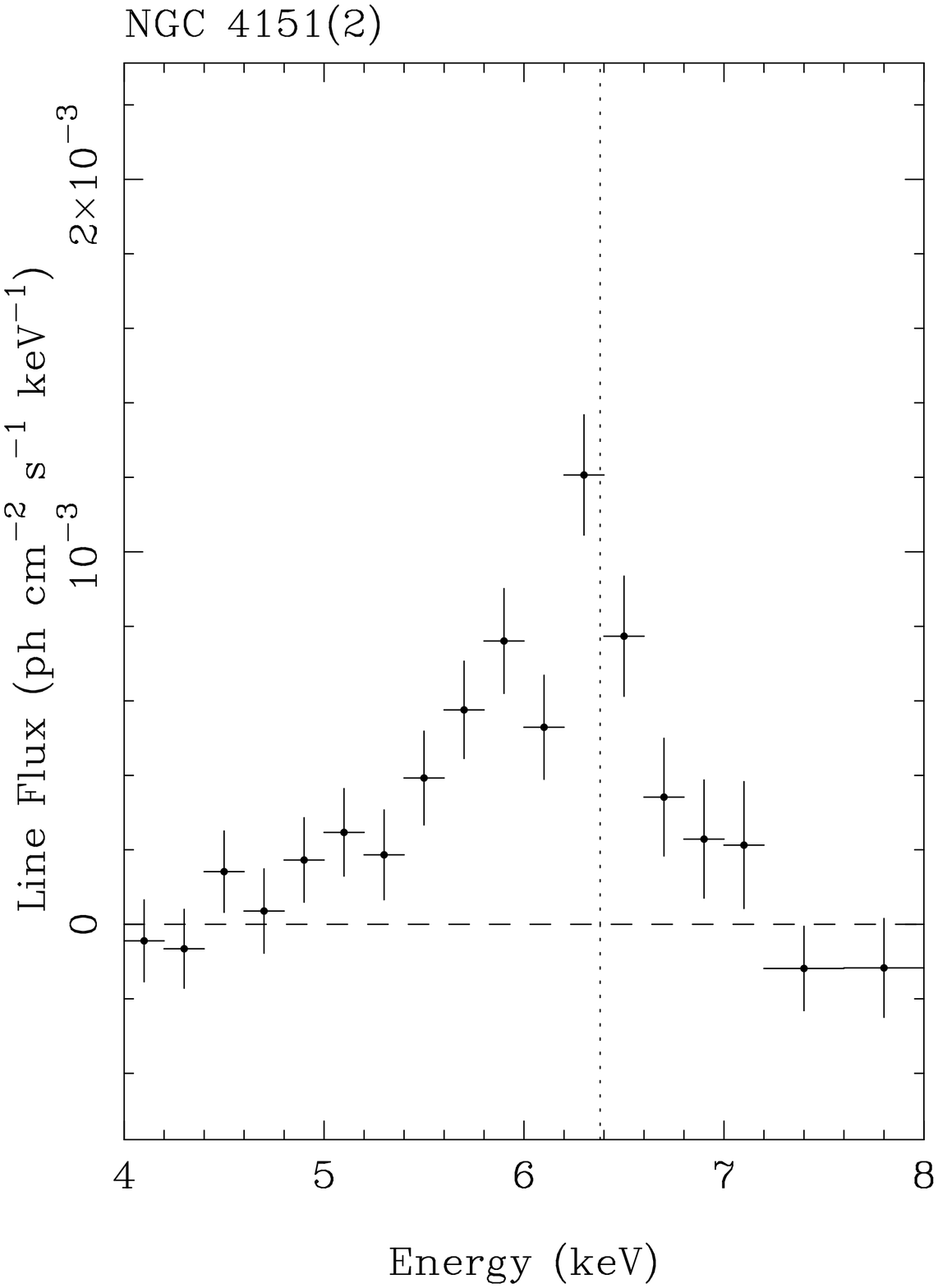}
\end{figure}

\begin{figure}
\epsscale{0.5}
\epsfysize=0.4\textwidth
\plotone{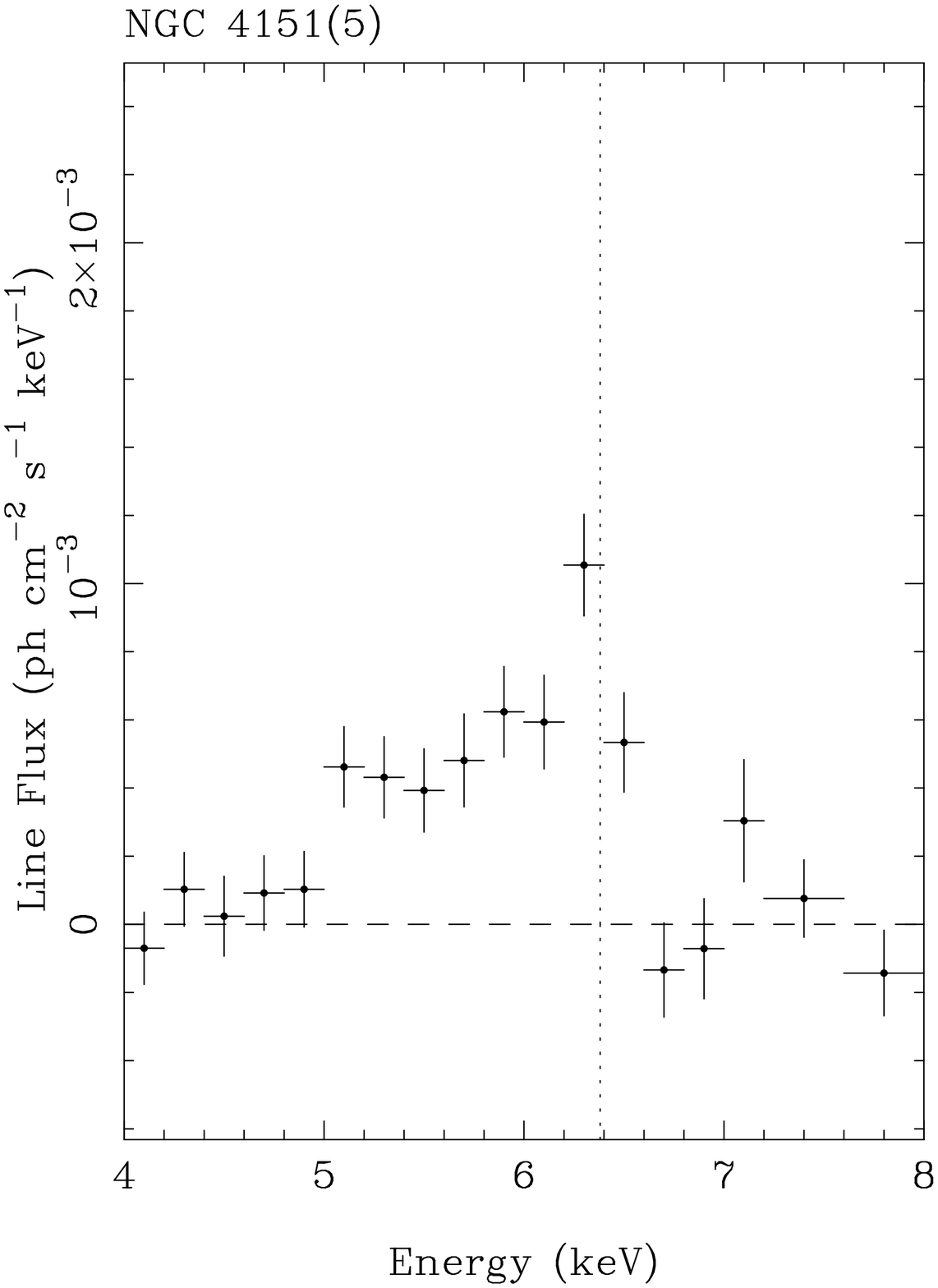}
\epsfysize=0.4\textwidth
\plotone{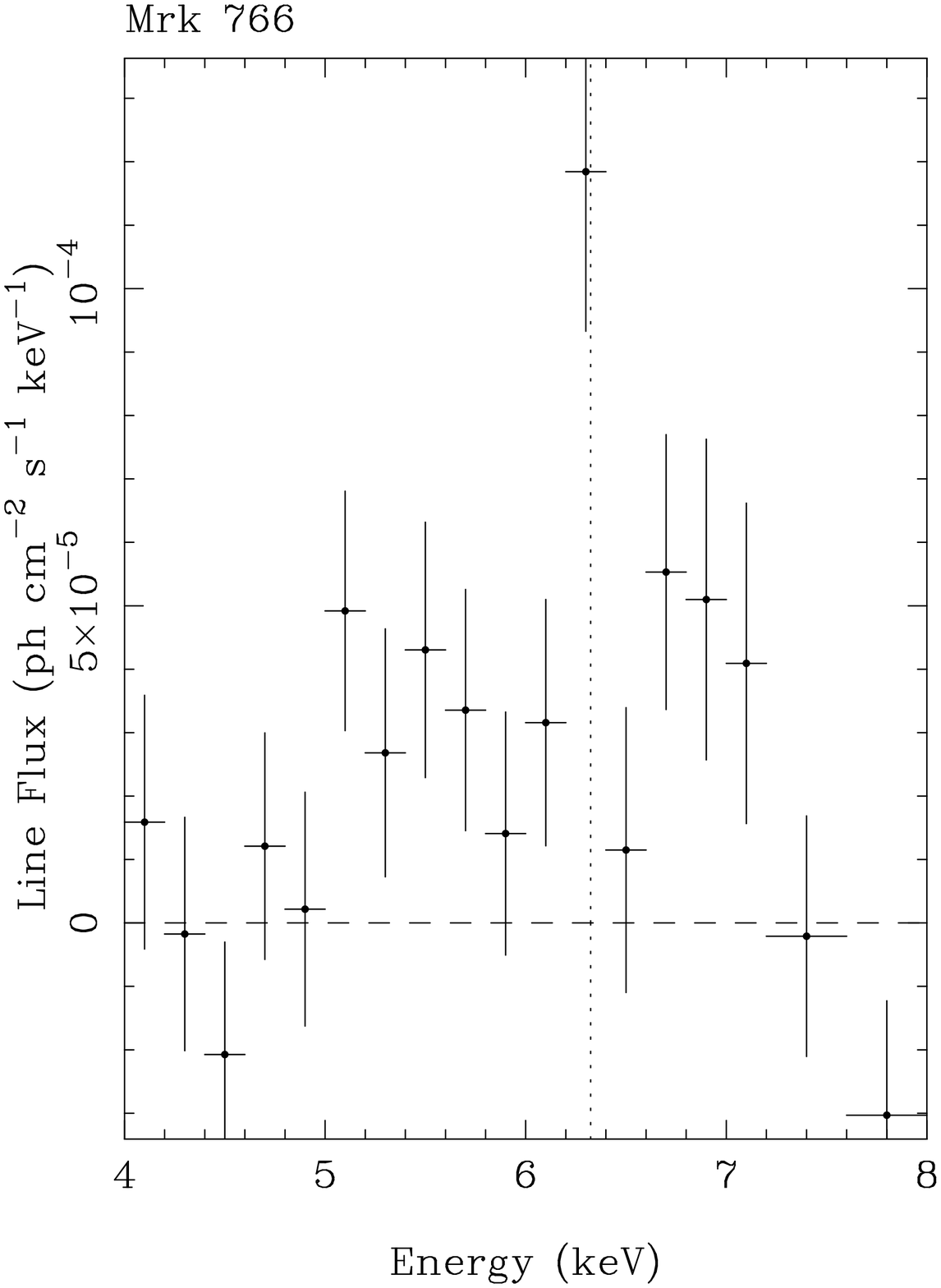}
\end{figure}

\begin{figure}
\epsscale{0.5}
\epsfysize=0.4\textwidth
\plotone{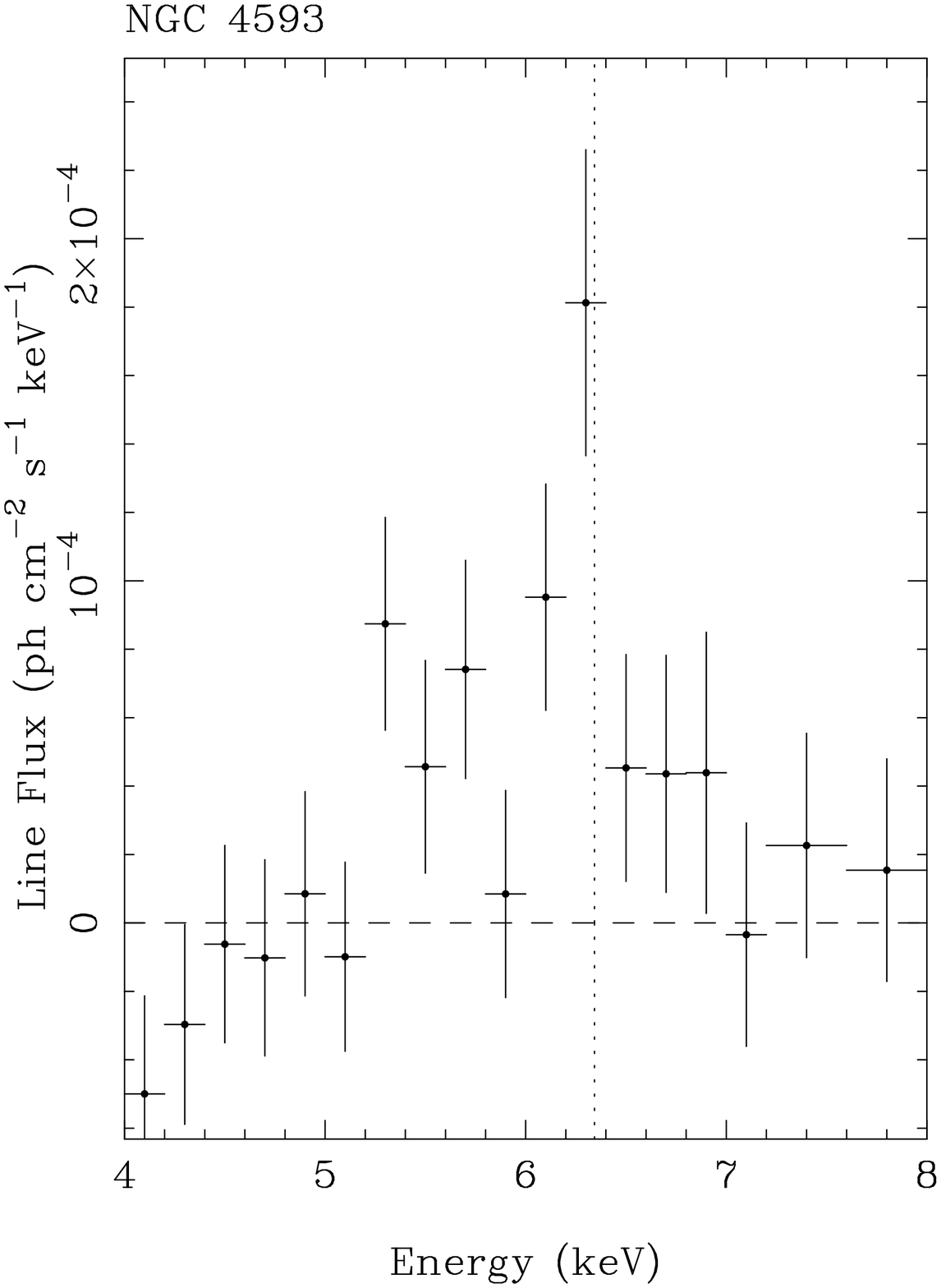}
\epsfysize=0.4\textwidth
\plotone{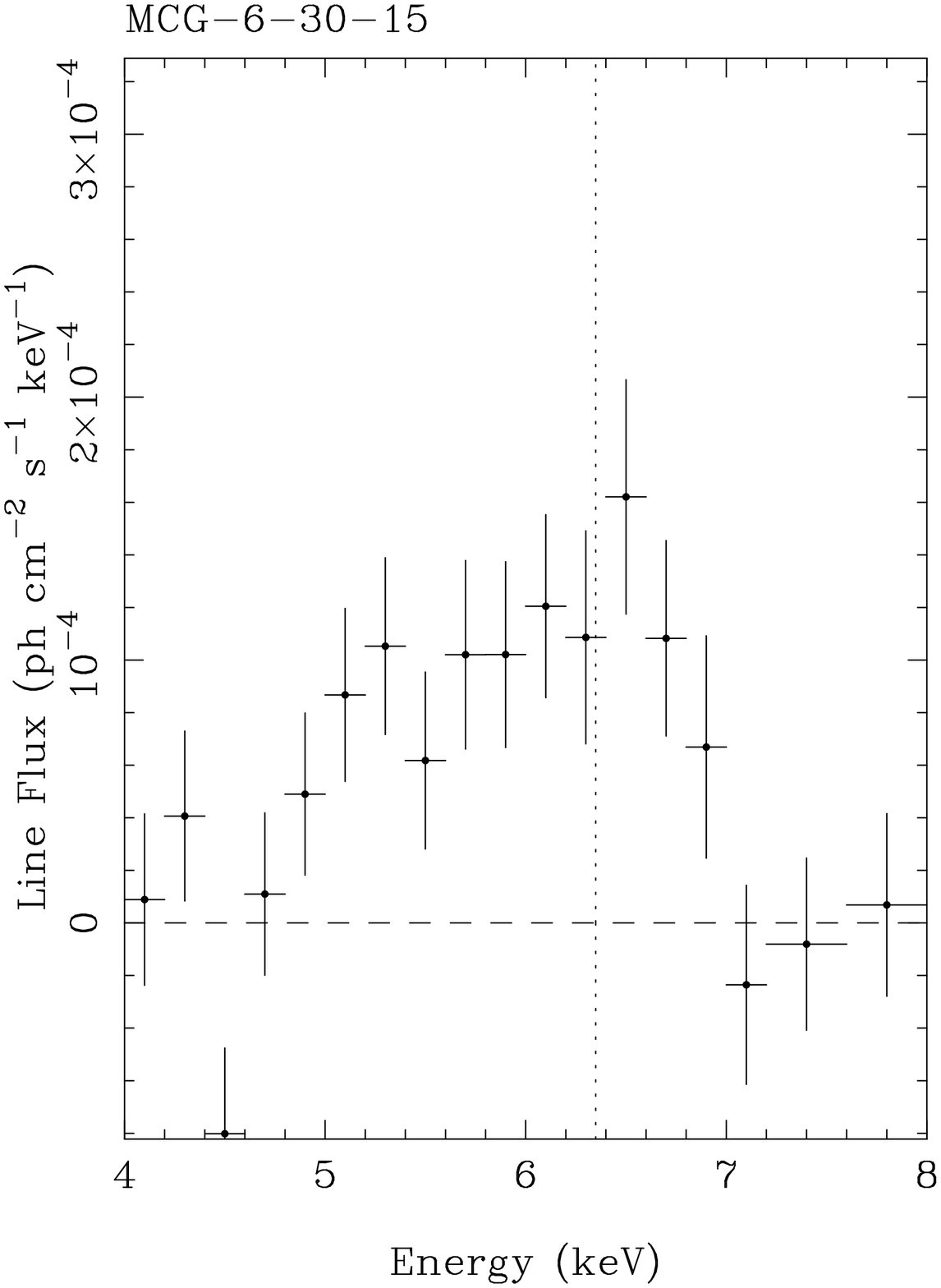}
\end{figure}

\begin{figure}
\epsscale{0.5}
\epsfysize=0.4\textwidth
\plotone{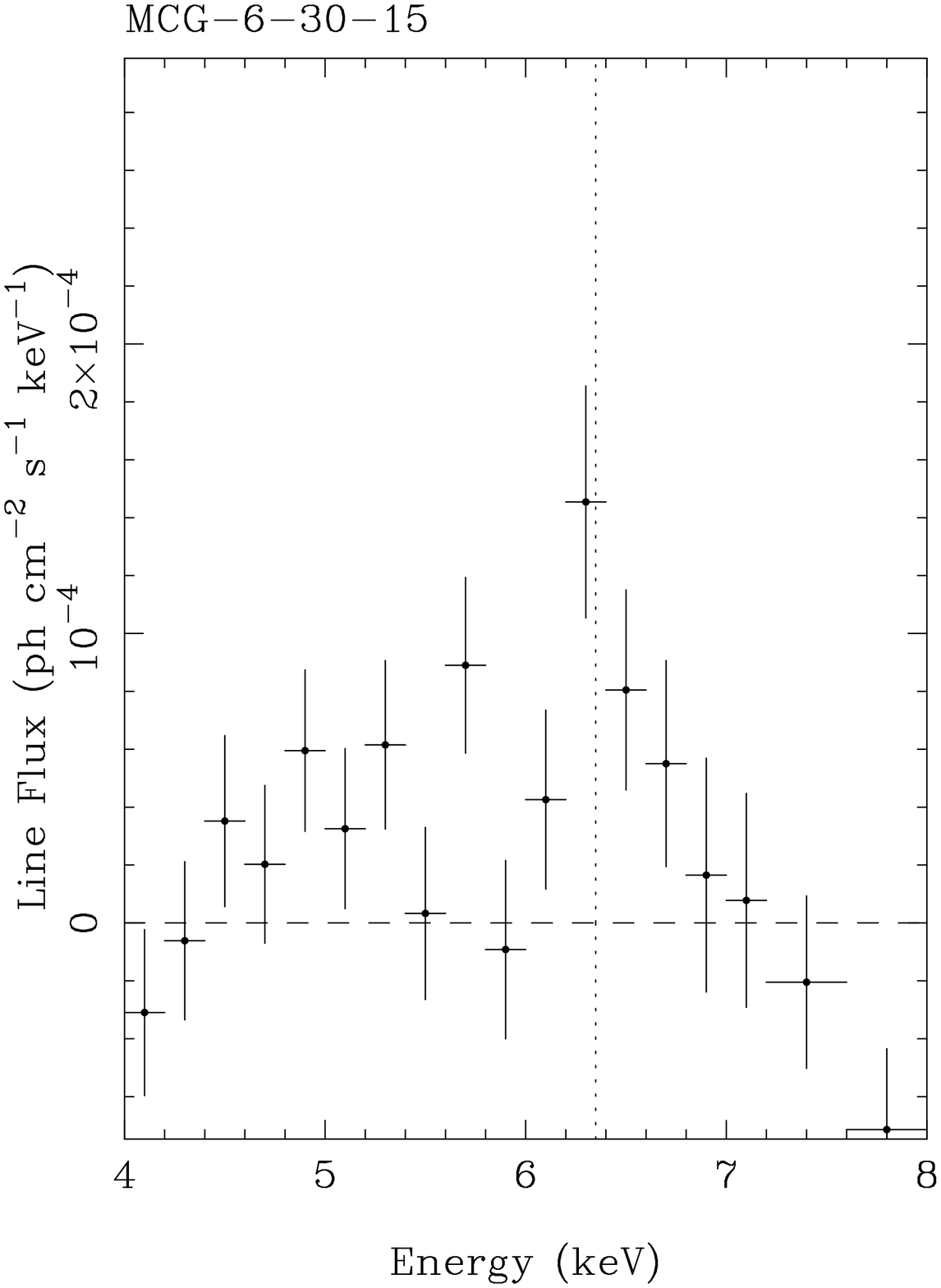}
\epsfysize=0.4\textwidth
\plotone{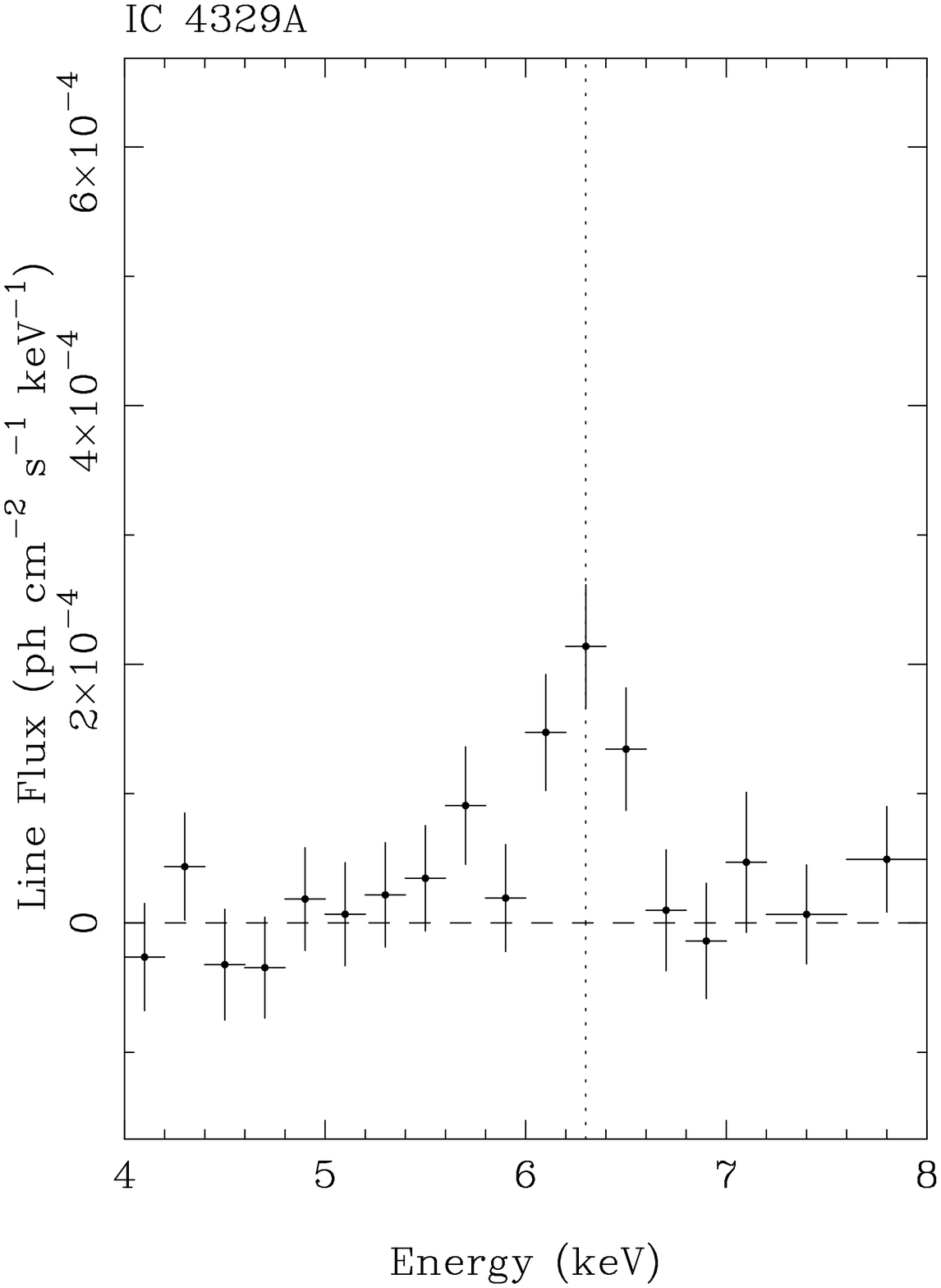}
\end{figure}

\begin{figure}
\epsscale{0.5}
\epsfysize=0.4\textwidth
\plotone{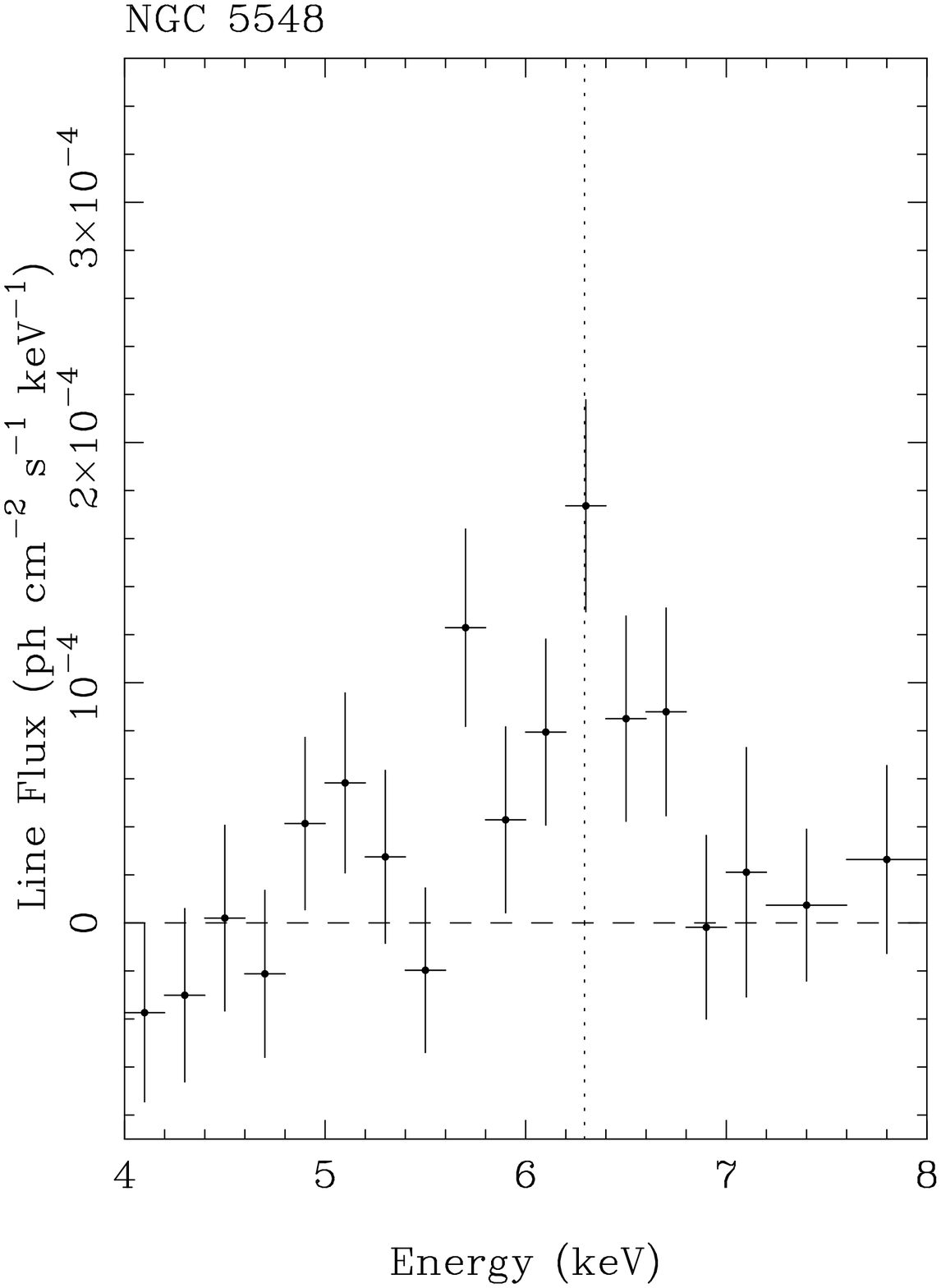}
\epsfysize=0.4\textwidth
\plotone{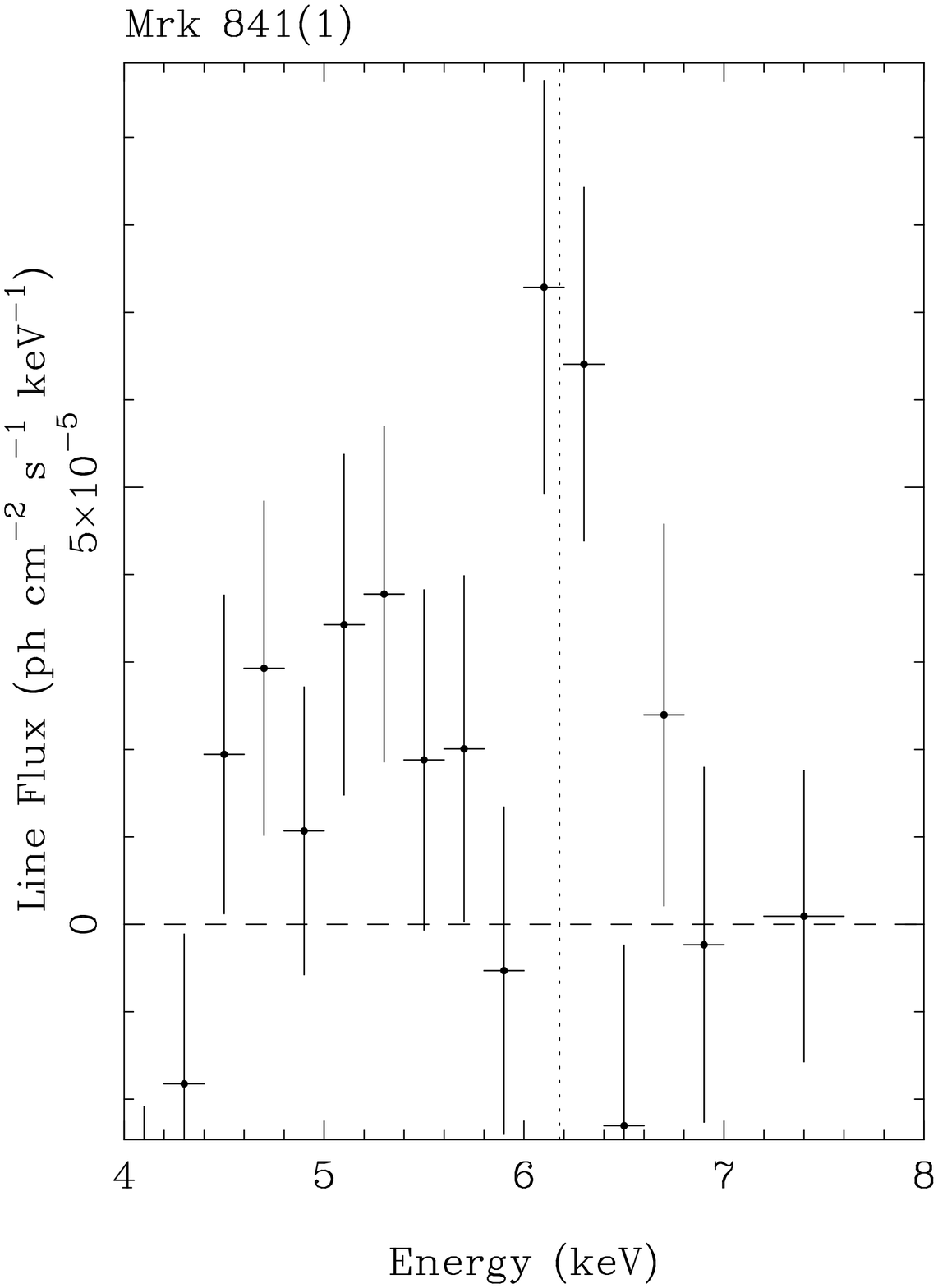}
\end{figure}

\begin{figure}
\epsscale{0.5}
\epsfysize=0.4\textwidth
\plotone{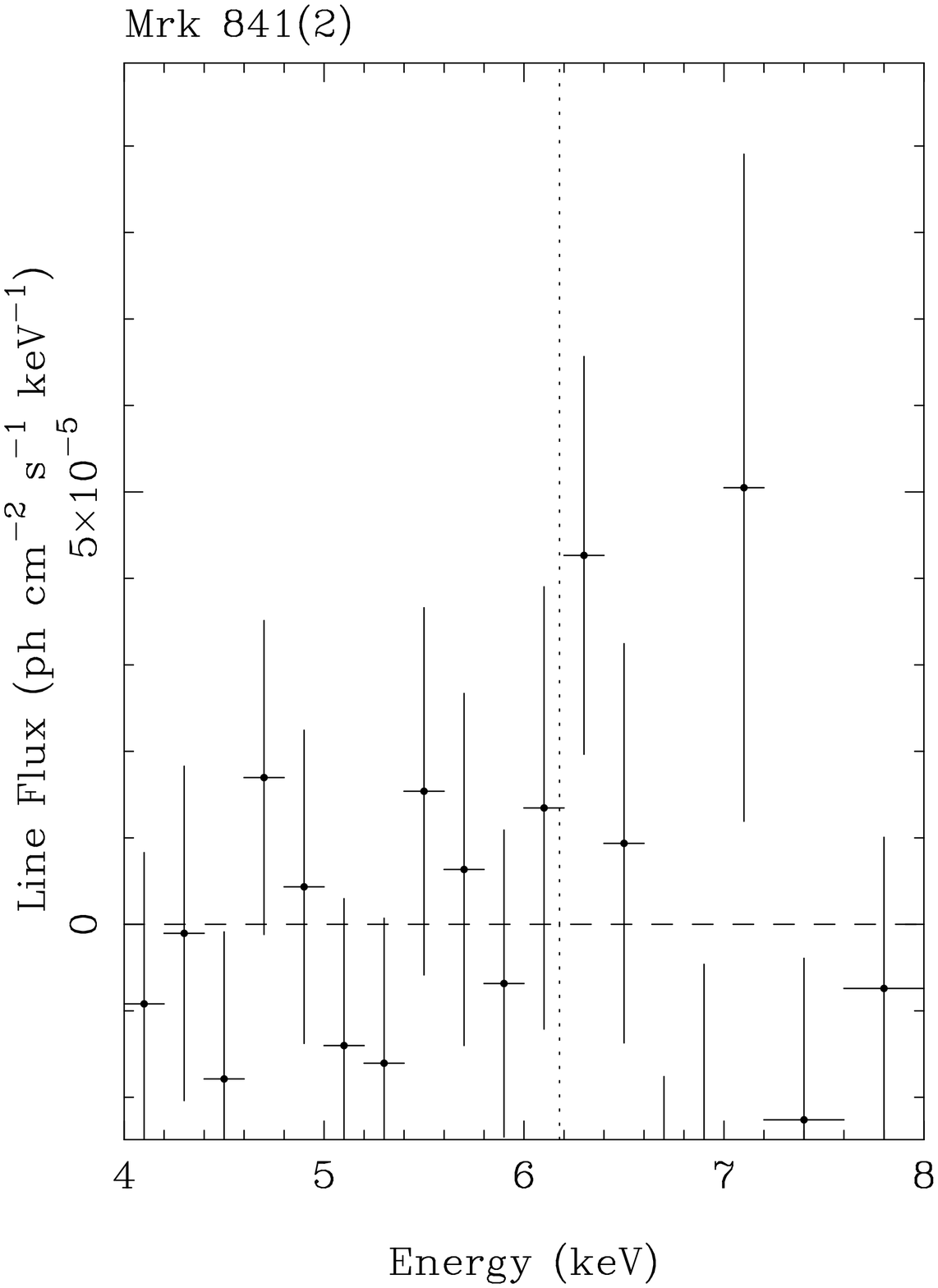}
\epsfysize=0.4\textwidth
\plotone{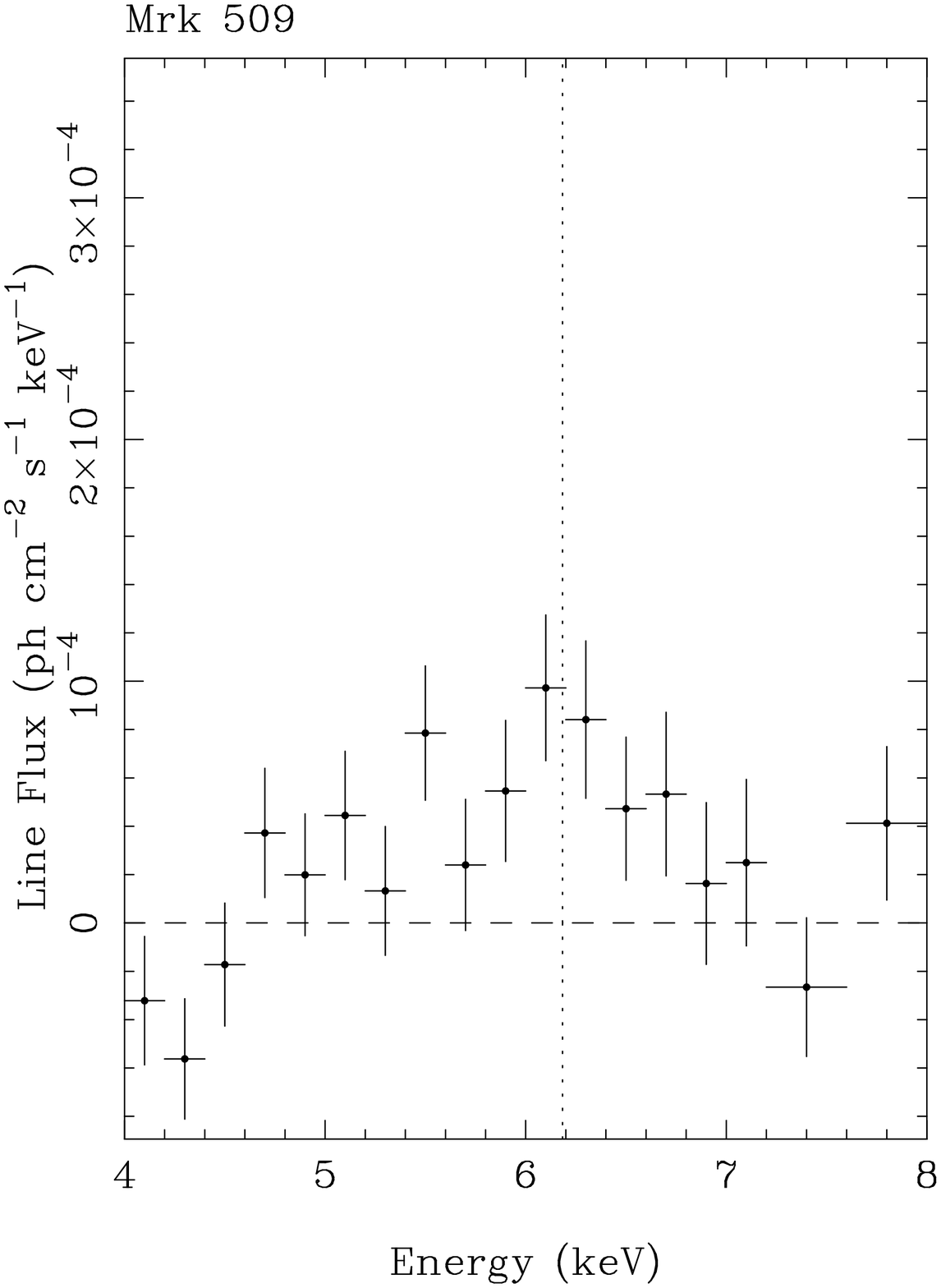}
\end{figure}

\begin{figure}
\epsscale{0.5}
\epsfysize=0.4\textwidth
\plotone{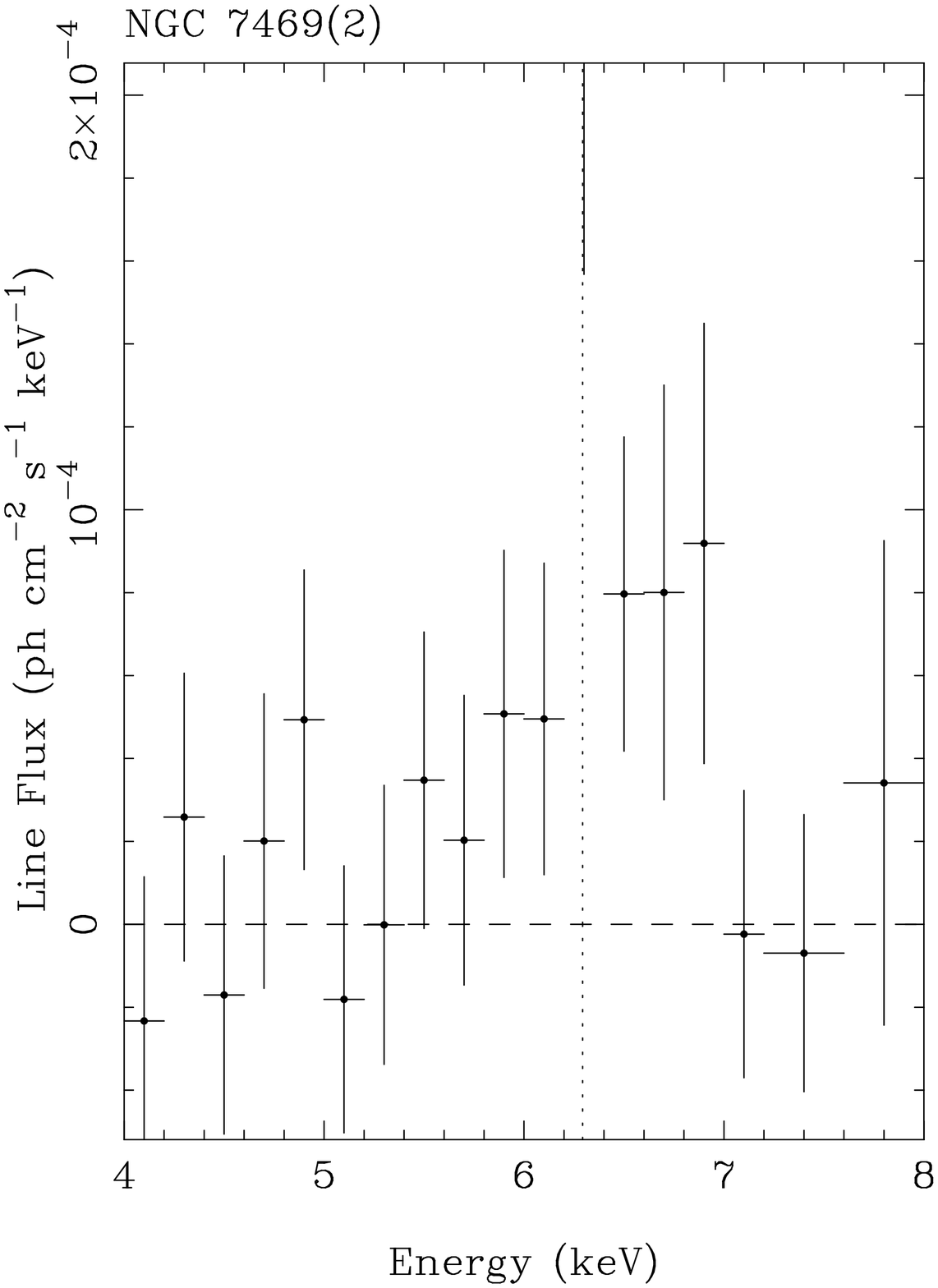}
\epsfysize=0.4\textwidth
\plotone{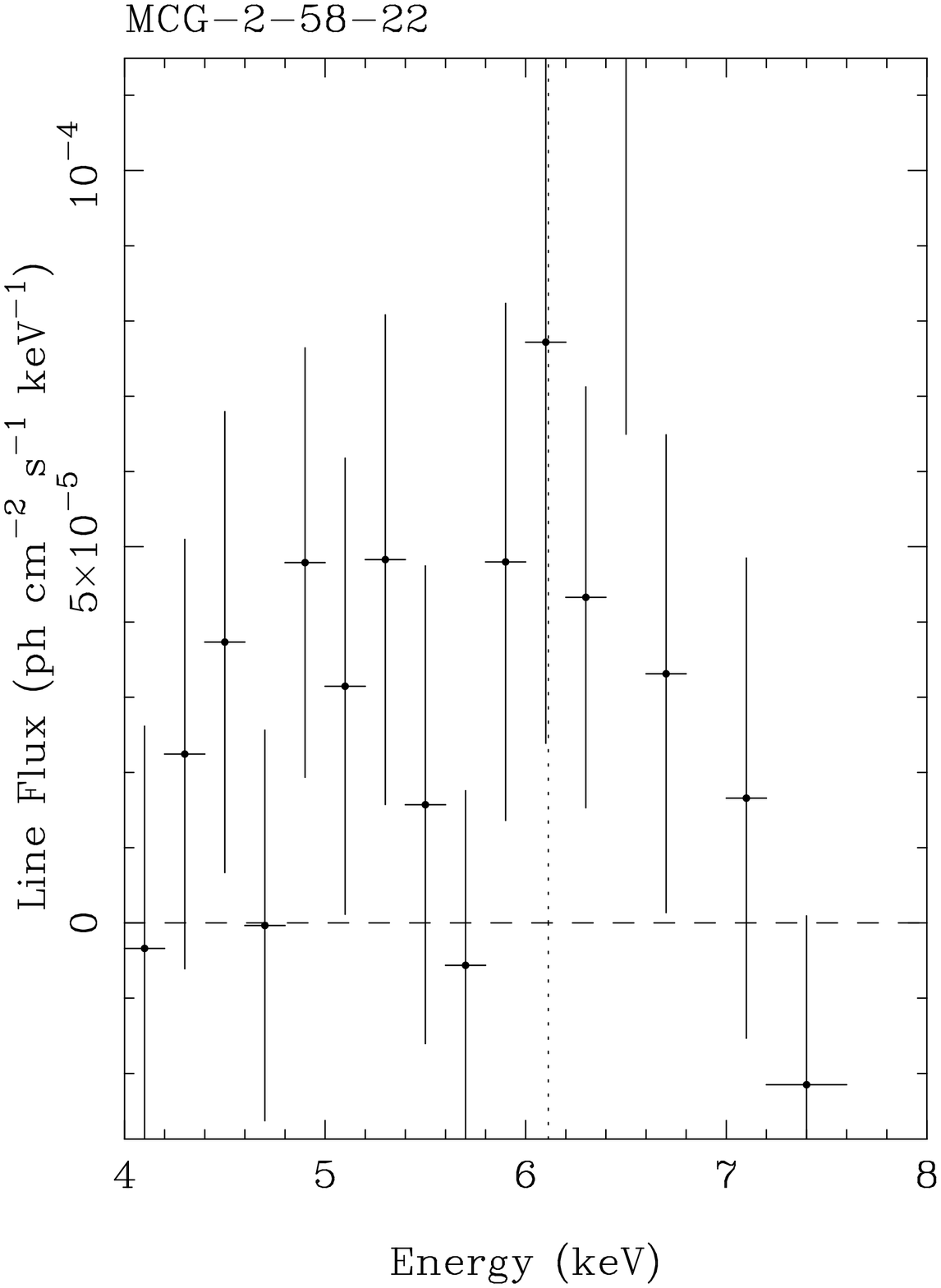}
\caption{The profiles of the Fe K$\alpha$ emission lines in our sample 
sources, measured by the SIS only. The continuum has been estimated by
fitting a power-law (with Galactic absorption) to the 3-5 and 7-10 keV SIS
data, and rebinning the resultant residuals. Most show significant broadening
beyond the instrumental resolution and some have the characteristic profile
expected if the line is produced close to a central black hole: an 
asymmetry skewed to the red. All plots have been rescaled to a fixed
fraction of the continuum at 6.4 keV. That energy (in the rest frame) is
marked by a vertical dotted line in each plot. The first plot (Mrk 335)
shows a horizontal scale in velocity space (corresponding to 40,000 km s$^{-1}$
and a vertical bar which indicates a 10~per cent variation from the continuum. 
\label{fig:profiles}}
\end{figure}

\begin{figure}
\epsscale{1.0}
\plotone{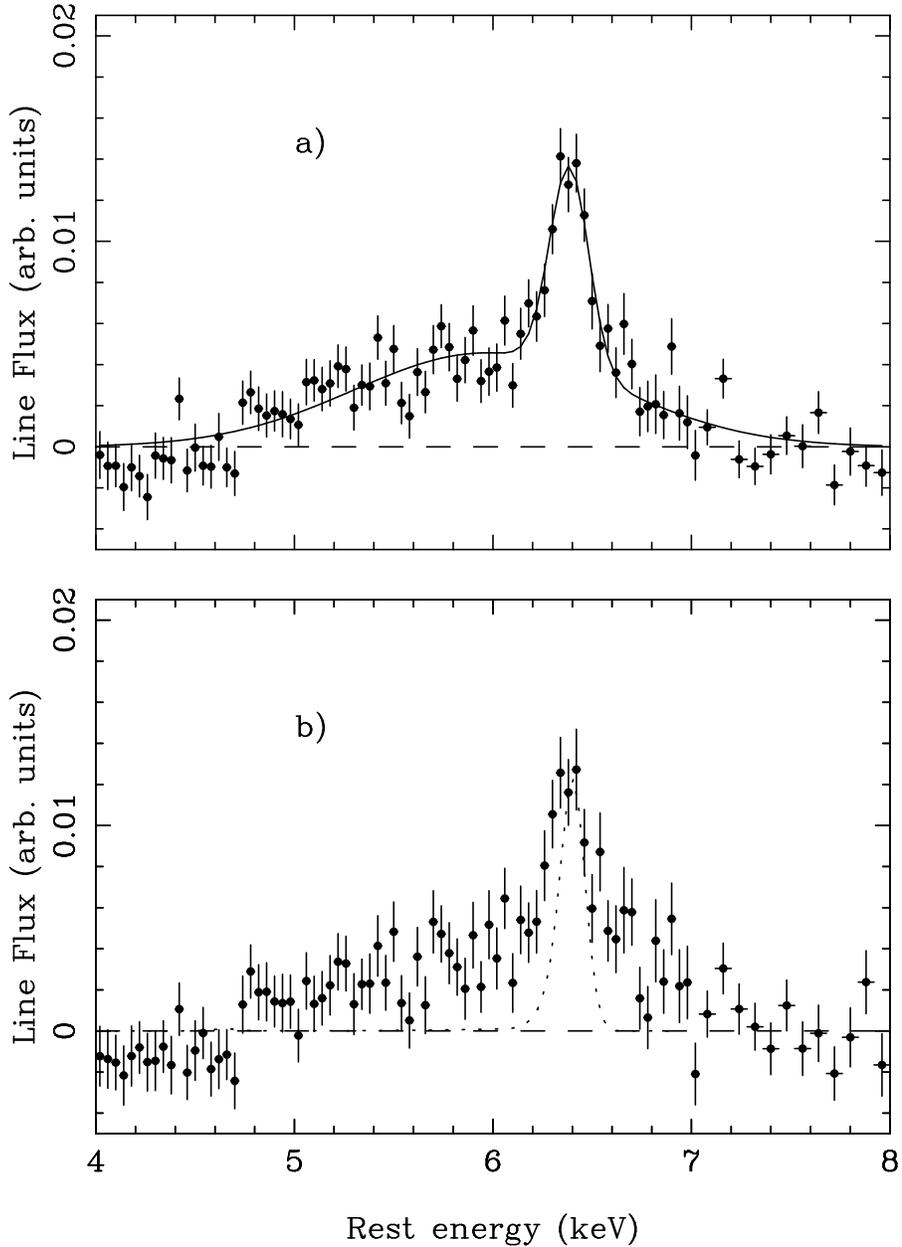}
\caption{Mean line profiles for the sample. These were produced by
transforming the data/model ratios for each sources into the
rest-frame of each sources, rebinning the resultant residuals, and
converting to flux space assuming the mean of the best-fit
continua. a) shows the results for the whole sample, while b) shows
the profile with the two previously-reported skewed lines, MCG-6-30-15
and NGC 4151, removed. The two plots are clearly remarkably
similar. The solid line in a) is a double-gaussian fit to the
profile. The narrower component peaks at 6.4~keV with a width of
$\sigma\sim0.1$~keV. The broader component is strongly redshifted,
with a centroid energy of 6.1~keV and a width of 0.7~keV.  It also
carries the bulk ($\sim 75$~per cent) of the flux. The dotted line in
b) is the spectral response of the ASCA SIS detectors at 6.4~keV.
\label{fig:mother}}
\end{figure}

\begin{figure}
\plotone{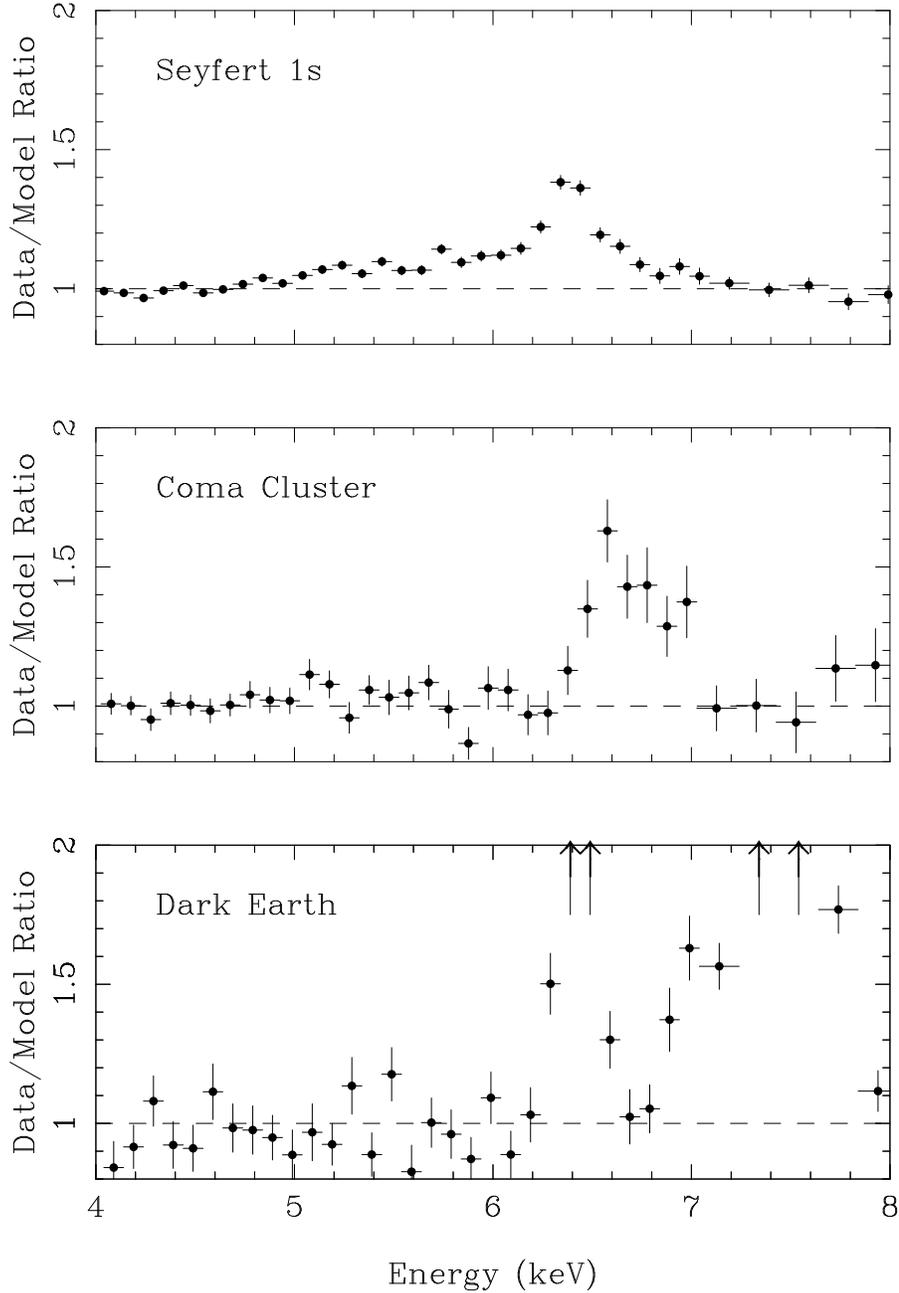}
\caption{Data/model residuals for a) our ASCA Seyfert 1 sample (top panel,
used to create Fig.~\ref{fig:mother}a); the Coma Cluster (middle panel). The
continuum here has been fitted with a bremsstrahlung model, excluding
the 5-7 keV region; and the Dark Earth. Here the continuum
has been modeled with a power-law excluding regions contaminated by lines,
the most prominent of which are iron K$\alpha$ and nickel K$\alpha$. 
Although all three panels show very clear evidence for emission lines,
only the Seyfert 1 ensemble shows a red wing.   
\label{fig:cal}}
\end{figure}

\begin{figure}
\plotone{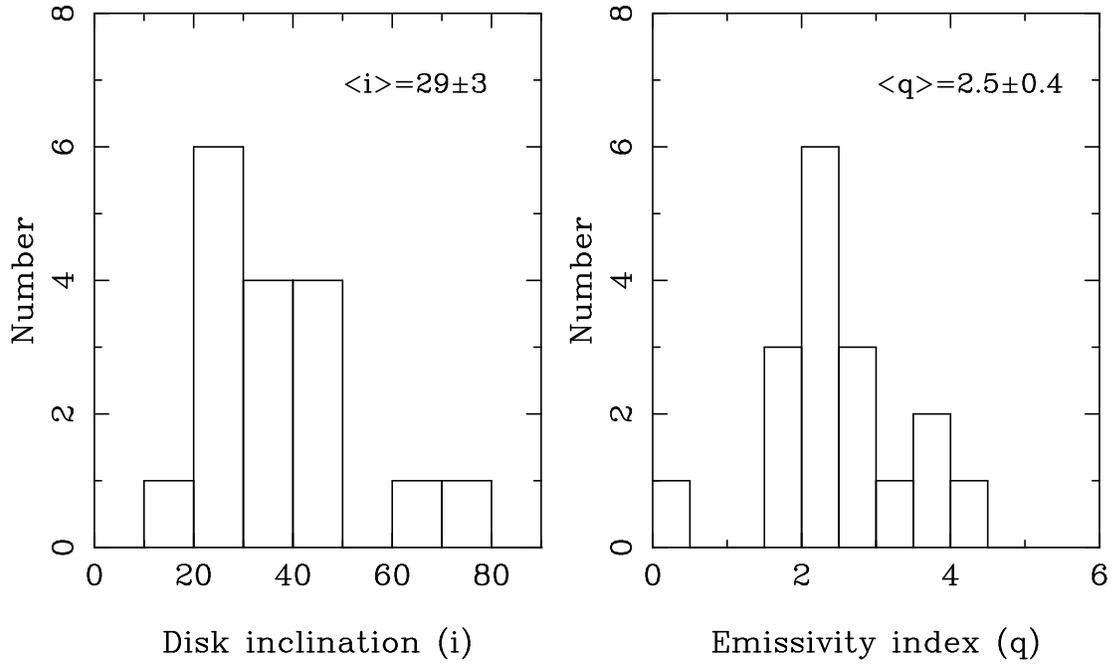}
\caption{Histograms of the parameters from the disk line
fits to the \asca\ data (see Table~\ref{tab:dl}). 
a) inclination $i$, b) index of the emissivity
function $q$ (see text). The spectra show a significant spread of
$q$, which parameterizes the geometry of the system, indicating
that this may change from source to source.}
\label{fig:dl}
\end{figure}

\begin{figure}
\plotone{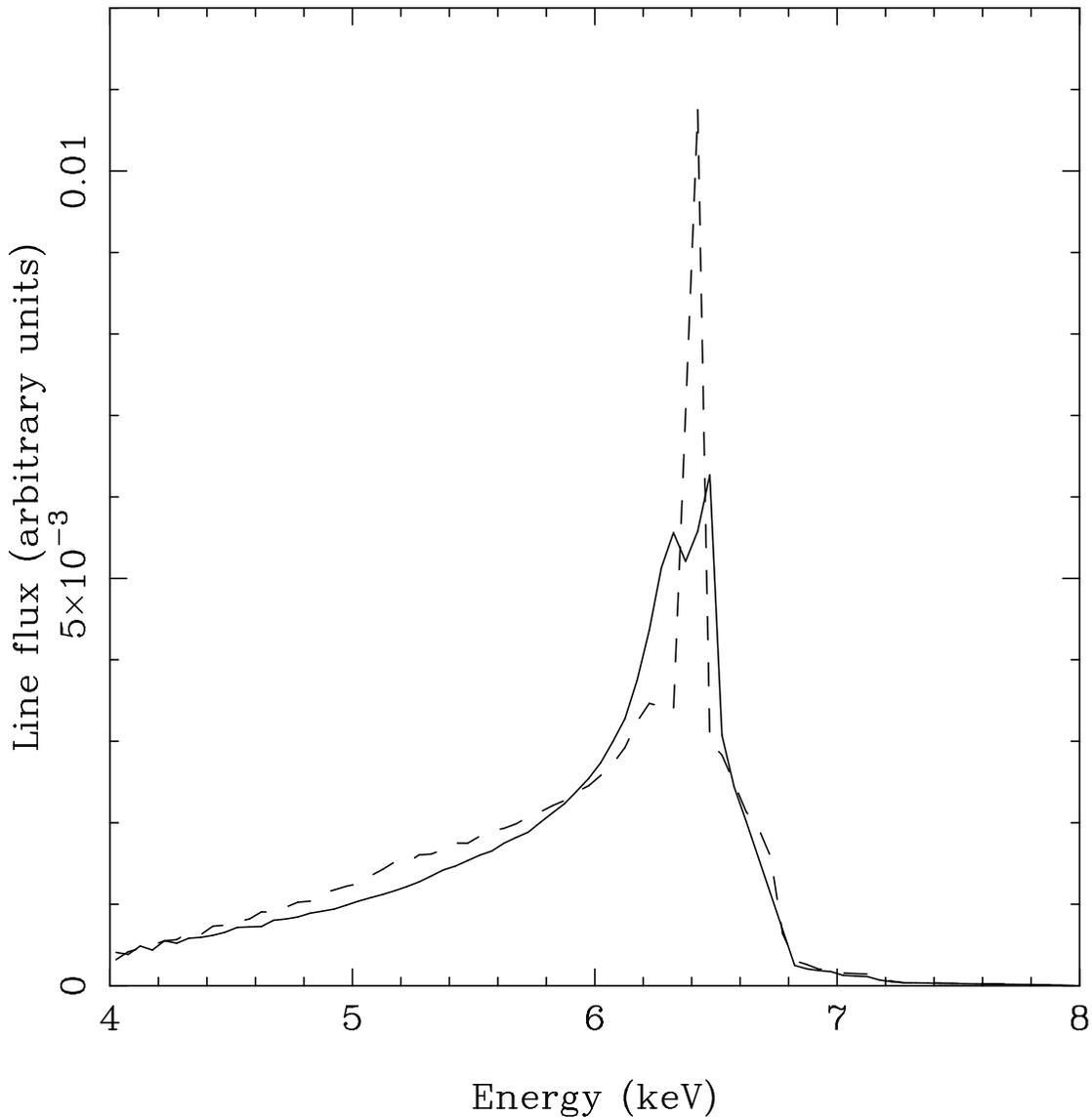}
\caption{Summed model profiles for the sample. The solid line shows the
pure disk-line model (Table~\ref{tab:dl}) which is very similar in shape
to the data shown in Fig~\ref{fig:mother}. The dashed line shows the effects
of a narrow component. 
With the
resolution of the ASCA detectors, it is difficult to state unequivocally
whether or not such a component is present.
However, including a narrow line
has little effect on the disk line profile. 
\label{fig:compmodels}}
\end{figure}

\begin{figure}
\plotone{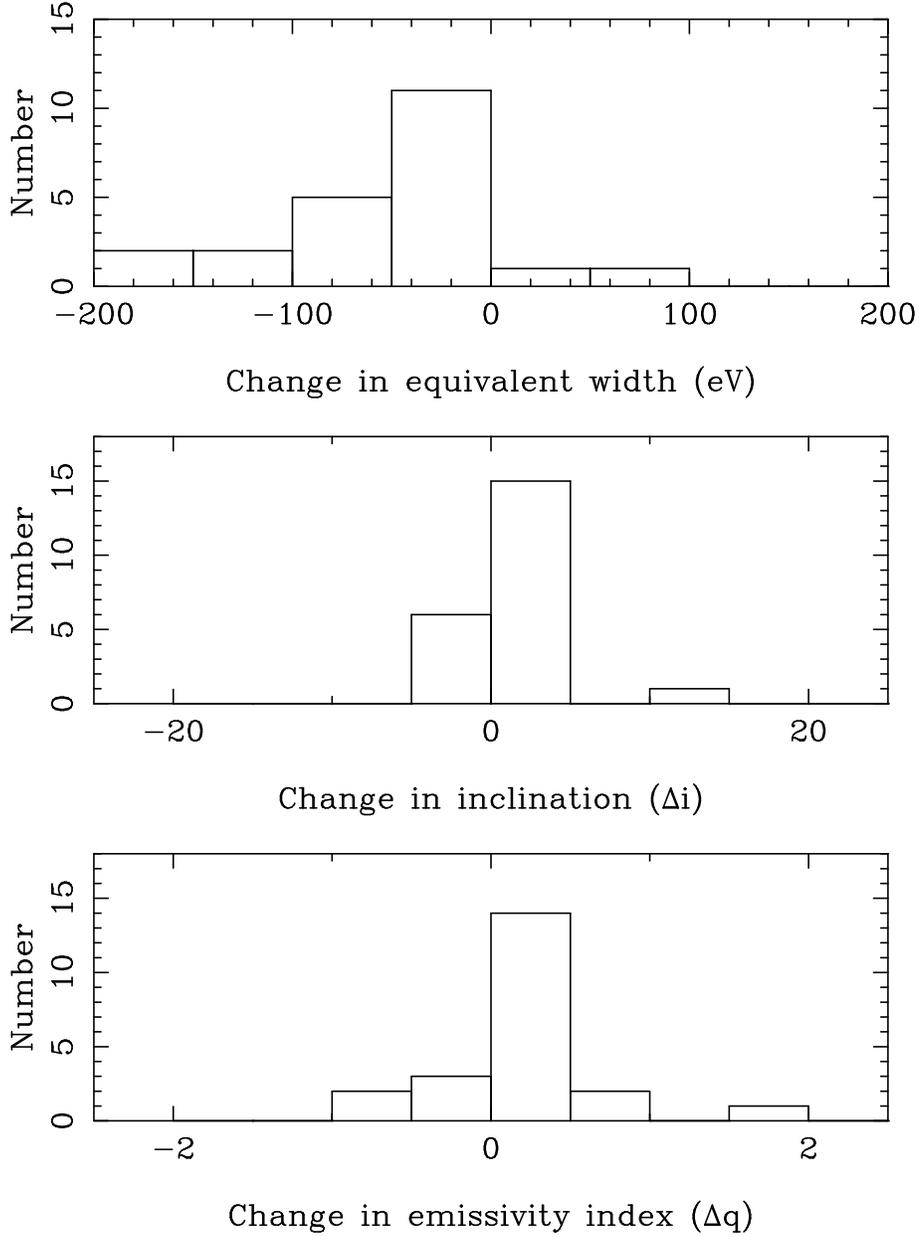}
\caption{Change in the disk line parameters when a narrow Gaussian,
with energy fixed at 6.4~keV, is introduced into the fit. There is a
small reduction in the equivalent width (top panel) corresponding
roughly to the equivalent width of the narrow line (with a mean
$\sim 30$~eV). $i$ (middle panel) is largely unchanged by the 
narrow component and the mean 
$q$ (bottom panel) increases insignificantly, to $2.8$.
\label{fig:dl-nl}}
\end{figure}

\clearpage
\begin{figure}
\plotone{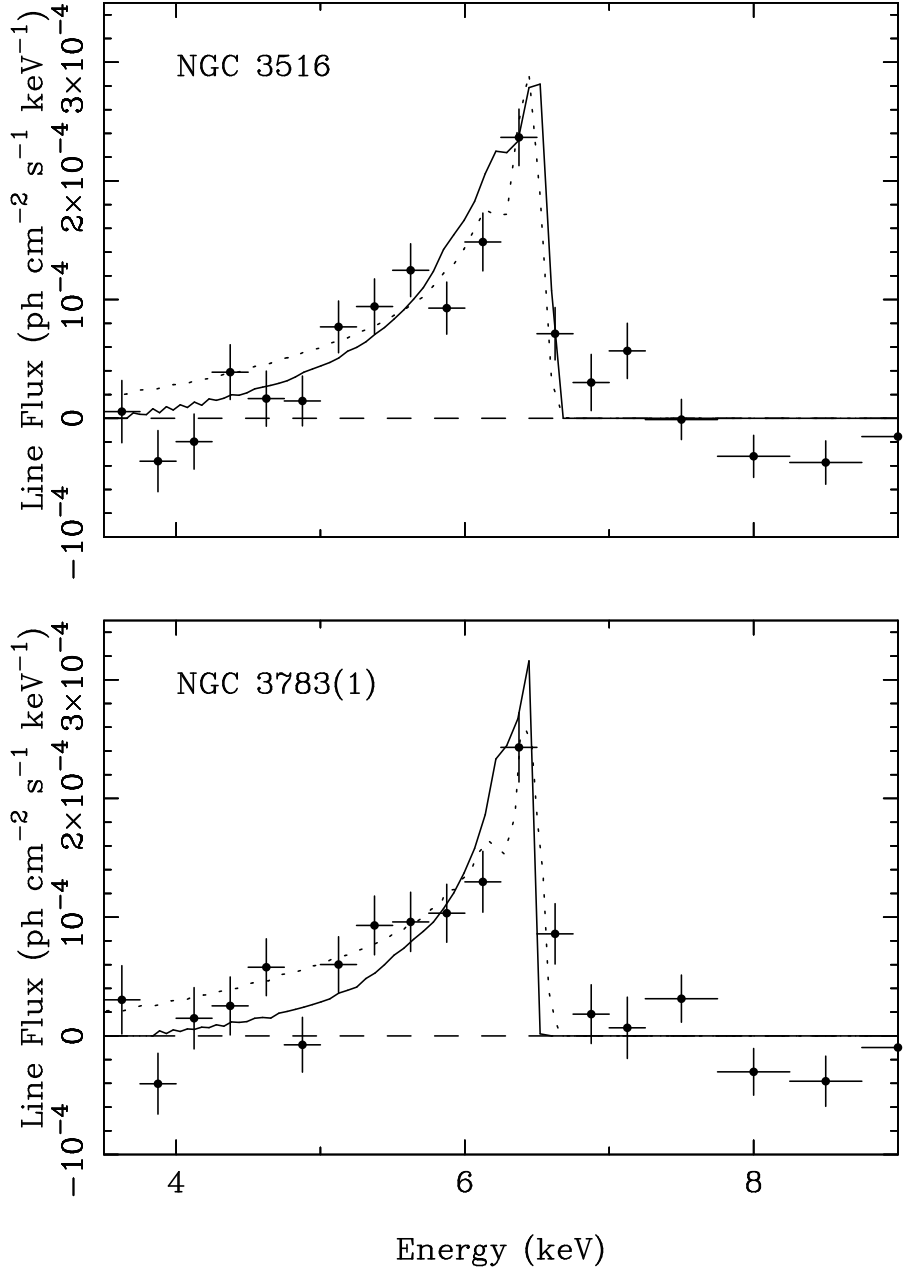}
\caption{Residual profiles for NGC 3516 (top panel) and NGC 3227 (bottom
panel). Data from all four instruments have been combined in bins
of 0.25~keV width. The best fit Schwarzschild (solid) and Kerr (dotted)
models are also shown.
\label{fig:model}}
\end{figure}

\begin{figure}
\plotone{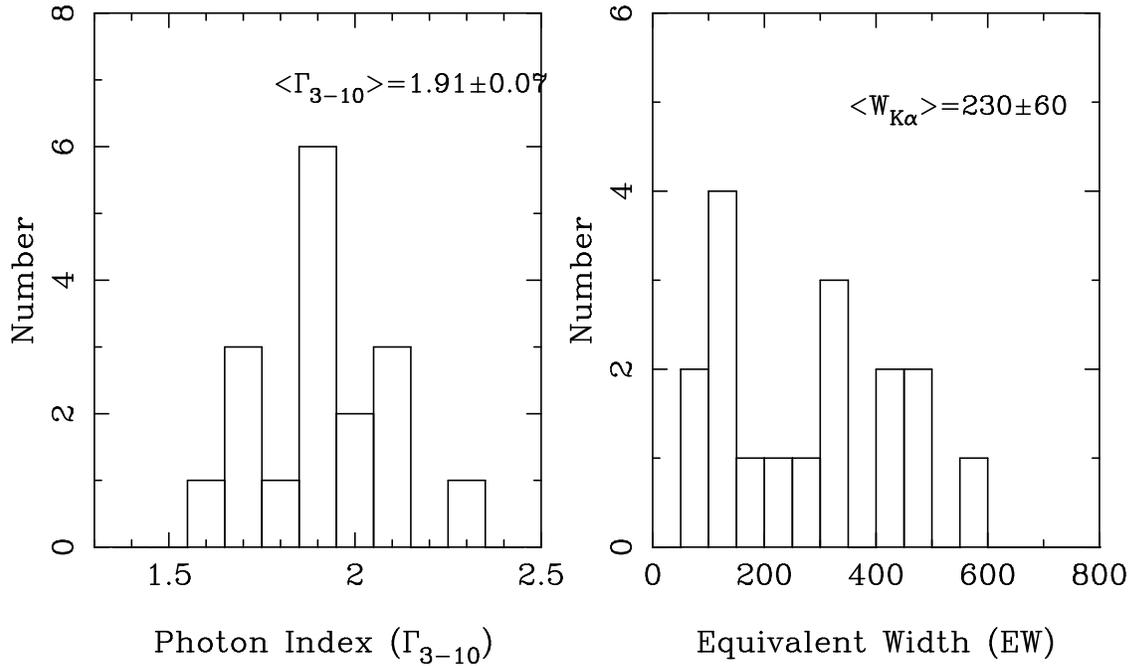}
\caption{Histograms of a) the photon index of the power law, 
$\Gamma_{3-10}$, and b) the line equivalent width, $W_{\rm K\alpha}$ from
disk line--plus--reflection fits to the \asca\ data (see 
Table~\ref{tab:dl-ref}). The inclination and emissivity index 
show negligible change when reflection is included. In contrast the
power-law index is steepened by $\Delta\Gamma=0.12$ and the mean 
equivalent width reduces 
by $\sim 50$~eV when the effects of reflection are accounted for.
\label{fig:ref}}
\end{figure}

\end{document}